\documentstyle[11pt,epsf]{article}
\def\lesssim{\le} 
\def\be{\begin{equation}}
\def\ee{\end{equation}}
\def\bea{\begin{eqnarray}}
\def\eea{\end{eqnarray}}

\begin{document}
\begin{titlepage}
\vspace*{-62pt} 
\begin{flushright}
CLNS 97/1512, NBI-HE-97-48, hep-ph/9709320\\
September 1997
\end{flushright}
\vspace{1cm}
\begin{center}
{\Large \bf Turning Around the Sphaleron Bound:

Electroweak Baryogenesis in an 

Alternative Post-inflationary Cosmology}\\
\vspace{1cm}
\normalsize
{Michael Joyce$^*$ and  Tomislav Prokopec$^\ddagger$
\vspace{1cm}

\centerline{$^*$ School of Mathematics, Trinity College, Dublin 
2, Ireland }
\vspace{.1cm}
\centerline{$^\ddagger$ 
Niels Bohr Institute, Blegdamsvej 17, DK-2100 Copenhagen, Denmark
}

\vspace{18pt}
}

\end{center}


\begin{quote} \hspace*{2em} 

\baselineskip 14pt

\noindent
\hspace{6.5cm} {\bf Abstract}

The usual sphaleron bound and the 
statement of the impossibility of baryon production at 
a second order phase transition or analytic cross-over are 
reformulated in the first part of the paper as requirements 
of the expansion rate of the Universe at the electroweak scale.  
With an (exact or effective) additional contribution to the energy 
density scaling as $1/a^6$, which dominates until just before 
nucleosynthesis, the observed baryon asymmetry may be produced at the
electroweak scale in simple extensions of the Minimal Standard Model,
even in the case that the phase transition is not first order.
We focus our attention on one such cosmology, in which the Universe 
goes through a period termed {\it kination} in which 
its energy is dominated by the kinetic energy of a scalar field.
The required kinetic energy dominated modes can occur either as
a field rolls down an exponential (or steeper) potential, or in
the oscillation of a field about the minimum of a steep power-law
potential. We implement in detail the former case with a single
exponential field first driving inflation, and then rolling into
a kinetic energy dominated mode. Reheating is achieved using
an alternative to the usual mechanism due to Spokoiny \cite{spokoiny},
in which the Universe is `reheated' by  particle creation in the 
expanding background. Density perturbations of the magnitude 
required for structure formation may also be generated. 
We show that the analogous model for the power-law potential cannot
be consistently implemented. In models with inflation driven by
a second field and the usual mechanism of reheating (by decay of the
inflaton) the required kinetic energy dominated cosmology is viable
in both types of potential.
\vspace{1cm}


\vspace{1cm}

\noindent
\small $^*$email: joyce@maths.tcd.ie

\small $^{\ddagger}$email: prokopec@nbi.dk

\end{quote}

%
\end{titlepage}
%

\section{Introduction}
\label{introduction}

\baselineskip 14pt

Nucleosynthesis provides a role model
for electroweak baryogenesis
to whose impressive heights it can still
only aspire. The great attraction of the idea
that the  baryon asymmetry of the Universe (BAU)
may have been created at the electroweak
epoch lies in the possibility that one day
an {\it ab initio}  calculation to rival
that of nucleosynthesis may be possible,
and that it will give a definitively
positive or negative answer. Rather than
simply providing an alternative to
scenarios for baryon creation at the GUT
scale, it has the fundamental interest
of relying on physics at a scale directly
accessible to experiments.
We can realistically hope to know the
correct theory of physics at the electroweak
scale, in particular the structure of the CP
violating and symmetry breaking sectors.
Just as in nucleosynthesis it is then
a question of putting this theory in an
expanding universe and finding the output.
Electroweak baryogenesis however faces more substantial
obstacles on the road to a reliable calculation than
did  nucleosynthesis, {\it e.g.\/}  
the determination  of the baryon asymmetry
involves all the details of departure from equilibrium at
the phase transition (if there is one), the crucial
baryon number violating processes arising from the
chiral anomaly at finite temperature involve many
difficult and still unresolved questions, {\it etc.} Much
progress has however been made, recently in particular
using lattice methods to study the phase transition
\cite{Mishareview},
and the problems seem not to be insurmountable.

The approach of this paper is somewhat orthogonal
to the direction of investigation of most work
on electroweak baryogenesis. Rather than investigating
some aspect of the particle physics, we consider
the cosmological side of the problem. The standard
and indeed most natural assumption about cosmology
at the electroweak epoch is that it is what one gets
by the simplest backward extrapolation from nucleosynthesis:
a homogeneous and isotropic radiation dominated universe.
In nucleosynthesis the assumption of such
a universe is relaxed to place limits on, for example,
the contribution of a magnetic field or of a
cosmological ``constant'' to the energy density.
In this paper we ask the analogous  question of
electroweak baryogenesis: how is the standard scenario
for production of the observed baryon asymmetry at the
electroweak epoch affected if we consider
cosmologies other than the standard one? And
are there simple alternative cosmologies which
lead to significantly different results for
electroweak baryogenesis?

The same sort of question has been previously addressed
in the context of calculations of the relic densities of
weakly interacting particles in work of  
Barrow \cite{Barrow} and
Kamionkowski and Turner (KT) \cite{KamionkowskiTurner}.
These relic densities depend on the temperature at
freeze-out which  occurs (approximately) when
the annihilation cross-section of the particular species
drops below the expansion rate of the Universe.
Barrow discussed the particular case of a non-anisotropic
universe, in which the average (volume) expansion rate
which determines the freeze-out has
an extra component driving it which scales as $1/a^6$
($a$ is the average scale factor).
Consistent with the requirement of
radiation domination at nucleosynthesis,
the expansion rate can thus be very much greater in the
anisotropic universe prior to nucleosynthesis
when dark matter relics typically freeze-out
($T \sim 100 MeV$), and the
requirement that such a particle be  the cosmological
dark matter may in principle place a bound on the
anisotropy. The important idea - that relic densities can
provide a probe of the Universe prior to nucleosynthesis,
which might be other than the standard radiation
dominated one  - was considered in a more general way
by KT, who discussed the case of
an anisotropic universe, as well as various others,
including a Brans-Dicke-Jordan theory of gravity.
In this latter case the effect  can  
also be modelled  as an extra contribution to the
energy density scaling as $1/a^6$, producing in the same
way a speeded up expansion rate before nucleosynthesis
without violating the nucleosynthesis constraints.
KT also mention an example
(which they describe as ``exotic'') of a
scalar field $\phi$ which oscillates in the minimum
of a potential $\phi^n$, for which the energy density
scales as $a^{-6n/(n+2)}$, {\it i.e.\/} faster than radiation
for $n>4$. 
Again the energy in such a mode can contribute
significantly before nucleosynthesis without disrupting the 
latter. As discussed in \cite{MJletter} the relevant
feature of this model is that it is the kinetic energy of
the scalar field which gives the dominant contribution to
the energy density of the Universe. As well as the oscillating
mode of the power-law potential, the scaling applies to a
scalar field rolling down a simple exponential potential.
Rather than being exotic (compared to the models which 
\cite{Barrow} and \cite{KamionkowskiTurner} focus on),
such models are minimal in the sense that they leave Einstein 
gravity intact and are consistent with the inflationary 
explanation of the homogeneity and isotropy of the Universe. 
In this paper we construct and study in detail a model for each of the
two cases in which the single field (exponential or power-law)
both inflates the Universe and then rolls into the kinetic
energy dominated mode. Reheating is achieved using a simple
alternative model of reheating proposed by Spokoiny several 
years ago \cite{spokoiny}. In the power-law potential density
perturbations are produced which are too large and the model
is not viable. Furthermore the coherent oscillating mode is 
unstable to decay due to parametric resonance. We also discuss
less constrained viable models in which the inflaton is a different 
field and reheating proceeds in the usual way (by decay of the condensate). 
The required potentials do in fact arise in many particle physics models:
Power-law potentials have been discussed, for example, in the context of 
supersymmetry motivated inflationary models in 
\cite{LazaridesShafi,BarreiroCopelandLythProkopec}.
(The lower order terms  can be excluded 
by imposing a discrete $Z_{n/2+1}$ symmetry 
on the superpotential). Exponential potentials arise quite generically 
in theories involving compactified dimensions, such as supergravity
and higher dimensional theories of gravity (for specific
examples, see \cite{exp-inflation1, exp-inflation2, halliwell-exp}).
The latter are also interesting in that they can play an important 
(potentially observable) role in the late-time cosmology of
structure formation \cite{pgfmj, vl}.

In the first part of the paper we address the question of how 
the expansion rate of the Universe affects the baryon asymmetry 
produced at the electroweak scale, without reference to any 
particular cosmological model.  
The two distinct cases -  a first order phase transition, and a
second-order phase transition or a cross-over - are
treated separately. In the first case the expansion rate
enters (i) in determining when the transition occurs, since
this depends on the cooling rate of the Universe below
the critical temperature, and (ii) in determining the
depletion of the baryon asymmetry produced by sphaleron
processes (and hence in determining the sphaleron bound).
In this case the baryon asymmetry is actually (at least
in certain extensions of the standard model) produced
on or near the bubble walls as they propagate through
the plasma, and does not depend directly on the
expansion rate. In the second case
the expansion rate is the sole parameter
which controls the departure from equilibrium
and the baryon asymmetry which can be generated depends
directly on it.

In the second part of the paper we turn to the discussion of 
alternative cosmologies, first reviewing those considered in
previous work and then turning to the detailed consideration of 
cosmologies dominated by the kinetic energy of a scalar field 
for a period between inflation and nucleosynthesis, 
concentrating in particular on the case  when  this 
phase (which, following \cite{MJletter} we term {\it{kination}}) 
persists until after the electroweak scale. 
In this case  the expansion rate at the
electroweak scale is increased, producing the
modifications to calculations for electroweak baryogenesis 
discussed in the first part.
As outlined above, the two types of models we consider are
exponential potentials
and power-law $\phi^n$ ($n>4$) potentials.
For both cases we first discuss a one field model,
in which the field both inflates the Universe and causes 
kination. Instead of decaying as in the standard explanation
of reheating, the inflaton rolls into a kinetic mode and
simply red-shifts away. The radiation created 
by the superluminal expansion  of the Universe 
at the transition between the two phases thermalizes and 
comes to dominate the energy density of the Universe
at a later time determined by the expansion rate at the end
of inflation \cite{spokoiny}. We show that in the exponential potential 
one can have (i) a transition to radiation domination as late as 
nucleosynthesis and (ii) thermalization of the radiation 
well prior to the electroweak scale, and further,
(iii) density perturbations of the right amplitude for structure formation. 
In the power-law potential, however, we find that the requirement that
the Universe become radiation dominated before nucleosynthesis 
leads to the production of density perturbations which are
much larger than is consistent with observations.
In any case the oscillating mode in this model typically 
decays non-perturbatively (through parametric resonance) unless
the self-couplings of the field are extremely small, and the
energy does not stay in the kinetic mode for long enough.
We conclude section 5 with a brief discussion of two fields 
models in which  one field is the inflaton and reheats the Universe 
in the standard way, and the second field is the 
`kinaton' which comes to dominate for a phase subsequent to inflation. 
In this case the second field can be either exponential or
power-law (provided the couplings are such that decay by parametric 
resonance does not occur until after nucleosynthesis).
In the last section we discuss the implications of our results 
for the testability of theories of electroweak baryogenesis, 
and consider briefly other ways in which pre-nucleosynthesis 
cosmology might be probed.

\section{Dependence on the Expansion Rate}

In electroweak cosmology the assumption is generally made
that the Universe is flat, homogeneous, isotropic, 
and radiation dominated.  
Hence all cosmological information is encoded
in the expansion rate $H_{\rm rad}$, which is given as a 
function 
of the plasma temperature $T$: 
\begin{equation}
H_{\rm rad}=h\frac {T^2}{M_{\rm P}} \,,\qquad 
h=\left ( \frac{\pi^2 g_*}{90}
 \right )^{\frac{1}{2}}
\label{radiation}
\end{equation}
where $g_* \sim 10^2$ 
is the number of relativistic degrees of freedom in 
the plasma and 
$M_{\rm P}=(8\pi G)^{-1/2}\simeq 2.4 \times 10^{18}\hbox{GeV}$ 
is 
the reduced Planck mass. 
The clean separation between the purely particle
physics and cosmological calculations occurs because of the
adiabaticity of the expansion. 
It is adiabatic because the  timescale characterizing the 
expansion -- 
$\tau_{\rm expansion}= M_{P}/(hT^2) \simeq 10^{16} 
(100\hbox{GeV}/T)T^{-1}$, 
taking  $g_*\simeq 10^2$ -- is much
greater than the timescales associated with the  
thermalization processes which have typical rates $\sim \alpha^2T$
(where $\alpha > {1/50}$ for all the 
interactions well above $100$GeV).
The phase transition can thus be studied using equilibrium 
methods
- the expansion of the Universe enters only in determining the 
cooling
rate  and, hence,  when the transition occurs (if it does).
In general we could of course consider any cosmology at
the electroweak scale, with the sole requirement
that it be consistent with nucleosynthesis. We have no
probe of the electroweak scale except that provided by
electroweak physics, and methodologically it makes
sense to ask how changing the standard assumption about
the Universe at this phase affects the predictions of the
remnants which result.  Here we limit ourselves to relaxing
only the assumption that the expansion rate is related to
the temperature by (\ref{radiation}). 
Instead we take 
\begin{equation}
H=H_{\rm ew} \left(\frac{T}{T_{\rm ew}}\right)^p
\label{tempdep}
\end{equation}
where $p$ is a number and the subscript `ew' means the
quantities are evaluated at some temperature characteristic of
the electroweak phase transition. 
Using $H=\dot{a}/a$ this corresponds to 
the time dependence $a \propto t^{1/p}$
for the scale factor $a$.  
All our results concerning baryogenesis are, we will see, 
essentially independent of $p$ because they depend only on 
temperatures very close to $T_{\rm ew}$.  
We will treat $H_{\rm ew}$ as a free parameter, 
only taking it to be such that the assumption
of adiabaticity is valid, which 
allows it to be different form the standard value by 
orders of magnitude. We will review in section 5 some
of the non-standard cosmologies which can be described by 
these assumptions. 
The model which we will discuss in detail is 
a homogeneous and isotropic universe dominated
by a kinetic mode of a scalar field rather than by radiation. 

Baryogenesis, the creation of baryons from an
initial zero baryon state, requires a departure from thermal
equilibrium. In the Big Bang Universe this is provided by the
expansion which causes the Universe to cool.
At the electroweak scale this cooling can lead to two
very different effects, depending on whether the electroweak
phase transition  is of first order or not.
Recall that at a first order phase transition as the Universe 
cools 
it becomes thermodynamically favourable for
the system to be in the ``broken'' state. 
The ``broken'' and ``unbroken'' phases are separated by a 
potential 
barrier which decreases as the Universe cools.
Once the barrier is low enough, the transition proceeds by the
nucleation and propagation of true vacuum bubbles.
This departure from equilibrium is
characterized by timescales 
which are much shorter
than that associated with the expansion. Almost all
proposed mechanisms for baryogenesis at the electroweak
scale make use of this dramatic departure from equilibrium,
using the interaction between the plasma and propagating walls
to generate the baryon asymmetry. In the case that the transition
is second order or cross-over there is no such effect. Everything
evolves continuously and the departure from equilibrium is 
controlled directly by the expansion rate. This usually 
leads one to conclude that anything but a first order phase
transition is inimical to baryogenesis at the electroweak scale. 
Once one relaxes the assumption that the expansion rate is its 
standard radiation dominated value, this conclusion does not 
follow and needs to be examined more carefully. We will thus 
treat these two cases separately in some detail.

\section{First Order Phase Transition}

In this case the baryons are created as the bubbles
of the true vacuum propagate through the false vacuum.
The net effect of the propagation of the bubble
through the medium in all proposed mechanisms
is the creation of a flux of baryons into the broken phase.
The expansion rate enters only indirectly through
other parameters involved in this calculation - 
through the temperature at which the transition occurs which
it determines, and, in certain regimes, through the bubble
wall velocity. On the other hand, it enters directly 
in the determination of the amount of the created asymmetry which 
survives once it is in the broken phase. We consider these
two dependences separately.

\subsection{Bubble Nucleation}

In this section we investigate how the 
bubble nucleation temperature depends on the expansion
rate of the Universe. We also briefly discuss how the 
bubble wall velocity may depend on this parameter.   
As the Universe supercools below the critical temperature 
$T_{\rm c}$, the fraction $f$ filled by nucleated bubbles
at a time $t$ is given by (see \cite{Mishareview} and
references therein):
\begin{equation}
f(t)=1- e^{-\Delta(t)}\,,\qquad
\Delta(t)=\int_{t_{\rm c}}^{t} dt' 
\frac{4 \pi}{3} v^3 (t-t')^3 {\cal R}(t')\,,
\quad {\cal R}=I_0 T^4 {\rm e}^{-S_{\rm b}/T}\,,
\label{eq: nucleation rate}
\end{equation}
where $S_{\rm b}$ is the bounce action, 
${\cal R}$ the nucleation rate per unit volume, 
$I_0$ is a prefactor which is
a slowly varying function of temperature
of order  one, given in more detail 
below, and $v$ is the bubble wall velocity.
Changing the integration variable to 
$x=(T_{\rm c}-T)/T_{\rm c}$ and using
the time-temperature relation $t \propto T^{-p}$
which follows from (\ref{tempdep}), one finds
\begin{equation}
\Delta(x)=\frac{4 \pi v^3}{3}I_0 
\left(\frac{T_{\rm c}}{H_{\rm c}}\right)^4
\!\int_{0}^{x} dx^\prime (1-x^\prime)^{3-p}
\left(\frac{1}{p(1-x)^p}-\frac{1}{p(1-x^\prime)^p}\right)^3 
\exp\left(
-\frac{S_{\rm b}(x^\prime)}{T_{\rm c}(1-x^\prime)}
\right)
\label{eq: Delta(x) I}
\end{equation}
where $H_{\rm c}$ is the expansion rate at $T_{\rm c}$.
We will see below that the nucleation temperature $T_{\rm nucl}$
defined by $\Delta(x_{\rm nucl})=1$ is always very close to the 
critical temperature 
so that we can take $0<x^\prime\le x\ll 1$ and expand to
linear order in (\ref{eq: Delta(x) I}) to get
\begin{equation}
\Delta(x)=\frac{4 \pi v^3}{3}I_0
\left(\frac{T_{\rm c}}{H_{\rm c}}\right)^4
\int_{0}^{x} dx^\prime 
\left(x-x^\prime\right)^3 
\exp\left(-\frac{S_{\rm b}(x^\prime)}{T_{\rm c}}\right)\,.
\label{eq: Delta(x) II}
\end{equation}
Keeping the first term in a derivative expansion of the bounce
action about $x$, {\it i.e.\/}
taking $S_{\rm b}(x^\prime)=S_{\rm b}(x)+
(dS_{\rm b}(x)/dx)(x^\prime-x)+{\cal O}(x^\prime-x)^2$,
the integral can be performed with the assumption that
$xd(S_{\rm b}/T)/dx (x\approx x_{\rm c}) \gg 1$, and gives the 
nucleation temperature implicitly as 
\begin{equation}
\frac{S_{\rm b}(T_{\rm nucl})}{T_{\rm c}}=
\ln\left[8\pi v^3I_0
\frac{(T_ {\rm c}/H_{\rm rad})^4}{(dS_{\rm b}/dT_{\rm nucl})^4}
\,\right]
-4\ln\frac{H_{\rm c}}{H_{\rm rad}}\,,
\label{eq: bounce action}
\end{equation}
where 
$H_{\rm rad}\simeq 1.2 \times 10^{-16} (T_{\rm c}/100\hbox{GeV})
T_{\rm c}$ is the expansion rate at 
$T_{\rm c}$ in the standard radiation dominated cosmology.

To check the consistency of our assumptions and 
evaluate this expression to give the nucleation temperature,
one must calculate the bounce action near the critical 
temperature in the particular model of interest. We consider 
the Minimal Standard Model (MSM) in the regime where it is 
described by the effective potential
\begin{equation}
V(\phi,T)=\frac{\gamma}{2}(T^2-T_0^2)\phi^2
-\frac{\alpha}{3}T\phi^3+\frac{\lambda_T}{4}\phi^4
\label{eq: effective potential}
\end{equation}
with the one-loop ring improved values 
\cite{DineLeighHuetLindeLinde,MooreProkopec} 
\begin{eqnarray}
\alpha &=& \frac{1}{2\pi}\frac{2m_W^3+m_Z^3}{v_0^3}
+\frac{1}{4\pi}\left(3+3^{\frac{3}{2}}\right)
\lambda_T^{\frac{3}{2}}\,,
\nonumber\\
\gamma &=& \frac{2m_W^2+m_Z^2+2m_t^2}{4 v_0^2}+
\frac{1}{2}\lambda_T\,,
\nonumber\\
\lambda_T &=& \frac{m_H^2}{2v_0^2}-
\frac{3}{16\pi^2 v_0^4}\left(
2m_W^4\ln\frac{m_W^2}{a_B T^2}+
m_Z^4\ln\frac{m_Z^2}{a_B T^2}-
4m_t^4\ln\frac{m_t^2}{a_F T^2}
\right)\,,
\nonumber\\
T_0^2 &=& \frac{m_H^2+8\beta v_0^2}{2\gamma}\,,\qquad
\beta=\frac{3}{64\pi^2v_0^4}
\left(4m_t^4-2m_W^4-m_Z^4\right)\,,
\label{eq: effective potential II}
\end{eqnarray}
where $v_0=246\hbox{\rm GeV}$, 
$a_B=(4\pi)^2{\rm e}^{-2\gamma_E}\simeq 50$, 
$a_F=(\pi)^2{\rm e}^{-2\gamma_E}\simeq 3.1$, and  
$\gamma_E$ is Euler's constant.
This treatment of the MSM is reasonably accurate up to  
$m_H \sim 60\hbox{GeV}$, 
when nonperturbative effects become important. 
With this effective potential the critical temperature $T_{\rm c}$
is given by 
\begin{equation}
T_{\rm c}=\frac{T_0}
{\left(1-\frac{2}{9}\frac{\alpha^2}{\lambda_T\gamma}\right)
^{1/2}}
\,,
\label{eq: critical temperature}
\end{equation}
and the latent heat $L$ and surface tension $\sigma$ by 
\begin{equation}
L=V(\phi_*,T)+T\frac{dV(\phi_*,T)}{dT}\,,\qquad
\sigma=\int_0^{\phi_*}d\phi \sqrt{2V}\,,
\label{eq: L and sigma}
\end{equation}
where $\phi_*$, defined by degeneracy of the minima ($V(\phi_*,T)=V(0,T)$),
is 
\begin{equation}
\frac{\phi_*}{T}=
\frac{2\alpha-\left[4\alpha^2-18\lambda_T\gamma(1-(T_0/T)^2) 
\right]^{1/2}}{3\lambda_T}\,.
\label{eq: phi_*}
\end{equation}
The (spherical) bounce action is given by
\begin{equation}
S_{\rm b}=4\pi\int r^2dr \left[
\frac{1}{2}\left(\frac{d\phi}{dr}\right)^2
+V(\phi,T)\right]
\label{eq: bounce action II}
\end{equation}
with the boundary conditions
$\phi(r=0)=\phi_*$, $d\phi/dr(r=0)=0$, and 
$\phi(r=\infty)=0$  
($r=|\vec x|$ is the radial coordinate). 
Rather than solving this exactly (which is numerically 
expensive),
or in the thin wall approximation (which is inaccurate for 
strong phase transitions when $\phi_*\sim T$),
we will use an approximation for $S_{\rm b}$ developed in 
\cite{DineLeighHuetLindeLinde}: 
\begin{eqnarray}
\frac{S_{\rm b}(T)}{T} &=& 9\pi\frac{\gamma^{3/2}}{\alpha^2}
\left[1-\left(\frac{T_0}{T}\right)^2\,\right]^{\frac{3}{2}}
f({\cal A})
\,,\nonumber \\
f({\cal A}) &=&1+\frac{{\cal A}}{4}\left[
1+\frac{2.4}{1-{\cal A}}+\frac{0.26}{(1-{\cal A})^2}
\,\right]\,,\quad
{\cal A}=\frac{1-(T_0/T)^2}{1-(T_0/T_{\rm c})^2}\,,
\quad 
\label{eq: bounce action approx}
\end{eqnarray}
which is valid for $0\le {\cal A}\le 0.95$. 

The prefactor $I_0$ in (\ref{eq: nucleation rate}) 
can be written as 
\begin{equation}
I_0=\frac{\kappa_{\rm dyn}}{2\pi}\left(\frac{S_{\rm b}}{2\pi}
\right)^{\frac{3}{2}}
\lambda_-^{-1/2}{\cal K}_{\rm bubble}^{-1/2}
\,,\qquad
\kappa_{\rm dyn}=\left[
\frac{2\sigma\rho}{R^3 L^2}
\right]^{\frac{1}{2}}
\label{eq: I_0}
\end{equation}
where $\kappa_{\rm dyn}$ is the dynamical prefactor as given 
in \cite{RuggeriFriedman}. $R$ is the radius of the 
nucleating bubble, which can be  
estimated in the thin wall approximation to be 
$R\simeq R_{\rm nucl}\sim 2\sigma/Lx_{\rm nucl}$,
$x_{\rm nucl}=1-T_{\rm nucl}/T_{\rm c}$, 
$\rho=\pi^2T^4 g_*/30$ is the energy density of the plasma 
with $g_*$ relativistic degrees of freedom.
The one loop fluctuation determinant consists of the `negative' mode  
$\lambda_-\simeq 0.05 g^{1/2} \phi(T)$, and we take 
${\cal K}_{\rm bubble}=1$
(for a more accurate value, see \cite{Baacke,KripfganzLaserSchmidt}).

We have solved  (\ref{eq: bounce action}) numerically to 
find the nucleation temperature $T_{\rm nucl}$, using
these values and approximations. We also used  
$m_t=175$GeV, $m_W=81$GeV, $m_Z=91$GeV, and 
took the bubble wall velocity $v = 0.4$ \cite{MooreProkopec}
\footnote{This is a friction limited upper bound. 
The results here
are of course not very sensitive to the details
of the prefactor in the nucleation rate.
As we will discuss below, this assumes that that the
Universe is not reheated significantly by latent heat release,
which is a reasonable approximation in the MSM.
In the case that significant reheating 
occurs, so that bubble nucleation stops, 
the correct value of $v$ in the early stages of nucleation
would be the speed of sound $v_s\approx 1/\sqrt{3}$ at 
which the shock fronts propagate. 
}.
\begin{figure}[htb]
\epsfxsize=5.5in
\centerline{\epsfbox{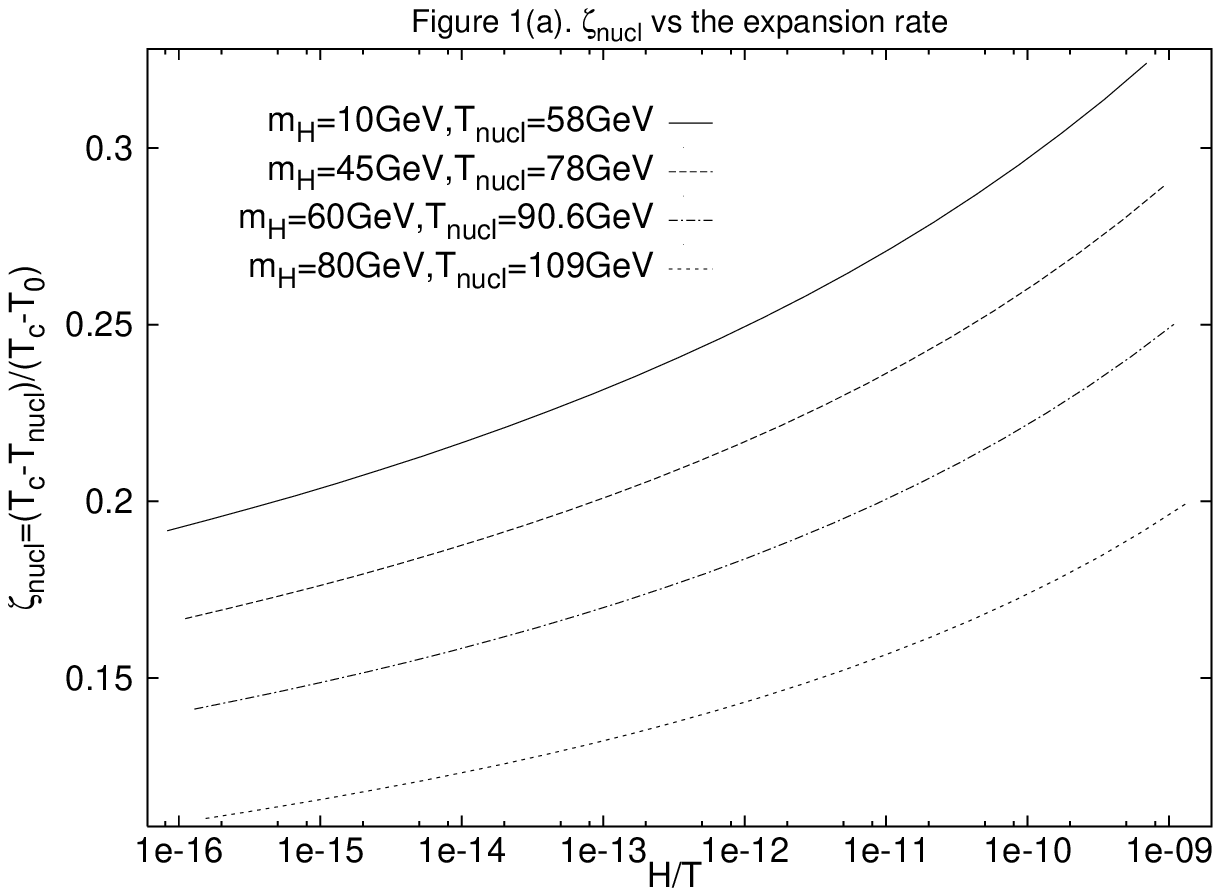}}
\end{figure}
In  Figure 1(a) we show a plot of 
$\zeta_{\rm nucl}=(T_{\rm c}-T_{\rm nucl})/(T_{\rm c}-T_0)$
against the logarithm of the expansion rate $H_{\rm c}$ 
at the critical temperature, 
for a range of Higgs masses $m_H$. 
We see clearly that the usual result in a radiation dominated universe
is qualitatively unchanged by varying the expansion rate over
orders of magnitude: $\zeta_{\rm c}$ is small, so the nucleation
temperature is very close to the critical temperature,
and much closer to $T_{\rm c}$ than $T_0$. There is a small
quantitative change,  $\zeta_{\rm c}$ varying by about $40\%$
as the expansion rate varies over five orders of magnitude,
but the change in absolute terms is tiny since
$\Delta T_{\rm nucl}=-(T_{\rm c}-T_0)\Delta \zeta_{\rm c} \le 5
\times 10^{-3}T_{\rm nucl}$.
Typically there is about the same change to $\zeta_{\rm c}$ over
this range at a fixed Higgs mass as is brought about by 
decreasing the Higgs mass by about $25$GeV in a radiation 
dominated universe\footnote{ $\zeta_{\rm nucl}$ 
increases as the Higgs mass $m_H$
decreases because the phase transition gets stronger, and
therefore more supercooling occurs.}.

The results are easy to understand both quantitatively and qualitatively.
Varying the bounce action (\ref{eq: bounce action approx}) and assuming 
$1-{\cal A}\lesssim 0.2$ (which is satisfied for most of the 
parameter space in figure 1(a)) so that the last term in the expression
dominates, one obtains 
$\delta(S_{\rm b}/T)\sim -2(\delta\zeta_{\rm nucl}/\zeta_{\rm nucl})
(S_{\rm b}/T)$\footnote{This estimate is just that obtained in the 
thin wall approximation in which 
$S_{\rm b\,, thin}\propto \zeta_{\rm nucl}^{-2}$.}, and hence 
(using (\ref{eq: bounce action}))
\begin{equation}
\frac{\Delta \zeta_{\rm nucl}}{\zeta_{\rm nucl}} \approx   
\frac{2\ln(H_{\rm c}/H_{\rm rad})}
{S_{\rm b}/T}\,. 
\label{eq: T_nucl correction}
\end{equation}
The bounce action $S_{\rm b}$  approximately halves,
going from $100T$ to $50T$ as the expansion rate
changes from $H_{\rm rad} \rightarrow 10^5 H_{\rm rad}$.
Taking an average value for  it in (\ref{eq: T_nucl correction})
gives good agreement with the estimates we made
above from the figure. Qualitatively the reason the
expansion rate changes the nucleation temperature 
so little is that,
as long as the Universe expands on a time-scale much
longer than $T_{\rm nucl}^{-1}$, the transition
always proceeds when the nucleation rate is very
suppressed, where the bounce action is an
extremely sensitive function of temperature.
The nucleation temperature decreases as the expansion
rate increases because the Universe must supercool more
to attain a less suppressed nucleation rate. 

Such a small change to the nucleation temperature 
leads to minor changes to the quantities which determine the
baryon asymmetry generated. We will see in section 3.2 
that between $T_c$ and $T_0$ the VEV of the Higgs field
changes by $50\%$, and its derivative with respect to $T$ 
by about a factor of three.
 From figure 1(a) this would mean an 
increase in the VEV at nucleation of $1\%$ or a little
more per order of magnitude increase in the expansion
rate. We would expect that this  result
will hold true in any typical electroweak
model and not just the MSM in the regime we have studied
it here. These minor changes to the VEV(s) and
the other macroscopic parameters which determine
the baryon asymmetry (bubble wall thickness, profile {\it etc.\/})
are essentially negligible in their effect on the
baryon asymmetry generated.

One condition must be attached to this conclusion, however: 
Other macroscopic effects can come into play as the bubbles propagate.
When the propagating bubbles
begin ``bathing'' in the hydrodynamic shock waves of the 
neighboring bubbles, {\it i.e.\/} when $\Delta(t)\sim 1$
in (\ref{eq: nucleation rate}), the plasma can heat up
and slow down the propagation 
of  the bubbles \cite{Heckler}. 
To determine how big this effect can be one compares
the latent heat release $L$
with the difference in the thermal energy density 
$\Delta \rho =4\rho(T_{\rm c}-T_{\rm nucl})/T_{\rm c}$ between
the nucleated phase and the unbroken phase. 
If $L /\Delta \rho \geq 1$ the system can reheat all
the way back up to $T_{\rm c}$. If such reheating occurs
the main effect on baryon generation at the bubble walls
is through the slowing down of the bubble walls.

\begin{figure}[htb]
\epsfxsize=5.5in
\centerline{\epsfbox{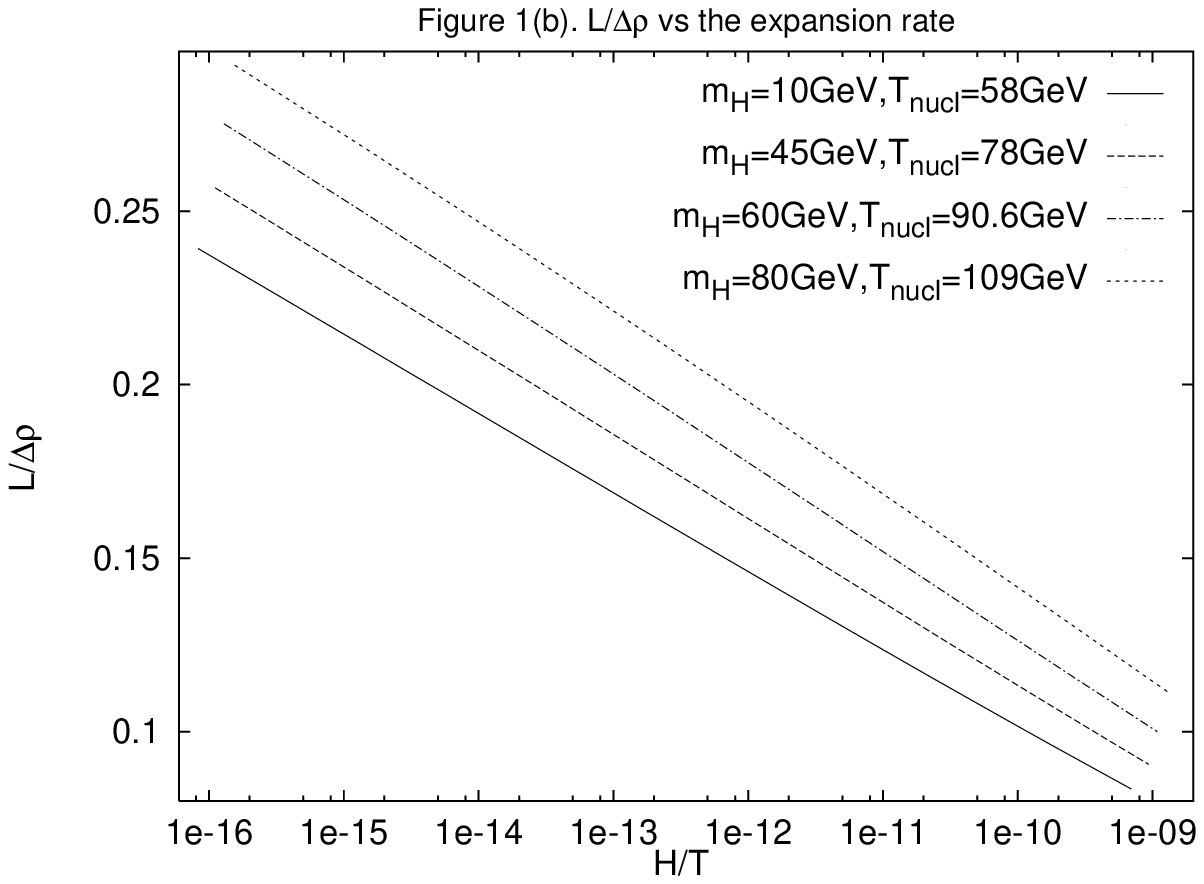}}
\end{figure}

In figure 1(b) we show a plot of this ratio $L/\Delta \rho$
as a function of the expansion rate in the MSM
with the same values and range of Higgs mass
as above. As the expansion rate increases 
the amount of re-heating  decreases - simply 
a result of the increased supercooling. 
In all the parameter space the ratio is less
than one, but of order one, so the effect of 
reheating may be significant, leading potentially
to a slow-down of the bubble walls relative to the
friction dominated regime. To determine the effect
precisely is a very involved problem, and we will
not attempt to tackle here the even more involved
one of looking at the effect on such slow-down of 
changing the expansion rate. We limit ourselves
to the qualitative observation that the amount of 
reheating decreases slightly as the expansion rate
increases, and that modifying the expansion rate by
orders of magnitude is likely to lead to only 
small corrections.

The one-loop effective action  which we use here
becomes an increasingly poor approximation for $m_H > 60$GeV,
and the latent heat is one of the quantities it estimates
very inaccurately. Lattice studies \cite{Mishareview}
have shown that there is a regime in the MSM where 
$L/ \Delta \rho > 1$. 
In other models also this is certainly a condition we 
can envisage being satisfied. As mentioned above,
in this case the bubble walls slow down to a 
final velocity determined by the expansion rate of the Universe, 
which has been estimated to be  
$v\sim 10^{-2} - 10^{-3}$ \cite{Heckler,Mishareview} 
in a radiation dominated universe. These estimates show
that this velocity  depends on the expansion rate only through the
combination $HR$ where $R$ is the average bubble radius,
which is  the radius at  which $\delta f/\delta R$
of (\ref{eq: nucleation rate})  peaks, {\it i.e.\/} 
$d^2\Delta / dR^2\approx 0$. A little algebra gives 
\begin{equation}
HR\simeq \frac{3 v_s}{|dS/dT|}\,,
\end{equation} 
where $v_s=1/\sqrt{3}$ is the speed of sound. 
Now using the expression (\ref{eq: bounce action}) above we
see that  $|dS_{\rm b}/dT|\propto S_{\rm b}/T$ decreases by 
about $10\%$ -- and hence  
the final wall velocity increases by the same amount --   
per order of magnitude change in the expansion rate. 
If $L/\Delta \rho > 1 $ for different expansion rates, this
will be the only change to the final bubble velocity.
Given the behaviour we have observed of $L/\Delta \rho$ it is clear 
that in certain models increasing the expansion rate considerably relative 
to its radiation dominated value 
could make the difference between this final wall velocity and a
(typically much larger) friction limited one corresponding to 
$L/\Delta \rho < 1$. In this particular case the change to the 
bubble wall velocity may not be so small.

The determination of how the baryon asymmetry
would be affected by such changes in the velocity of the 
bubble wall is a model dependent problem. The velocity 
dependence of the baryon asymmetry depends on what precise 
mechanism is operative, which depends on both the microscopic
and macroscopic physics. At low velocities ($v < 10^{-3}$)
the result always goes to zero at least linearly, and
at larger velocities the most sensitive dependence is
$\sim 1/v^2$. Using this dependence and assuming the greatest
possible effect due to a change in the expansion rate 
(from an upper bound in the friction dominated regime
$v \sim 0.4$ to the lower bound of the adiabatic `complete reheat' 
regime with $v \sim 10^{-3}$) would give a change in the
calculated BAU by (at most) $10^5$. As we have discussed however,
in most models the change will be much smaller and  
probably very small. 
A more detailed investigation of
this question would however be required to draw stronger
quantitative conclusions. 

\subsection{Washout and the Sphaleron Bound}

The baryons created at the bubble walls are
subject to decay after they enter the broken phase
if the baryon number violating processes are not sufficiently
suppressed. The requirement that this attenuation not
reduce the created asymmetry to less than 
that required for nucleosynthesis
leads to the sphaleron bound \cite{mishasphal} 
in a radiation dominated universe. 
Here this bound will be restated as a requirement of 
the expansion rate of the Universe in a given theory.  
In the course of this discussion we will also
draw attention to imprecisions in commonplace
statements of the sphaleron bound (with the usual
assumption of radiation domination) which can be
of considerable importance. 
 
Since the time-scale associated with the baryon number
violating sphaleron processes is much longer than the
time-scale for thermalization processes, the baryon
number after the completion of the electroweak phase transition
is given by its local equilibrium value
\begin{eqnarray}
\langle B\rangle=\frac{1}{Z} 
Tr\left[ B{\rm e}^{-\beta (H-\mu_B B- \sum_A \mu_A Q_A)}\right]
\label{eq: B}
\end{eqnarray}
where $\mu_B$ and $\mu_A$ are chemical potentials for 
baryon number and the other charges $Q_A$  
conserved on the relevant time-scale, {\it i.e.\/}
exactly conserved, or violated at a  rate slower than 
the baryon number violating processes. On the time scale over
which the violation of baryon number is relevant, the system
relaxes to equilibrium at a rate 
$\dot{B}= -\bar\Gamma_{\rm sph} (\Delta F/T) \Delta B$, 
where $\bar\Gamma_{\rm sph}$  is the rate  per
unit volume
of sphaleron processes in which the Chern-Simons number 
$N_{\rm cs}$ changes 
by {\it one\/} unit, and $\Delta F$ is the free energy
change per process. Since 
$\Delta B= N_F \Delta N_{\rm cs}$ per 
process ($N_F$ is  the number of fermion families), 
we get 
\begin{equation}
\dot B=-N_F^2 \frac{\bar \Gamma_{\rm sph}}{T} \mu_B
\label{eq: dot B}
\end{equation}
where we assume that other charges are defined so
that they are conserved in these processes ($\Delta Q_A=0$).
The sphaleron rate is given by \cite{ArnoldMcLerran} 
\begin{equation}
\bar\Gamma_{\rm sph} =
{\cal C}\; g\frac{\phi^7}{T^3} 
{\rm e}^{-\frac{E_{\rm sph}}{T}}\,,
\quad E_{\rm sph}=
{\cal B}\left(\frac{4\pi}{\alpha_w}\right)^{\frac{1}{2}}
\phi(T) 
\,,\quad
{\cal C}=\frac{\omega_-}{2\pi\,g\phi(T)}
{\cal{N}}_{\rm tr}{\cal{N}}_{\rm rot}{\cal{V}}_{\rm rot}
{\cal K}_{\rm sph}
\label{eq: sphalrate}
\end{equation}
where $\alpha_w=g^2/4\pi \approx 1/29$, 
${\cal B}$ is a monotonically increasing function of 
$\lambda/g^2=m_H^2/8m_W^2$ ranging between 1.5 and 2.7 as 
$\lambda/g^2$ varies from 0 to $\infty$ 
\cite{KlinkhamerManton} and $\cal C$ 
is a temperature independent `constant', given
fully below (with ${\omega_-}\sim g\phi(T)$ the frequency 
of the negative mode of the sphaleron,  
${\cal K}_{\rm sph}$ the one loop fluctuation 
determinant, ${\cal{V}}_{\rm rot}=8\pi^2$ a 
group volume factor, and ${\cal{N}}_{\rm tr}$ 
and ${\cal{N}}_{\rm rot}$ the number of translational
and rotational degrees of freedom).

The conserved charges $Q_A$ are just the primordial values of
the exactly (or, in some cases, approximately) conserved charges 
in the electroweak model with which we are calculating.  In 
scenarios for electroweak baryogenesis these are always 
taken to be zero. From (\ref{eq: B}) it then follows that
$\mu_B$ can be expressed
in terms of $B$, so that (\ref{eq: dot B}) becomes simply
\begin{equation}
\dot B= - \alpha_n \Gamma_{\rm sph} B\,,\qquad
\Gamma_{\rm sph}=6N_F^2\frac{\bar\Gamma_{\rm sph}}{T^3}
\label{eq: relaxation}
\end{equation}
where $\alpha_n$ is a number of order {\it one\/} 
whose precise value depends on the model and its 
corresponding set of charges $Q_A$. In section 4 
below we carry out the constraint calculation
explicitly and find $\alpha_n \approx 0.4$ 
for typical electroweak models.
Integrating (\ref{eq: relaxation}) gives the baryon asymmetry
$B_{\rm freeze}$ which survives to partake of nucleosynthesis:
\begin{equation}
B_{\rm freeze}= 
B(T_{\rm b})
\exp\left [- \int_{t_{\rm b}}^{\infty} dt
 \alpha_n \Gamma_{\rm sph}(t)\right ]
=B(T_{\rm b})\exp\left [-H_{\rm b}^{-1}
\int_{0}^{T_{\rm b}}dT
\frac{\alpha_n \Gamma_{\rm sph}}{T}
\big(\frac{T_{\rm b}}{T}\big)^{p}\right]\,,
\label{eq: depletion}
\end{equation}
where $B(T_{\rm b})$ is the
baryon asymmetry at the completion of the transition,
at temperature  $T_{\rm b}$
and $H_{\rm b}$ is the expansion rate 
at that time. As discussed in the previous section
the appropriate value of $T_{\rm b}$ depends on the details of the
of the phase transition and lies in the range 
$T_{\rm b} \in [T_{\rm c}, T_{\rm nucl}]$. 
To obtain the latter form of (\ref{eq: depletion})
we have used the time-temperature relation
$t \propto T^{-p}$ which follows from (\ref{tempdep}).
Changing variables to $y=T_{\rm b}/T$ we can write 
the {\it{depletion factor}} $\cal D$ as
\begin{equation}
{\cal D}\equiv -\ln\frac{B_{\rm freeze}}{B_{\rm b}} =
\frac{T_{\rm b}}{H_{\rm b}}
\times 6\alpha_n {\cal C} N_F^2\, g
\int_1^{\infty} \left(\frac{\phi(T)}{T_{\rm b}}\right)^7
y^{5+p} {\rm e}^{-\frac{E_{\rm sph}(T)}{T_{\rm b}}y}dy\,.
\label{eq: Dfirstchange}
\end{equation}
Over the range of integration the factor $E_{\rm sph}/T_{\rm b}$ 
in the exponential increases from its minimum value at $y=1$, 
which is quite a large number $\sim 30$. This means that the 
dominant contribution to this integral comes from temperatures 
very close to $T_{\rm b}$ with $y \sim   1+ 1/30$. In fact we will see
below that the rate of change of the VEV is typically
large enough to narrow the range of temperatures 
which dominate the integral even more than this. 
The $p$ dependence in the integral is therefore 
very weak and the only significant effect of the change in the 
expansion rate from its radiation dominated value $H_{\rm rad}$
is to change the depletion factor $\cal D$ by the factor 
$H_{\rm rad}/H_{\rm b}$. Increasing the expansion rate decreases
the depletion because the sphaleron rate decouples at a
higher temperature. 

Is this change significant? For a given theory (with all
parameters determined) the depletion factor is (in principle) 
calculable. There is essentially no depletion for any 
expansion rate greater than the expansion 
rate $H_{\rm sph}$ given by setting ${\cal D}=1$ in 
(\ref{eq: Dfirstchange}). For $H < H_{\rm sph}$, however,
a baryon asymmetry produced at the first order phase 
transition is attenuated by a factor $e^{-H_{\rm sph}/H}$.
Whether a change in the expansion rate from that in a radiation 
dominated universe to a different value is important therefore 
depends on what the critical expansion rate $H_{\rm sph}$ is 
in the particular model. If a model has 
$H_{\rm sph}=10^{n}H_{\rm rad}$, 
the baryon asymmetry left behind in the universe with 
$H \sim H_{\rm sph}$ may be compatible with observation, 
and that in a radiation dominated universe too small
by a factor $e^{-10^n}$. If, on the other hand, 
$H_{\rm sph} < H_{\rm rad}$
the asymmetry will survive unattenuated in either universe.

We now turn to determining the effect of treating the expansion
rate as a variable on the sphaleron bound in its more familiar
forms, in which the requirement 
${\cal D} \le 1$ is converted
to a bound on parameters in a particular model. The bound
is usually stated as a lower bound on the sphaleron energy,
or as a lower bound on the ratio of a VEV to the temperature 
at the nucleation or critical temperature, and then converted 
into a bound on parameters in the specific model concerned.
We will follow through the derivation of such bounds in detail,
particularly because we wish to note certain points which are
often overlooked in this context. We then analyse the
case of the MSM in detail using the same effective potential 
(\ref{eq: effective potential}) and (\ref{eq: effective potential II})
used in the previous section.

Using the sphaleron energy $x={\cal B}(4\pi/g)(\phi(T)/T)$ 
as the variable in (\ref{eq: Dfirstchange}) we obtain
the sphaleron bound in its new form as 
\begin{eqnarray}
H_{\rm b} \geq  H_{\rm sph} &=&
6\alpha_n N_F^2 {\cal C} 
\left(\frac{\alpha_w}{4\pi}\right)^{4}\, g
{\cal B}^{-8}
\int_{x_{\rm b}}^{\infty}\left(\frac{T_{\rm b}}{T(x)}\right)^p
\left\vert\frac{d(\phi/T)}{dT}(x)\right\vert^{-1}
dx x^7{\rm e}^{-x}\nonumber\\
&\approx&
\left(\frac{\alpha_w}{4\pi}\right)^{\frac{1}{2}}
{\cal B}^{-1}
\left\vert T\frac{d(\phi/T)}{dT}\right\vert_{\rm b}^{-1}
\alpha_n\Gamma_{\rm sph}(T_{\rm b})
\label{eq: sphaleron bound}
\end{eqnarray}
where, to derive the latter expression, we assumed 
that over the range of temperatures which contribute to 
the integral the derivative term is approximately constant, 
and  $(T_{\rm b}/T(x))^p\approx 1$. 
Let us assess the validity of this approximation in more
detail in the case of the MSM.
At any temperature at which the two phases coexist, i.e.
between $T_{\rm c}$ and $T_0$,
\begin{equation}
T\frac{d}{dT}\left(\frac{\phi}{T}\right)\approx
-\frac{2\gamma(T_0/T)^2}
{2\lambda_T(\phi/T)-\alpha}
-\frac{T}{\lambda_T}\frac{d\lambda_T}{dT}\left[
\frac{\phi}{T}
+\frac{\gamma\left(1-\left(T_0/T\right)^2\right)}
{2\lambda_T\phi/T-\alpha}\right]
\,,
\label{eq: dx/dT}
\end{equation}
where $Td\lambda_T/dT=-8\beta$, and 
\begin{equation}
\frac{\phi}{T}=\frac{\alpha+\left[
\alpha^2-4\lambda_T\gamma(1-(T_0/T)^2)
\right]^{1/2}}{2\lambda_T}
\,.
\label{eq: phi/T}
\end{equation}
For simplicity, in (\ref{eq: dx/dT}) we neglected the 
temperature dependence of $\alpha$ and $\gamma$, which would 
result in numerically irrelevant corrections.  
Even though both 
$\phi/T$ and its derivative in (\ref{eq: dx/dT}) and 
(\ref{eq: phi/T}) are very sensitive functions of 
$T$, in the temperature interval $T_{\rm c}\ge T\ge T_0$
both are monotonically decreasing, and  
we can write their lower and upper bounds as follows:
\begin{equation}
\frac{\phi_{\rm c}}{T_{\rm c}}=
\frac{2}{3}\frac{\alpha}{\lambda_T}
\,,\qquad
\frac{\phi_0}{T_0}=\frac{\alpha}{\lambda_T}\,
\label{eq: bounds on phi/T}
\end{equation}
and 
\begin{equation}
\left[T\frac{d}{dT}\frac{\phi}{T}\right]_{c}\!=\!
-\frac{6\gamma}{\alpha}
\left(\frac{T_0}{T_{\rm c}}\right)^2
+\frac{4\alpha}{3\lambda_T}\left(
1+\frac{8\beta}{\lambda_T}\right)\!
\approx -\frac{6\gamma}{\alpha}
\,,\quad
\left[T\frac{d}{dT}\frac{\phi}{T}\right]_{0}\!=\!
-\frac{2\gamma}{\alpha}+\frac{8\alpha\beta}{\lambda_T^2}
\approx\! -\frac{2\gamma}{\alpha}
\,,
\label{eq: dx/dT bounds}
\end{equation}
where $\frac{\gamma}{\alpha} \approx 18$.
The large value of the derivative means that the pre-factor in
front of the sphaleron rate in (\ref{eq: sphaleron bound})
is $\ge 10^3$. This is essentially just the (inverse) fraction of
an expansion time in which the sphaleron freezes out (leading
to the difference from the naive freeze-out estimate
$H_{\rm sph}\sim \Gamma_{\rm sph}(T_{\rm b})$). The range of
temperatures which contributes in the integral is therefore 
much less than between $T_{\rm c}$ and $T_0 \simeq 0.99 T_{\rm c}$, 
and the constant derivative approximation used in evaluating it
is indeed very accurate. 
Further, as $T_{\rm b}$ varies in this range the change in the result 
associated with the derivative is at most this factor of three.
In what follows we will  keep track of this dependence of
the sphaleron bound on $T_{\rm b}$, 
and quantify it in comparison to the other
effects on the bound in which we are interested here. 

The sphaleron bound as given in (\ref{eq: sphaleron bound}) 
can be converted, for a given expansion rate $H_{\rm b}$, into
a lower bound on the ratio $\phi_{\rm b}/T_{\rm b}$ (where 
$\phi_{\rm b} \equiv \phi(T_{\rm b})$). A numerically 
convenient and instructive way to write the lower bound on
this quantity is in the implicit form
\begin{equation}
\!\,\frac{\phi_{\rm b}}{T_{\rm b}}  
= \frac{1}{\cal B}
\left(\frac{\alpha_w}{4\pi}\right)^{\frac{1}{2}}\!
\left[ 
\ln \frac{6 N_F^2\alpha_n
\left(
\frac{\omega_-}{2\pi g\phi(T)}
{\cal N}_{\rm tr}{\cal N}_{\rm rot}
{\cal V}_{\rm rot}{\cal K}_{\rm sph}\right)
\alpha_w}
{\left\vert T\frac{d}{dT}\frac{\phi}{T} 
\right\vert_{\rm b}{\cal B}\frac{H_{\rm rad}}{T_{\rm b}}}
- \ln\frac{H_{\rm b}}{H_{\rm rad}} 
+ 7\ln \frac{\phi_{\rm b}}{T_{\rm b}}
\right]
\label{eq: sphaleron bound II}
\end{equation}

\begin{figure}[htb]
\epsfxsize=5.5in
\centerline{\epsfbox{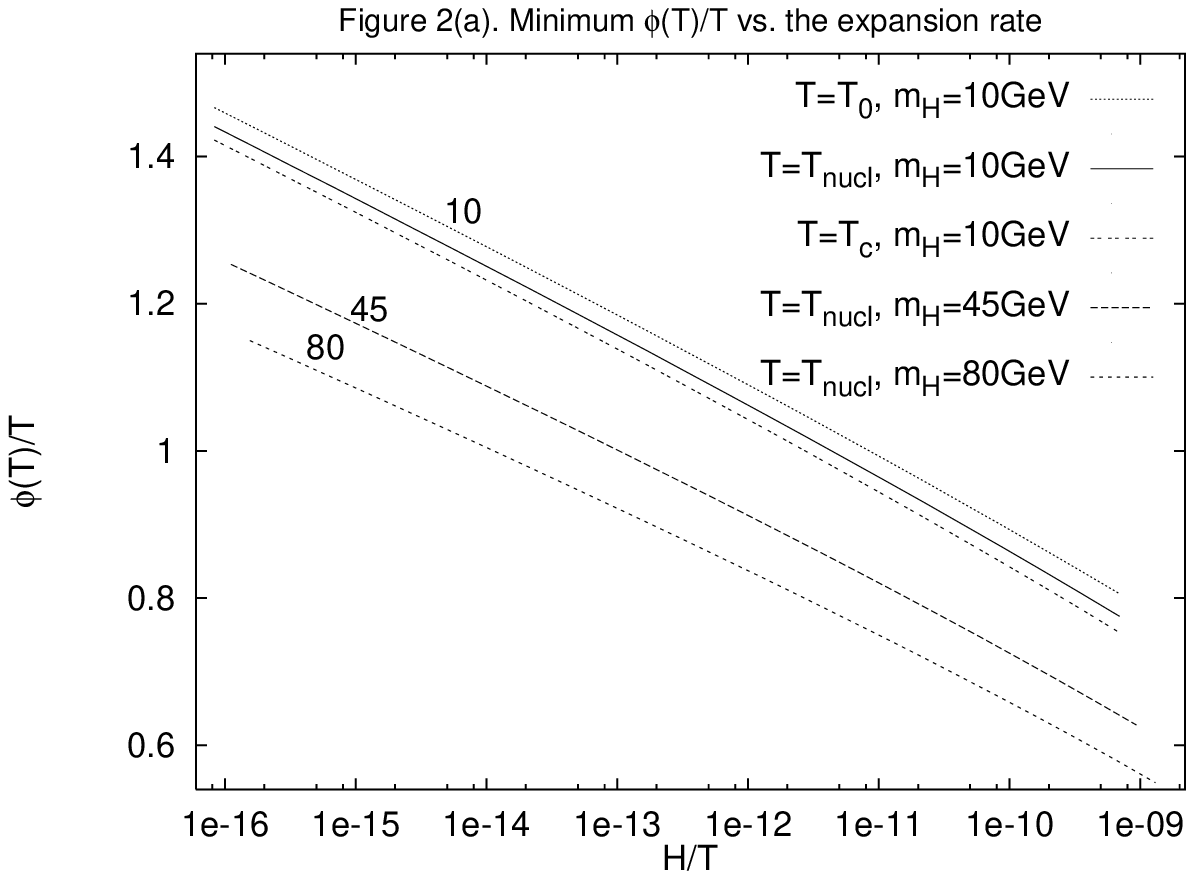}}
\end{figure}
In figure 2(a) we show the solutions to this
equation obtained from an iterative evaluation of 
(\ref{eq: sphaleron bound II}), for the MSM.
We have taken ${\cal V}_{\rm rot}=8\pi^2$, and
fit ${\cal N}_{\rm tr}{\cal N}_{\rm rot}
\simeq 86-5\ln(m_H^2/8m_W^2)$ \cite{CarsonMcLerranWang}.
The one loop result for ${\cal K}_{\rm nucl}$ we took from
\cite{BaackeJunker}: ${\cal K}_{\rm sph}=
\{7.54,5.64,4.57,3.89,3.74\}$
for $m_H=\{0.4,0.5,0.6,0.8,1\}m_W$, and  
extrapolated or interpolated for other values  
$m_H\in [10,80]$GeV.
$\omega_-$ we took from \cite{CarsonMcLerranWang}, where
it was found that $\omega_-/g\phi(T)\in [0.4,0.55]$ for  
$m_H\in [10,80]$GeV. We neglected the plasma effects
on $\omega_-$.
Finally, we took $\alpha_n=0.4$, and 
${\cal B}=\{1.52,1.61,1.83,2.10\}$ 
for $m_H^2/m_w^2\in\{0,8\times 10^{-3},8\times 10^{-2},
0.8,8\}$ \cite{KlinkhamerManton}, and 
quadratically interpolated for the intermediate values.

Besides varying the expansion rate over the range shown
in the figure, we have taken a wide range of Higgs masses 
and different values for $T_{\rm b}$. It is instructive to do this 
because the sphaleron bound is often stated as a bound 
on this ratio of VEV to temperature as if this were a
model-independent and temperature independent statement of it. 
We see from figure 2(a) that this is very far from being true. 
For a range of Higgs masses from 10GeV to 80GeV the bound on $\phi/T$ 
decreases by about $20\%$. That most of the dependence comes from 
the factor ${\cal B}$, which varies
non-negligibly with the Higgs mass, can be verified 
easily\footnote{$\delta(\phi/T)/(\phi/T)\simeq 
-(\delta{\cal B}/{\cal B})
\left[{1-7\cal B}^{-1}(\alpha_w/4\pi)^{1/2}/(\phi/T)
\right]^{-1}$. For $m_H=10$GeV to $80$GeV, 
$\delta{\cal B}/{\cal B}\simeq 1/8$, and hence 
$\delta(\phi/T)/(\phi/T)\simeq 1/6$, accounting 
for most of the dependence on $m_H$ on figure 2(a).}. 
\begin{figure}[htb]
\epsfxsize=5.5in
\centerline{\epsfbox{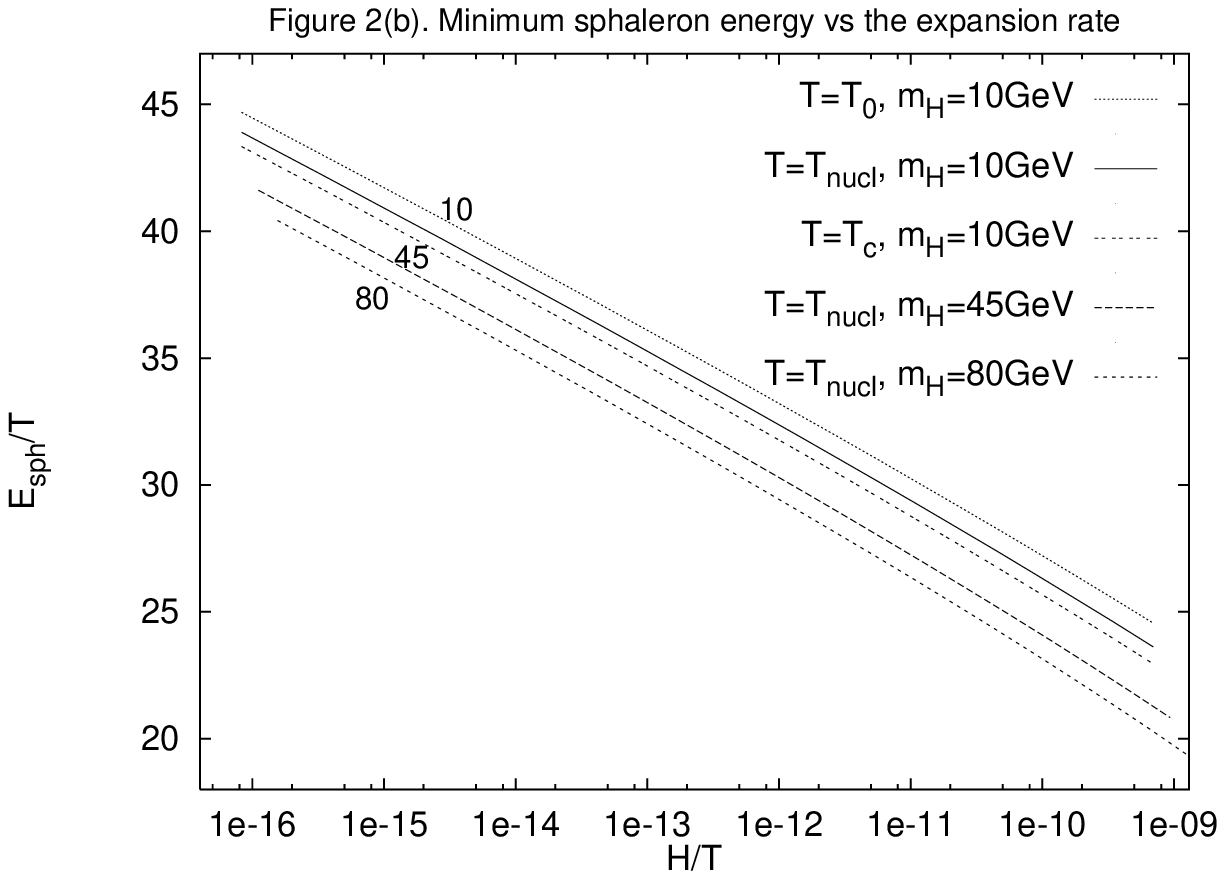}}
\end{figure}
This is also clear from figure 2(b), 
which shows the sphaleron
bound as a lower bound on the sphaleron energy 
$E_{\rm sph}^{\rm b}/T_{\rm b}$ 
(where $E_{\rm sph}^{\rm b} \equiv E_{\rm sph}(T_{\rm b})$)
as a function of the expansion rate.
There is however still a significant mass dependence
(approximately $8\%$ over the mass range considered)
in the bound stated this way.
The temperature dependence of the bounds  is comparatively
smaller - as $T$ increases from $T_0$ to $T_{\rm c}$, the bound 
on $\phi/T$ decreases by $3-4 \%$, and for 
$T_{\rm b}\in [T_{\rm nucl},T_{\rm c}]$ by less than $1\%$.
This dependence comes from the derivative of the VEV
inside the logarithm, which as we saw above can vary
by a factor of three over the range from $T_{\rm c}$ to $T_0$. 
In analyzing any particular model in detail the parameter 
dependence of the sphaleron bound stated this way in terms
of these quantities should clearly be borne in mind and
carefully examined.

The dependence we are primarily interested in here
is that seen in both figures 2(a) and 2(b)
on the expansion rate of the Universe.  
Both $\phi_{\rm b}/T_{\rm b}$ and 
$E_{\rm sph}^{\rm b}/T_{\rm b}$ show an almost exact linear 
dependence on the logarithm of $H$ which is evident from  
(\ref{eq: sphaleron bound II}). For a small fractional
change in the bound on $\phi_{\rm b}/T_{\rm b}$ or
$E_{\rm sph}^{\rm b}/T_{\rm b}$
due to a change in the expansion rate from $H_{\rm rad}$ 
we have the approximate formula
\begin{equation}
{\cal B}\left(\frac{4 \pi}{ \alpha_w}\right)^{1/2}
\Delta\left(\frac{\phi_{\rm b}}{T_{\rm b}}\right) 
\equiv\Delta\left(\frac{E_{\rm sph}^{\rm b}}{T_{\rm b}}\right) 
\approx -\frac{ \ln \frac{H}{H_{\rm rad}}}{1-7  
\left(\frac{E_{\rm sph}^{\rm b}}{T_{\rm b}}\right)^{-1}}
\label{approximate-shift}
\end{equation}
which, given $(4\pi/\alpha_w)^{1/2} \approx 20$,
agrees well with the numbers read off from the figures.
Over five orders of magnitude in the expansion rate
we see a decrease in the bound on $\phi_{\rm b}/T_{\rm b}$ by
about $0.4$, or approximately $0.08$ per order of magnitude
increase in the expansion rate.

The usual starting point for analysis of most extensions
of the standard model departs from the sphaleron bound 
given as a lower bound on the ratio of the sphaleron energy or 
the appropriate VEV to the temperature, and then converts this
to a bound on the parameters of the model. We have noted 
that such a procedure should be considered more carefully
as there can in fact be significant model dependence in
the bounds on these quantities. We have derived nevertheless
how such bounds are changed as a function of the expansion 
rate of the Universe, and the approximate form 
(\ref{approximate-shift}) is essentially model independent.
Using this formula one can therefore turn the usual sphaleron bound
for any given model into a lower bound on the expansion rate as a function
of model parameters, provided one has the correct form of the
bounds on $\phi_{\rm b}/T_{\rm b}$ 
(or $E_{\rm sph}^{\rm b}/T_{\rm b}$) in the radiation dominated
case: For each set of parameter values one calculates the value
of the given ratio, and then solves using (\ref{approximate-shift})
for the expansion rate which reduces (or increases) the radiation 
dominated value to the calculated critical value. 

However, the most direct way to calculate the sphaleron 
bound as a lower bound on the expansion rate is  simply 
to evaluate the integral (\ref{eq: sphaleron bound})
directly to find $H_{\rm sph}$ for each value of the parameters 
in the theory. We have done this for the MSM  
using the same parameter values and effective potential 
as above, and for the temperatures 
$T_{\rm b}=T_{\rm c},T_{\rm nucl},T_{0}$.
\begin{figure}[htb]
\epsfxsize=5.5in
\centerline{\epsfbox{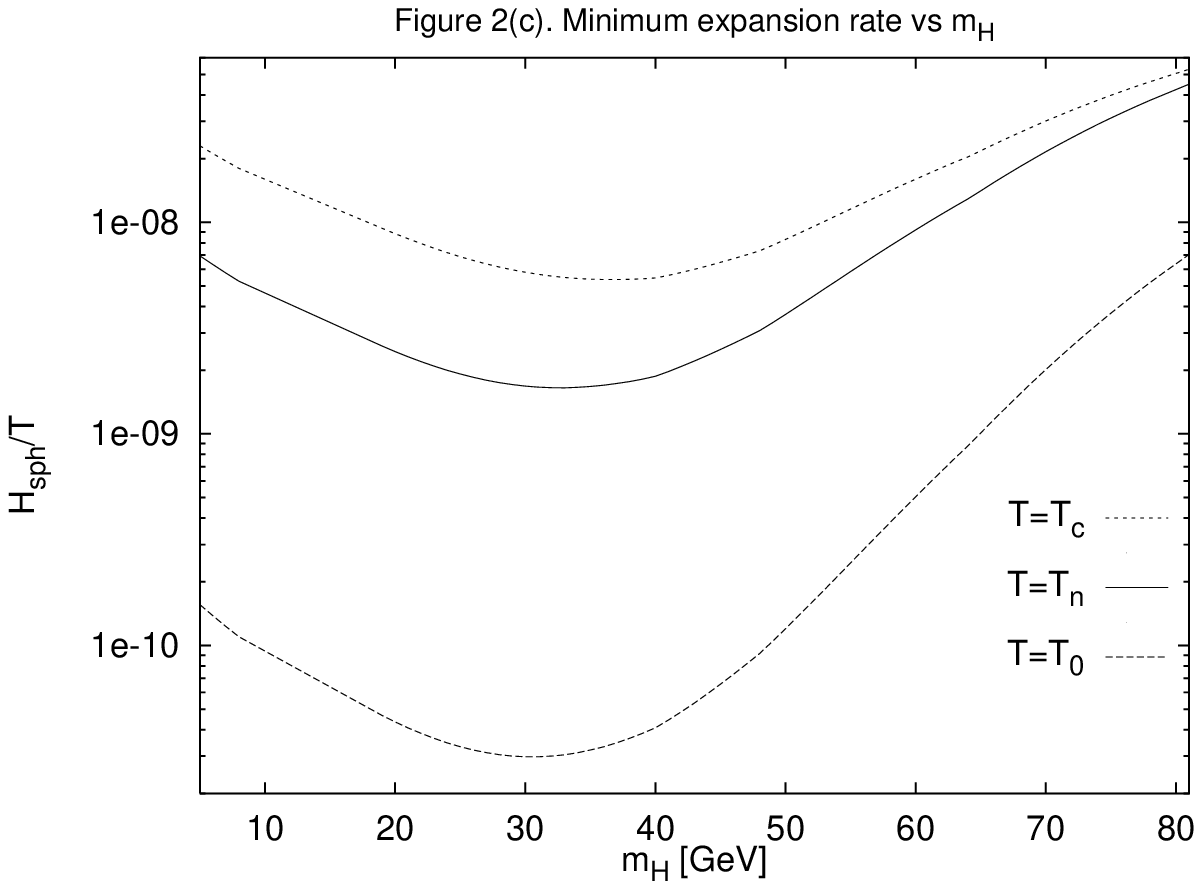}}
\end{figure}
The result is shown in figure 2(c), where the sphaleron
bound is given as a plot of the minimum expansion rate 
required as a function of the Higgs mass $m_H$.
The dependence on the temperature seen in the figure 
is greater than in the bound on $\phi/T$,
since it also enters in relating $m_H$ to $\phi/T$,
as shown in figure  2(d).  
\begin{figure}[htb]
\epsfxsize=5.5in
\centerline{\epsfbox{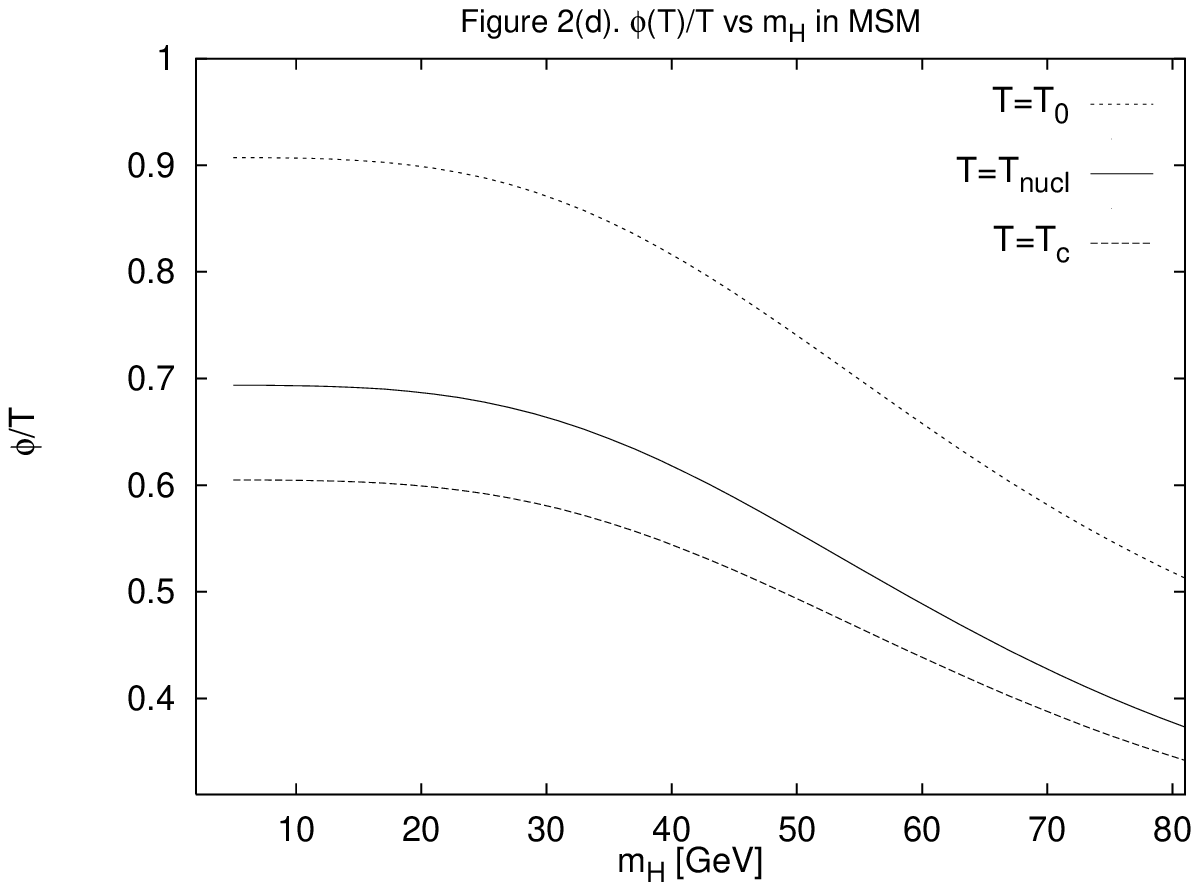}}
\end{figure}
Figure 2(c) shows dramatically how badly the usual sphaleron
bound is violated in the MSM \footnote{Studies of the two loop
effective potential and lattice studies show that the 
one-loop ring improved effective potential we are using
underestimate the strength of the phase transition, but not 
enough to significantly alter the conclusions drawn here.}.
For {\it no} physical Higgs mass is the 
minimum required expansion rate within orders of magnitude 
of that in a radiation dominated universe. 
The discrepancy of this result with the early sphaleron bounds
calculated for the MSM \cite{mishasphal} is explained
by the much larger (now physical) top quark mass $m_t = 175$GeV
used here.  For small $m_H$ one can see from
(\ref{eq: effective potential II}) that the 
one-loop thermal contribution from the top quark dominates
$\lambda_T$, and therefore, from (\ref{eq: phi/T}), that $\phi/T$  stops 
increasing and levels off as seen in figure 2(d).
The increase in the minimum expansion rate seen 
in figure 2(c) as $m_H$ decreases below this value
comes simply from the dependence on $m_H$ of
the sphaleron energy through $\cal{B}$ (which
decreases, increasing the sphaleron rate).

In many extensions of the standard model it has been shown that,
in contrast to the MSM, there are physically allowed regions of 
the parameter space where the usual sphaleron
bound is satisfied. The way of stating the sphaleron bound
we have illustrated for the MSM can be easily generalized 
to any such model. Besides being, as we argue in this paper,
a more correct way to state the sphaleron bound (given that
the expansion rate really is an unconstrained parameter),
our discussion also shows that it is an instructive way to
state it, because it quantifies how well or badly the bound
is satisfied or violated. If we state the sphaleron
bound in this way, it is easy to determine the effect 
on the calculated bounds in any change to input parameters (e.g. to 
any of the pre-factors in the sphaleron rate).

Having discussed how the sphaleron bound should be restated
as a lower bound on the expansion rate, 
let us ask finally what electroweak baryogenesis at a first order
phase transition can potentially tell us about 
the expansion rate at that epoch. A priori we do not
know what it is and can use baryogenesis as a probe.
If the correct electroweak theory turns out to be one
in which there is a first order phase transition 
which successfully produces exactly the right 
amount of baryons during the phase transition, we would
have compelling evidence that the expansion rate is
greater than the corresponding critical value $H_{\rm sph}$.
But it can tell us no more. If the model satisfies the
`old' sphaleron bound with the assumption of radiation 
domination, but has $H_{\rm sph} < H_{\rm rad}$ 
(as it typically will),
the success of the model provides no evidence that
the Universe expands at $H_{\rm rad}$. It could even 
potentially expand orders of magnitude slower than 
$H_{\rm rad}$. 
We will now see that in contrast electroweak baryogenesis
in a homogeneous universe provides a much more sensitive 
probe of the expansion rate at that scale.

\section{Baryogenesis in a Homogeneous Universe}

Analysis using the effective potential constructed in
perturbation theory indicates a first order phase transition
but is only of validity for Higgs masses up to about 
$60\hbox{GeV}$. 
Recent non-perturbative results \cite{kajantiecrossover}
indicate that for heavier Higgs masses the line of first order
phase transitions ends in a second order phase
transition at about $80\hbox{GeV}$  in the Minimal Standard
Model. For larger masses the transition is an
analytic cross-over, {\it i.e.\/} there is actually no phase 
transition since all physical quantities vary continuously 
(and differentiably) as a function of temperature. This sort of 
behavior is typical of a system in which there is no order
parameter which can define the symmetry state of
the system - the gauge symmetry is never strictly
speaking broken or unbroken. 

The only departure from equilibrium in this case is
that caused directly by the expansion of the Universe.
All physical quantities vary on a timescale characterized
by the cooling rate $\sim H$.
Unlike the case of bubble nucleation there is no
separation between the mechanism by which the baryons are
created and the part of the calculation involving the
expansion rate directly, a separation which allowed us to take 
the created asymmetry simply as an input without specifying
how it was created. Here we must make use of a particular model
in order to answer the question of how the baryon asymmetry
depends on the expansion rate.

Most work on mechanisms for electroweak baryogenesis
has considered extensions of the Standard Model with
an additional source of CP violation beyond the KM
matrix. On bubble walls formed at a first order phase
transition the CP violation produces in various ways
a term biasing the anomalous sphaleron processes, causing the
creation of baryons on or around the wall.  These source terms,
which are present when there is space or time dependence of 
the condensate fields, can equally be used to bias the anomalous 
processes and produce baryons when the phase transition is not
first order. In the case of a second-order or cross-over 
transition
we expect the evolution to be homogeneous with time dependence 
only
of the condensate fields, and we will model the problem this way.
In fact the validity of the analysis is broader than just the 
regime
where the phase transition is not first order. It also describes 
well the period after the completion of a first order phase 
transition.
In particular, as we will discuss below, it describes the case 
where
the phase transition is too weakly first order to satisfy the
sphaleron bound.

We will now consider separately two types of source terms 
for baryogenesis discussed in the literature.

\subsection{Potentials for Baryon Number}

The first apparently viable mechanisms for electroweak
baryogenesis, discussed in \cite{tz} and \cite{dhss}, 
considered potentials for baryon number.
The models differ in their particular realisations
of this potential. 
In various theories -- two
doublet extensions of the minimal standard
model \cite{tz} and supersymmetric theories
with or without an additional singlet \cite{dhss} --
there are CP odd terms in the effective action
for the gauge-Higgs sector, of the form
$(g^2/16 \pi^2)  \chi F \tilde{F}$, where $F$ and $\tilde{F}$
are the $SU(2)$ field strength tensor and
its dual, ${\chi}$ is some field or combination
of fields which acquire VEVs at the phase transition,
times a numerical factor (typically
a suppression). When these terms are integrated
by parts and the anomaly equation 
$(g^2/16 \pi^2) F \tilde{F}= \partial_\mu j^\mu_B$ is used, 
in the homogeneous case (with time dependence only)
they produce terms calculationally equivalent to a chemical
potential for baryon number $\dot{\chi} B$.
Specifically in two doublet models there are terms
with \cite{tz}
\begin{eqnarray}
\dot{\chi}_B &=& -\frac{7\zeta_3}{4}
\left (\frac{m_t}{\pi T}\right)^2 \frac{2}{v_1^2}
i\left(\Phi_1^\dagger{\cal D}_0\Phi_1-
({\cal D}_0\Phi_1)^\dagger\Phi_1\right)\,
 \nonumber \\
&\approx& {7\zeta_3}
\left (\frac{m_t}{\pi T}\right)^2 
\frac{v_2^2}{v_1^2+v_1^2}\dot{\theta}\,, 
\qquad \zeta_3\simeq 1.202\,,
\label{eq: tz} 
\end{eqnarray}
where $\theta$ is the relative phase between the two doublets,
with VEVs of magnitude $v_1$ and $v_2$ (where the former couples 
to the top quarks).
In theories with CP violation characterized by
some scale $M$ \cite{dhss} the equivalent quantity $\dot{\chi}_B$
is 
\begin{equation}
{\frac{1}{3M^2}} \partial_0  |\phi|^2 \,,
\qquad
{\frac{1}{3M}} \partial_0  s\,,
\label{eq: dhss}
\end{equation}
where the first case is a theory with doublets
only, the second one with a singlet $s$.

Up to higher derivative corrections to the VEVs
the system in this background tries to thermalize
to the equilibrium in the presence of this extra term,
in which the baryon number is given by the expression
in (\ref{eq: B}), with $\mu_B=0$ and  $H$ including
the additional term due to the background.

It is perhaps instructive to note exactly why an equilibrium 
calculation of this sort can give a baryon asymmetry. 
The Sakharov requirement of a departure from
equilibrium is sometimes shown using this type of
expression by acting with 
$\Theta=$CPT on baryon number as follows
\begin{eqnarray}
\langle B\rangle
=Tr\left[{\Theta}B{\Theta}^\dagger{\Theta}
{\rm e}^{-\beta H}{\Theta}^\dagger\right]
=-\langle B\rangle
\label{eq: sakharov}
\end{eqnarray}
and therefore $\langle B\rangle =0$. The same will hold true 
if we allow non-zero chemical potentials for charges which are 
CP even and, of course, it will not hold if we impose a chemical 
potential for CP odd charges like $B$ or $B-L$.
In the case we are considering the reason it does not
hold is that the additional effective term in $H$ is not
CPT invariant, as the time varying condensate field violates
CPT spontaneously. The underlying Hamiltonian is of course 
CPT invariant, but in the expanding Universe this symmetry is
spontaneously violated. 

The constraints which are to be imposed are those
setting all the charges which are conserved over
the relevant timescale to zero.
Compared to the case of the same source term used
to generate a biasing of baryon number on the bubble
walls during a first order phase transition, there are
thus two important differences:

(i) The `relevant timescale' on the bubble wall
(of thickness $L$ moving with velocity $v_w$)
is the wall passage time $L/v_w$, typically
$\sim 10^2/T$. Here it is that characterizing the
time rate of change of the field; in the 
homogeneous case this will be $\sim H^{-1}$. 
Thus the numerous processes ({\it e.g.\/} chirality
flipping processes of the lighter quarks and leptons), 
which are effectively inoperative on the bubble wall, 
are equilibrated in the present case. The set of 
relevant conserved charges is therefore much smaller
(and the calculation therefore simpler).
For a radiation dominated universe the only
conserved charges at the electroweak scale are
the exactly conserved charges - hypercharge $Y$,
electric charge $Q$, and $\frac{1}{3}B-L_i$
($L_i$ the lepton number in generation $i$).
The charge which is violated slowest is
right-handed electron number $e_R$, since it
is coupled to other species only through its
small Yukawa coupling $y_e$, by processes
with rate $\sim y_e^2 T \sim 10^{-12} T$
(times a number of order $0.1 - 0.01$
- see \cite{CKOrighthanded}, \cite{JPTparttwo}).
Thus for $H > 10^{-12}T$ we will
also need to add $e_R$ as a conserved charge,
for $H > 10^{-8} T$ both $e_R$ and $\mu_R$,
{\it etc}.

(ii) On a bubble wall the constraints forcing the
conserved charges to zero are appropriate only
when negligible charge can be transported onto the wall
over the relevant time scale. (The charges are
conserved only globally, not locally on the bubble wall
unless this is true \cite{JPTconstraints}.)
This places a condition on the thickness of
the wall $L > D/v_w$  for the applicability
of this simple form of the calculation.
This condition follows from 
the requirement that the wall passage time 
be greater that the diffusion time 
($L/v_w>D/v_w^2$), in order for transport to be 
inefficient.  In the present case the Universe is 
(assumed) homogeneous
and the global constraints forcing the charges
equal to zero are always appropriate.

Because the electroweak phase transition is
not a symmetry breaking phase transition, we
cannot define an exact criterion for 
whether the `broken' or `unbroken' basis of particle states
presents the correct description 
in the equilibrium calculation. A correct calculation
would assume neither basis. A simple
example of such a calculation has been
discussed recently in \cite{KhlebnikovShaposhnikov} and a 
cross-over
from one limit  to the other explicitly  shown to occur
at $m_W \sim m_D$  (the vacuum mass and Debye mass of the
gauge bosons respectively). Converting this to
a constraint on the ratio of the VEV to temperature,
it turns out that the symmetric phase calculation is
a better approximation when the sphaleron freezes out.
Thus we will calculate here in this approximation,
using the `unbroken' phase classification of the states.
In section 4.2 we will also see that either basis of
states gives almost identical results.

We take the case when $e_R=0$ as this will turn out
to be appropriate in the regime of expansion rates
of interest for the generation of the
observed BAU\footnote{We will neglect the 
potentially interesting effect 
discussed in \cite{mjms}.
Incorporating it could lead in
certain cases to minor changes to the final baryon asymmetry.}.
However, the numerical difference induced by this 
additional constraint will turn out to be insignificant. 
Expressing the charges in terms
of particles densities $n_\alpha$,
and using the linear approximation
$n_\alpha=(T^2/12)k_\alpha \mu_\alpha$,
where $\mu_\alpha$ is given in terms of
the chemical potentials $\mu_A$ for the
charges $Q_A$ by $\mu_\alpha= q_\alpha^A \mu_A$
(where $q_\alpha^A$ is the $Q_A$ charge of the species $\alpha$)
and $k_\alpha$ is 
a statistical factor which
is equal to 1(2) for fermions (bosons) in the
massless approximation\footnote{ Including the lowest order mass
correction to this simple formula results in: 
$n_\alpha=(T^2/6)\mu_\alpha(1-(3/2)(m_\alpha/\pi T)^2)$
for fermions, and
$n_\alpha=(T^2/3)\mu_\alpha(1-(3/2)(m_\alpha/\pi T))$
for bosons. In this paper we ignore these mass corrections since 
they appear as a subleading correction
to the result presented in the text.}, we find 
\begin{eqnarray}
Y &=& \frac{T^2}{6}\bigg[ (10+n)\mu_Y + 2 \mu_B
+ \frac{8}{3} \Sigma_j \mu_{j} -\mu_{e_R} \bigg]
\nonumber \\
\frac{1}{3}B-L_i &=&\frac{T^2}{6}\bigg[\frac{8}{3}\mu_Y
+ \frac{4}{3}
\mu_B + \Sigma_j (\frac{4}{9}+ 3 \delta_{ij}) \mu_j -
\delta_{1i}\mu_{e_R} \bigg]
\nonumber\\
e_R &=& \frac{T^2}{6}\bigg[ -\mu_Y  - \mu_1 + \mu_{e_R} \bigg] \\
B &=& \frac{T^2}{6}
\bigg[2\mu_Y+4\mu_B+\frac{4}{3}\Sigma_j\mu_{j} \bigg]
\end{eqnarray}
where $\mu_i$ is the chemical potential for $B-L_i$,
and $n$ is the number of Higgs doublets.

Setting the first three charges equal to zero\footnote{In 
the `unbroken' phase one can choose to
constrain any two linear combinations of
hypercharge $Y$ and isospin $T_3$. The choice of $Y$
and $T_3$ is simple, because $T_3$ is then
proportional to its own chemical potential and
$T_3=0$ is trivial.} we find
\begin{equation}
B_0= c_n\, T^2\mu_B\,,\qquad
c_n=\frac{1}{6}\; 
\frac {36(29+6n)}{399 + 82n}
\label{eq: B_0 vs mu_B}
\end{equation}
Note that $c_n=B_0/(T^2\mu_B)$ is
almost insensitive to $n$, the number of Higgs doublets,
varying only  between $0.436$ and $0.439$,
as $n$ changes from $0$ to $\infty$ 
\footnote{If we had not assumed
the right-handed electron to be in equilibrium, the change
would be small. In this case $c_n=0.455\rightarrow 0.462$, as 
$n=0\rightarrow \infty$.}. 
Taking the source term to be as given in (\ref{eq: tz}) or
(\ref{eq: dhss}), we set $\mu_B=\dot{\chi}_B$ in 
(\ref{eq: B_0 vs mu_B})
to obtain the baryon asymmetry in the ``equilibrium'' to which 
the baryon number violating processes will try to bring the 
plasma in the slowly varying background. To calculate the rate
at which these slow processes bring the system to this
state, we impose a chemical potential $\mu_B$ on baryon number
and include the source term. The baryon number $B$ in this 
state is then given by (\ref{eq: B_0 vs mu_B}) with the
replacement $\mu_B \rightarrow \mu_B+\dot\chi_B$,
and therefore
\begin{equation}
\mu_B= c_n^{-1}
\left(\frac{B}{T^2}-c_n\dot\chi_B\right)\,.
\end{equation}
Using (\ref{eq: B}) this gives the rate at which $B$ approaches
its ``equilibrium'' as
\begin{equation}
\dot{B}= - \alpha_n\Gamma_{\rm sph}\big(B-c_n\dot{\chi}_BT^2  
\big)
\,,\qquad
\Gamma_{\rm sph}=
6N_F^2\frac{\bar\Gamma_{\rm sph}}{T^3}\,,\qquad
\alpha_n=\frac{1}{6c_n}\,,
\label{eq: dot B mu B}
\end{equation}
where $\alpha_n=0.382\rightarrow 0.380$ as $n=0\rightarrow\infty$ 
and $\bar\Gamma_{\rm sph}$ is given 
in (\ref{eq: sphalrate}).

Before calculating the final baryon asymmetry we consider 
another treatment of induced source terms for baryogenesis
in a time dependent background.

\subsection{Potential for Hypercharge}

A different treatment of the biasing of baryon number
was given in \cite{CKNsponbaryo}\footnote{The account given 
here is not precisely that of the
original version of the idea given in \cite{CKNsponbaryo},
which treated a potential for {\it fermionic} hypercharge. It was
pointed out in \cite{DineThomas} that the
rotation should also be performed on the Higgs fields,
giving a potential for total hypercharge which in the
unbroken limit (VEVs $\rightarrow 0$) is pure gauge, 
and therefore can have no physical effect.
The leading baryon production is in this case 
mass-squared suppressed.
For a discussion of this point,  
see also section 3 of \cite{JPTpartone}. }.
In the broken phase of a two Higgs doublet model the relative
phase $\theta$ of the neutral components of the
Higgs fields enters in the fermionic mass terms.
A hypercharge rotation of the fields to remove
this phase from the mass term produces {\it at tree level}
a real mass term and an additional term in the Lagrangian
which, in the homogeneous case can be written simply
as  $\dot{\chi}_Y Y$, where $Y$ is the hypercharge operator
\footnote{We follow the convention used in \cite{CKNsponbaryo}.
There is nothing special about the choice of hypercharge.
The essential element is that it is an anomaly free charge which
is spontaneously broken by the mass term. A rotation proportional
to isospin, for example, or any charge which is a linear 
combination 
of hypercharge and a charge exactly conserved in 
the broken phase is equally good. It is not difficult to check 
that
the extra induced `source' term always drops out in the 
calculations
given below.}, and
\begin{equation}
\dot{\chi}_Y= -\frac{2v_2^2}{v_1^2+v_2^2}\dot{\theta}.
\label{ysource}
\end{equation}
In the unbroken phase this is just a gauge term,
but in the broken phase it can have physical
significance because hypercharge is not conserved,
being violated by VEV suppressed terms.  

Again, as discussed in section 4.1, we can calculate either
in the broken phase or unbroken phase, but the latter
is probably more appropriate for the temperature range
of relevance. In this case of course
we must include the information about hypercharge violating
processes to get a non-zero answer, so calculating in the 
unbroken phase  means taking the basis of chiral states of 
the unbroken phase and treating the mass terms as interaction 
vertices which can violate hypercharge by flipping chirality.  
The correct constraint calculation is therefore one in which
we take the same global conserved charges as in the previous
calculation, but 
instead of the constraint on hypercharge, we must impose
the constraint on the (conserved) electric charge $Q$,
and we get 
\begin{eqnarray}
Q &=& \frac{T^2}{6}\bigg[(10+n)\mu_Y+2(10+n)\mu_Q  
+2\mu_B+\frac{8}{3}
\Sigma_j \mu_{j} -\mu_{e_R} \bigg]\,,
\nonumber \\
\frac{1}{3}B-L_i &=&
\frac{T^2}{6}\bigg[\frac{8}{3}\mu_Y+\frac{8}{3}\mu_Q 
+\frac{4}{3}\mu_B + 
\Sigma_j (\frac{4}{9}+ 3 \delta_{ij}) 
\mu_j - \delta_{1i}\mu_{e_R} \bigg]\,,
\nonumber \\
e_R &=& \frac{T^2}{6}\bigg[-\mu_Y-\mu_Q-\mu_1+ 
\mu_{e_R}\bigg]\,,
\nonumber \\
B &=&\frac{T^2}{6}
\bigg[2 \mu_Y+2\mu_Q+4\mu_B+\frac{4}{3}\Sigma_j
\mu_{j} \bigg]\,.
\label{eq: hypercharge equations}
\end{eqnarray}
The ``chemical potential'' $\mu_Y$ for hypercharge here is 
the effective one which arises from the source term for
hypercharge, {\it i.e.\/}
$\mu_Y=-\dot{\chi}_Y$. Setting the conserved 
charges 
to zero in (\ref{eq: hypercharge equations}) 
we can solve for $B$
to find
\begin{equation}
B=\frac{T^2}{6}\left(
\frac{24(10+n)}{1219+164n}\dot{\chi}_Y
+\frac{36(89+12n)}{1219+164n}\mu_B
\right)\,.
\label{eq: B vs mu_Y mu_B}
\end{equation}
Using (\ref{eq: dot B}) 
we obtain the equation describing the relaxation of baryon
number to the ``equilibrium'' in presence of the hypercharge 
source term:
\begin{eqnarray}
\dot{B} &=& - 
\alpha^\prime_n\Gamma_{\rm sph}
\big(B-c^\prime_n\dot{\chi}_YT^2\big)\,,
\nonumber\\
\Gamma_{\rm sph} &=& 
6N_F^2\frac{\bar\Gamma_{\rm sph}}{T^3}
\,,\qquad 
\alpha^\prime_n=\frac{1219+164n}{36(89+12n)}
\,,\qquad 
c^\prime_n=\frac{24(10+n)}{6(1219+164n)}\,.
\label{eq: dot B mu Y}
\end{eqnarray}
Note that $c^\prime_n=0.033\rightarrow 0.024$, and 
$\alpha^\prime_n=0.3805\rightarrow 0.3796$ 
for $n=0\rightarrow \infty$. The rate of 
relaxation is essentially independent of the number
of doublets, 
and it is almost identical to 
the rate in the presence of a potential for baryon number
in section 4.1.  

It is noteworthy that the coefficient $c_n'$ 
is significantly smaller than the corresponding 
coefficient in the case of a potential for baryon number
($c_n/c^\prime_n\simeq 13 - 18$ as $n=0\rightarrow \infty$).
It follows therefore that, even though the baryon number
source (\ref{eq: tz}) is suppressed by a factor of mass over
temperature squared relative to that in (\ref{ysource}),
the former may give the dominant source term for 
baryogenesis.
The reason for this is a suppression due to strong sphaleron
processes in the case of a hypercharge source term 
\cite{GiudiceShaposhnikov}. 
In the massless quark approximation these force the densities
of right and left-handed baryons equal,
{\it i.e.\/} $B_L=B_R$.
On the other hand, it is easy to show that, with source terms
for a charge such as hypercharge which is conserved in baryon 
number
violating processes, $\dot{B} \propto (3B_L+L_L)$.
With $B-L=0$ this implies $\dot{B} \propto (\frac{5}{2}B-L_R)$
(where $L_R$ is the density of all right-handed leptons).
Therefore, setting $L_R=0$ would lead to an equilibrium with
$B=0$, {\it i.e.\/} a vanishing source term for baryon number.
The non-zero
result we have obtained is therefore proportional to the charge
in the right-handed leptons, which gives a small statistical 
factor
related to the fraction of the total number of degrees of freedom
they represent.

When the VEVs approach zero, the result in 
(\ref{eq: dot B mu Y})
does indeed vanish, but not explicitly.  In this limit 
the rate of the `hypercharge violating' processes goes to zero, 
so the ``equilibrium'' calculation is no longer appropriate
as it 
is only valid on a time scale longer than one which diverges
as one sends the VEVs to zero. It is not difficult to
check that one does indeed get zero for the baryon number
in the presence of this source term when we impose the extra
constraint appropriate in this limit
\footnote{
Choosing to impose the conservation on  $T_3$ with chemical 
potential
$\mu_{T_3}$, we get $T_3=\frac{T^2}{6}(10+n)(\mu_{T_3}+\mu_Q)$
and $Q$ also picks up the extra term $(10+n)\mu_{T_3}$.
Imposing $T_3=0$ leaves only the linear combination
$\dot{\theta} + \mu_Q$ in the other equations. The solution
is the trivial unperturbed equilibrium.}. 

\subsection{The Baryon Asymmetry}
 
The equations (\ref{eq: dot B mu B}) and 
(\ref{eq: dot B mu Y}) have the same form for both source terms, 
and so we can analyse them together.
Integrating (\ref{eq: dot B mu B}) gives the baryon asymmetry
as a function of time as
\begin{equation}
B(t)=
 -\int_{t^\prime}^t 
\int_{t_i}^{t}dt^\prime 
c_n\dot\chi(t')T^2(t')
\frac{d}{dt'}\exp\left [ -\int_{t^\prime}^t 
dt^{\prime\prime}\alpha_n\Gamma_{\rm sph}(t^{\prime\prime})
\right ] 
\simeq [c_n\dot\chi T^2]_{\rm freeze}\,,
\label{eq: bsolution}
\end{equation}
where $t_i$ is an initial time chosen
before the phase transition or cross-over 
takes place such that the source
term may be taken to be zero.
This expression is simply the source term 
integrated against the appropriate Green's function.
The freeze-out time (temperature) $t_f$($T_f$) is that
at which the integral in the exponent is equal to one, i.e. 
\begin{equation}
\int_{t_f}^t 
dt^{\prime\prime}\alpha_n
\Gamma_{\rm sph}(t^{\prime\prime})\equiv 1\,.
\end{equation}
The approximation in (\ref{eq: bsolution})
follows since we would expect
that the time-scale characterizing the variation in
${\dot \chi}$ should be of the same order as that characterizing
the change in $\phi/T$. However, as discussed in section 3.2,
there is an exponential dependence in the sphaleron rate 
on $\phi/T$ with a large pre-factor ($\sim 1/\alpha_w$).
This means that the derivative inside the integral in 
(\ref{eq: bsolution}) can be approximated by a delta-function
at $t_f$ down to a time scale much shorter than that over
which $\dot{\chi}$ varies, and the result follows.
The sphaleron rate $\Gamma_{\rm sph}$ enters only
in determining the freeze-out value for the source
$[c_n\dot\chi T^2]_{\rm freeze}$.
Optimally, the sphaleron processes switch off when  the 
source is at its maximum, leading to an estimate of the 
maximum production of baryons at a second 
order phase transition or cross-over:
$B_{\rm max}\simeq [c_n\dot\chi T^2]_{\rm max}$.
To a very good approximation the final
baryon to entropy ratio is 
\begin{equation}
\frac{B}{s} = -\frac{45 c_n}{2\pi^2 g_*} 
\left(\frac{H}{T}\right)_{\rm freeze}
\bigg(T\frac{ d\chi }{dT}
\bigg)_{\rm freeze}\,,
\label{eq: B/s}
\end{equation}
using the fact that $d\chi/dt= -H T d\chi/dT$ 
and that the entropy density  
$s=(2 \pi^2/45)g_* T^3$ 
($g_*$ the number of relativistic degrees of freedom). 
The subscript
denotes that all these quantities are to be
evaluated at the sphaleron `freeze-out'. 

In the case of baryon production in a homogeneous Universe
with source terms of this type the final baryon 
asymmetry is therefore proportional to the expansion
rate at freeze-out. This contrasts completely
with the case of baryogenesis at a first order transition, for 
which the
baryon asymmetry can be effectively the same for an expansion 
rate differing by many orders of magnitude.   

We can invert (\ref{eq: B/s}) to get the range of 
expansion rates consistent with the baryon to entropy 
required by nucleosynthesis\footnote{This range corresponds to
the conservative bounds from direct observations of element
abundances given in \cite{sarkar}. Tighter bounds, corresponding
roughly to the range $(3-9)\times 10^{-11}$ in 
(\ref{eq: expansionbound}), are given in 
\cite{CopiSchrammTurner} and \cite{Fieldsetal}..}
\begin{equation}
\left(
\frac{H}{T}\right)_{\rm freeze} 
\simeq (2-12) \times 10^{-11}\, g_*\, \frac{0.44}{c_n}  
\frac{1}{|(Td\chi/dT)_{\rm freeze}|}
\label{eq: expansionbound}
\end{equation}

How big is $(Td\chi/dT)_{\rm freeze}$ in any given theory? 
A full treatment of the phase transition in any of
the models mentioned would be required to actually
calculate this, a task however considerably beyond
the methods used to date in the study
of the phase transition\footnote{Perturbative methods
apply when the phase transition is fairly strongly first
order. The methods which have been employed to describe
the opposite regime do not include the evolution of the
CP odd fields relevant here.}.   
A {\it naive\/} guess would be 
\begin{equation}
T\frac{d\chi}{dT}\sim\frac{d \phi }{dT}\epsilon
\sim\epsilon
\label{eq: guess}
\end{equation}
taking the field $\chi$ to trace the VEV (or combination
of VEVs), which is itself then assumed to evolve roughly
in proportion to the temperature ({\it i.e.\/} on a timescale 
given by the expansion rate).
The parameter $\epsilon$ is one characterizing CP violation,
which we might expect to be constrained by CP violation
phenomenology of the relevant model.

A full calculation of any given model at finite temperature
would be required to turn the bound (\ref{eq: expansionbound})
into a precise one on the expansion rate alone. However, short
of such a calculation, we can do better than the very naive 
estimate
given by (\ref{eq: guess}).  

(i) In section 3.2 we examined the Minimal Standard Model
and saw that, near the critical temperature, the VEV $\phi$
is a very sensitive function of the temperature, 
with $Td(\phi/T)/dT \approx d\phi/dT \sim 
(100 -30)$ in the
range of temperatures $T_{\rm c}$ to $T_0$, and 
about $60$ at the nucleation temperature. 
Typically the sphaleron
will freeze-out in this range of temperatures, as the
sphaleron rate changes by many orders of magnitude.
The same sort of behavior can be seen to continue
at larger Higgs masses in the non-perturbative 
treatment of the phase transition, 
in the case when the phase transition is 
a ``sharp - but regular - cross-over''
\cite{kajantiecrossover}.
This means that the range of temperatures over which physical 
measurables like the susceptibility vary is a small fraction of 
the temperature at which the change occurs.
(It is, of course, this ``sharpness'' which allows one still
to talk about a phase transition when, strictly speaking, there
is none.) From the data in \cite{kajantiecrossover} we 
see that there
is a range of temperature of a few GeV which compares with
a ``transition temperature'' anywhere between $60$ and
$200$GeV. Thus the standard model estimate of
$d\phi/dT \sim 60$ seems reasonable, much larger than
our naive estimate in (\ref{eq: guess}). 

(ii) We can also learn something about the constraints on
$\epsilon$ by looking at the effective potential for
a particular case. 
Consider a two Higgs doublet model. One interesting regime
is that in which the evolution of the CP violating angle
is determined dominantly by terms which break CP {\it 
spontaneously}.
In a CP invariant Higgs potential \cite{LiuWolfenstein}
only the terms
\begin{equation}
\lambda_5 \big((\phi_1^\dagger \phi_2)^2 +  h.c. \big) \qquad
(\lambda_6 \phi_1^\dagger \phi_1 + \lambda_7 \phi_2^\dagger
\phi_2)(\phi_1^\dagger \phi_2 +  h.c.)
\label{eq: CPinvariant}
\end{equation}
are functions of the relative angle of the two VEVs.
Taking the real parts of the VEVs to be determined
by the rest of the potential ({\it i.e.\/} working in the
approximation that the terms (\ref{eq: CPinvariant})
are small) this gives a quadratic potential for
the cosine of the relative phase, which (taking
$\lambda_5 -\lambda_7$ positive), is minimized at
\begin{equation}
\cos{\theta} = - {\rm min} \left( 1,
\frac{\lambda_6}{4\lambda_5}\frac{v_1}{v_2} +
\frac{\lambda_7}{4\lambda_5}\frac{v_2}{v_1} \right)
\label{eq: CP violating terms}
\end{equation}
There are two (CP conjugate) solutions which will be split
by additional explicit CP violation. 
How the angle changes as 
the VEVs do depends on the values of the ratios of the
couplings $\lambda_6/\lambda_5$ and $\lambda_7/\lambda_5$.
A necessary condition for  $d\theta/dT\neq 0$ at 
the phase transition is $\cos\theta\neq -1$, which 
is the case when 
$\lambda_7/2\lambda_5<v_1/v_2< 2\lambda_5/\lambda_6$.
For  couplings such that
the first term in  (\ref{eq: CP violating terms})
dominates, and $v_1$ changing faster than
$v_2$ as a function of temperature, we have 
\begin{equation}
T \frac{d \theta}{dT} 
\approx -\frac{1}{\tan \theta}\frac{T}{v_1}\frac{dv_1}{dT}
\,,\quad {\rm for} \quad \frac{v_1}{v_2}> 
\left(\frac{\lambda_7}{\lambda_6}\right)^{\frac{1}{2}}\,,
\quad \frac{d\ln v_1}{d\ln T}>\frac{d\ln v_2}{d\ln T}
\label{change-of-theta}
\end{equation}
Typically we have $T{d \theta}/{dT} \approx {dv_1}/{dT}$,
but there are also parts of parameter space 
(near $\cos \theta =-1$) 
where the phase can change much faster than this. 
The only role of the explicit CP violation here is
to split the two degenerate minima so that the same sign is
chosen everywhere. This illustrates that the  constraint on the 
parameter which we called $\epsilon$ from CP violating phenomena 
at zero temperature may be extremely weak. 
With a moderate fine tuning it can be
considerably larger than one, and not related directly to any
small parameter associated with the smallness of CP violation.
In fact in theories such as the Minimal Supersymmetric 
Standard Model (MSSM) it is naturally the case that 
the terms which break CP spontaneously
(which are induced in the plasma through thermal corrections) 
are dominant over the terms which break CP explicitly (which are 
suppressed by a loop factor) \cite{ComelliPietroniRiotto}.
We conclude, on the basis of a simple analysis of the two Higgs
doublet model, that the naive estimate 
(\ref{eq: guess}) for $T\frac{d \chi}{dT}$ 
with $\epsilon \sim 1$ is too small by 
about two orders of magnitude.
A result of this magnitude is  obtained for a large portion 
of the parameter space, without any tuning. 
With a moderate fine tuning, the effective CP violation can
be further enhanced. To make a more precise statement would
require a detailed analysis of the Higgs sector of the 
particular model.

\section{Non-Standard Cosmologies}

Having established quantitively the dependence of
the baryon asymmetry on the expansion rate in
two possible scenarios for baryon production at the
electroweak scale,
we now turn to the discussion of physical mechanisms
which could lead to such a different expansion rate at the
electroweak scale.

As mentioned in the introduction this kind of question has previously
been treated in the context of calculations of relic densities
of weakly interacting particles in \cite{Barrow}
and \cite{KamionkowskiTurner}. 
The relic density of a weakly interacting species is determined by the
temperature at which the species decouples from the
ordinary (visible) matter, which depends,
just as in the case of the sphaleron decoupling discussed above,
on a comparison between the appropriate interaction rate and
the expansion rate of the Universe. In typical models this
decoupling occurs before nucleosynthesis, and therefore one
is led to consider, just as we are doing here, possible 
alternatives
to radiation domination at that epoch. The alternative which is
considered
exclusively in \cite{Barrow} and in most detail in
\cite{KamionkowskiTurner} is:

(i) An Anisotropic Universe: A universe which is homogeneous but
not isotropic is described by a metric with three scale factors,
one for each spatial dimension. With an adiabatic approximation
({\it i.e.\/} expansion slow enough to allow thermalization)
it is the effective volume expansion rate $\overline{H}$ 
associated with an
average scale factor $\overline{a}$ which determines how the 
temperature changes in the same way as in the isotropic FRW
spacetime. The equation of motion for $\overline{H}$ is just 
that of the FRW space, but with an additional term which is
equivalent to a component of the energy density scaling as
$1/\overline{a}^{\,6}$.

A further possibility considered in \cite{KamionkowskiTurner} is

(ii) Non-standard theories of gravity. The case studied
in \cite{KamionkowskiTurner} is a Brans-Dicke theory, which 
again turns out to effectively produce an extra component
in the energy density scaling as $1/a^6$.  There are also of course many
other variants on standard Einstein gravity which can be 
considered.

The simple possibility we will concentrate on is:

(iii) Einstein gravity with isotropy and homogeneity, 
but with an extra contribution to the energy density which is important
prior to nucleosynthesis. As noted by one of us (MJ) in
\cite{MJletter} any mode of a scalar field dominated by its kinetic 
energy has the required property, as its energy density can scale as
fast as $1/a^6$. The electroweak phase transition could
potentially occur during a phase of the Universe dominated by the
kinetic energy of a scalar field, 
termed  {\it{kination}} in \cite{MJletter}, 
which can end before nucleosynthesis as
the kinetic energy density red-shifts away relative to the
radiation.
Below we will discuss several ways in which such a phase
can come about within the context of inflationary cosmology,
which explains the assumed isotropy and homogeneity.
In particular we will discuss models which come
naturally out of an alternative to the usual theory
of reheating discussed by Spokoiny in \cite{spokoiny}.

A clear motivation for considering such models 
follows from the calculations in the previous section. If we
have such a component scaling as $1/a^6$, the expansion rate
is given by
\begin{equation}
H^2= \left(\frac{\dot{a}}{a}\right)^2= \frac{8 \pi
G}{3}\frac{\rho_e}{2}\left[
\left(\frac{a_e}{a}\right)^6 + 
f(a)\left(\frac{a_e}{a}\right)^4
\right]\,,
\label{eq: einsteinb}
\end{equation}
where $a_{e}$ is the scale factor when the density
in the mode becomes equal to that in radiation and
$\rho_e$ is the total energy density at that time. 
The factor $f(a)$ accounts for the effect of decouplings,
and, assuming adiabatic decouplings,
$f(a)=[g(a_e)/g(a)]^{1/3}$, where $g(a)$
is the number of relativistic degrees of freedom as a function
of the scale factor $a$.
Nucleosynthesis constraints place a lower bound
on $T_e$, the temperature at the time of equality
of radiation-kinetic energy density, which can be
inferred from the corresponding bounds on additional
relativistic particle degrees of freedom.
This is the case since the predominant effect of 
such extra degrees of freedom is also in the change they cause to 
the expansion rate at the beginning of nucleosynthesis,
which determines the crucial ratio of neutrons to protons
when the weak interactions drop out of equilibrium at
$\sim 1$MeV. We take here the conservative bounds of
\cite{sarkar}, which allow the equivalent
of 1.5 extra Dirac neutrino degrees of freedom over
the three degrees of freedom of the Standard Model,
{\it i.e.\/}
we allow an additional energy density at $1$MeV which is 
$3(7/8)/10.75\approx 0.25$ of the standard
model one (with $10.75$ degrees of freedom). From
(\ref{eq: einsteinb}) this means
\begin{equation}
\frac{1}{\sqrt{f(a_{\rm ns})}}\frac{a_e}{a_{\rm ns}}
\lesssim 0.5
\label{nucleo-bound}
\end{equation}
Using $Ta=f(a)T_e a_e$ this gives 
the upper bound on the expansion rate at the electroweak
scale 
\begin{equation}
\frac{H}{T} \lesssim 1.8 \times 10^{-11}
\left(\frac{T_{\rm freeze}}{100\hbox{\rm GeV}}\right)^2
\label{nucleosynthesis-upperbound}
\end{equation}
The result differs only by $\sqrt{3/2}$ if we take the less
conservative nucleosynthesis bound of \cite{CopiSchrammTurner}
and \cite{Fieldsetal}.

Taking $H_{\rm max}$ to be the expansion rate corresponding to
the upper bound (\ref{nucleosynthesis-upperbound}),  
the requirement (\ref{eq: expansionbound}) for 
generation of the observed BAU at the electroweak scale in
a homogeneous Universe can be expressed
as a requirement of the relevant CP violating parameter
\begin{equation}
\left\vert T\frac{d\chi}{dT}\right\vert_{\rm freeze} 
\approx  (1-6) g_*
\left(\frac{100 \hbox{\rm GeV}}{T_{\rm freeze}}\right) ^2 
\left(\frac{H_{\rm max}(100 \hbox{\rm GeV})}
{H(100 \hbox{\rm GeV})}\right)
\label{eq: CP violation bound}
\end{equation}
Absorbing the nucleosynthesis limit, {\it i.e.\/}
taking $H = H_{\rm max}$
we have the strict lower bound \footnote{ 
Note that taking the upper bound on the expansion rate 
in (\ref{nucleosynthesis-upperbound}) corresponds to
absorbing the upper bound on effective number of degrees of freedom 
at nucleosynthesis, which is only consistent with the
lowest baryon to entropy ratio (increasing the expansion rate 
increases the fraction of Helium).} 
\begin{equation}
\left\vert T\frac{d\chi}{dT}\right\vert_{\rm freeze} 
\geq g_* \left(\frac{100 \hbox{\rm GeV}}
{T_{\rm freeze}}\right)^2 
\label{eq: CP violation bound strict}
\end{equation}

  From the analysis in section 4.3 it follows that this bound
may indeed be satisfied in extensions of the MSM such as those 
we have discussed, without any fine tuning. 
If the upper bound (\ref{nucleosynthesis-upperbound}) 
on the expansion rate is not saturated, the CP violation parameter
is required to be larger as given by (\ref{eq: CP violation bound}).
As discussed in section 4.3, with some fine-tuning of parameters
in the potential, this parameter can indeed  
be enhanced to considerably greater than 
the typical value which just satisfies the lower bound 
(\ref{eq: CP violation bound strict}). An exact statement of
how large it can be would require a detailed examination of 
the model in question.

The important result is that in a cosmology with an additional
component scaling as $1/a^6$ which dominates prior to
nucleosynthesis, the creation of the baryon asymmetry is possible
({\it i.e.\/} consistent with all observations) at the 
electroweak scale in a homogeneous expanding Universe. 
The fact that generation of the BAU in this case has generally
been dismissed as impossible provides clear motivation for the
consideration of such cosmologies in greater detail. Certainly
also as experiment pushes the bounds on scalar particles upwards,
the usual sphaleron bound for generation of the BAU is becoming
increasingly severe and alternative mechanisms for the production
of the BAU within the context of electroweak cosmology become more
relevant.

\subsection{Kination}

Consider the dynamics of a real scalar field $\phi$
with potential $V(\phi)$. Varying 
the action
\begin{equation}
{\cal{S}}= \int d^4 x \sqrt{-g} \left[ \frac{1}{2} g^{\mu
\nu}(\partial_{\mu} \phi)^{\dagger} (\partial_{\nu} \phi)
- V(\phi)\right]
\end{equation}
and taking the FRW metric $ds^2=dt^2-a(t)^2d{\vec x}^{\,2}$ with scale
factor $a(t)$, gives the equation of motion for the homogeneous modes
which, after multiplication by $\dot{\phi}$, can be written as follows
\begin{eqnarray}
\frac{d}{dt}\left[\frac{1}{2}\dot{\phi}^2+V(\phi)
\right] + 3 H 
\dot{\phi}^2=0\,.
\label{eq: eompotl}
\end{eqnarray}
Defining $\zeta(t) = {V(\phi)}/{\rho(\phi)}$, 
where $\rho(t)=\dot{\phi}^2/2 + V(\phi)$, we find
\begin{eqnarray}
\rho(t) = \rho(t_o) \exp{-\int_{t_o}^{t} 6[1 -\zeta(t)] 
H(t) dt}
= \rho(t_o) \exp{-\int_{a_o}^{a} 6[1 - \zeta(a)] \frac 
{da}{a}}\,.
\label{eq: scaling}
\end{eqnarray}
When the kinetic energy dominates,
$\zeta \rightarrow 0$ and
\begin{equation}
 \rho  \propto \frac{1}{a^6}\,.
\label{eq: sixscaling}
\end{equation}
If a potential
possesses a flat direction, for example, the energy in the
associated coherent goldstone mode scales in this way.
In this case the scaling
can be seen to follow directly
from the conservation of the Noether current
associated with the symmetry. Consider for example
a complex scalar $\Phi$ with a potential invariant
under the global symmetry $\Phi \rightarrow e^{i \theta}\Phi$
\begin{equation}
 {\lambda}(\Phi^\dagger \Phi - v^2)^2\,.
\end{equation}
The mode $\Phi=ve^{-i \theta}$ with $\dot{\theta}=const$
is a solution of the equations of motion for which the
conserved Noether charge is 
\begin{equation}
j^{o}\equiv\rho_\theta= a^3 i \Phi^{\dagger}
\stackrel{\leftrightarrow}{\partial^{\,o}}\Phi
=2a^3 v^2 \dot{\theta}\,.
\end{equation}
Thus $\dot \theta \propto 1/a^3$ and $\rho=v^2 
\dot\theta^2/2 \propto 1/a^6$.

Such kinetic energy dominated modes  represent
the opposite limit to inflation \cite{inflation}
which is driven by  potential
energy so that $\zeta \rightarrow 1$ and $\rho(t) \approx
\rho(t_o)$.
Indeed for any homogeneous mode (assuming only that $V(\phi)$
is positive) we have that
\begin{equation}
\rho(t_o)\left(\frac{a_o}{a}\right)^6 
\leq \rho(t)\leq\rho(t_o)\,,\qquad t\geq t_o\,.
\label{eq: scalinglimits}
\end{equation}
Instead of superluminal expansion in
inflation, a kinetic energy dominated mode of a
scalar potential drives a subluminal expansion
very similar to that of radiation 
($a\propto t^{{1}/{2}}$)
or matter ($a \propto t^{{2}/{3}}$).
Writing the stress energy tensor in terms
of the pressure $p$ and the energy density $\rho$ 
in the standard way, the equation of state is
$p=\rho$ for the kinetic mode,
in contrast to $p=({1}/{3})\rho$ (radiation),
$p=0$ (matter), and $p=-\rho$ (inflation).

We now consider various ways in which a phase
of kination could come about.
Inflation is the standard paradigm which explains
isotropy and homogeneity of the Universe as it appears
today. A scalar field drives a period of
inflation and subsequently decays,
filling the Universe with radiation and matter.
We will assume that a period of inflation
produces the isotropic and homogeneous
Universe, but ask how it might come about that
after inflation a reheated universe
would be dominated by a kinetic scalar mode.

Two questions can be separated:

$\bullet$
How can a scalar field potential 
support a mode that is kinetic energy dominated? This 
question is twofold.
Firstly, what shape must the potential have to keep the kinetic 
energy dominant? Secondly, what is required of the field in 
order that
energy does not leak out of the coherent mode?

$\bullet$
How can kinetic modes come to dominate  
the energy density, {\it i.e.\/} how can they be
excited?

Let us start with the {\it first\/} question.
The most trivial case
of potential energy domination is the example used above of
an exactly flat potential. This case is not of interest here, 
since the energy in such a mode is negligible at the 
end of inflation,
as it also red-shifts away as $1/a^6$ during inflation. 
Any kinetic energy domination must therefore occur through the 
roll  of a field in a potential after inflation. The  dynamics of a
homogeneous real scalar field $\phi$  with potential $V(\phi)$
in an expanding FRW universe are described by the equations
\begin{eqnarray}
\ddot{\phi} + 3 H \dot{\phi} +
V'(\phi) &=&
\frac{1}{a^3}\frac{d}{dt}\left(a^3\dot{\phi}\right)+V'(\phi)=0
\label{eq: potleoma}\\
H^2 &=& \frac{1}{3M_{\rm P}^2}
\left[\frac{1}{2}\dot{\phi}^2 + V(\phi)  + 
\rho_r\right]
\label{eq: potleomb}\\
\dot{\rho_r} + 4H \rho_r &=& 0
\label{eq: potleomc}
\end{eqnarray}
where $\rho_r$ is the energy density in radiation, to which we 
assume the
scalar field to be coupled only
through gravity. This is a roll damped 
by the expansion of the Universe and the first question is 
therefore: how steep 
must a potential be in order that the roll be more effective 
in creating kinetic energy than the damping is in attenuating it?
A hint of the solution is immediately given by considering again 
the trivial  case
$V(\phi)=0$, which gives the solution
\begin{equation}
\dot\phi(t)= \dot\phi_o \left(\frac{a_o}{a}\right)^3
=\dot\phi_o \left(\frac{t_o}{t}\right)\,,
\qquad \phi(t)= \phi_o +\dot\phi_o t_o \ln \frac {t}{t_o}
\label{eq: phikin}
\end{equation}
when $\rho_r=0$.
If, with this solution for $\phi$, the potential is such that  
the terms which depend on it decrease 
faster than the other terms in the equations of motion, 
the kinetic energy  domination will continue once established. 
Given that the time dependence is logarithmic, it is clear
that an exponential potential is what is required.
That exponential potentials define what is steep
enough for kination can be seen by taking the explicit
attractor solutions given in \cite{JHalliwell} for
the potential 
$V(\phi)=V_o {\rm e}^{-\lambda\phi}$
\begin{equation}
\phi(t)=
\frac{2}{\lambda}\ln t\,,
\qquad a\propto t^{\frac{2}{\lambda^2}}\,,
\qquad \zeta=1-\frac{\lambda^2}{6}\,,
\quad \lambda^2<6\,,
\label{eq: jhsoln}
\end{equation}
and the origin of $\phi$ is redefined so that
$V_o=2\lambda^{-2}\left(6\lambda^{-2}-1\right)$.  
(For simplicity we took $M_{\rm P}=1$.)
The context within which \cite{JHalliwell}
discussed this potential was ``power-law inflation'', 
for which the superluminal expansion
occurs when $\lambda<\sqrt{2}$. From (\ref{eq: scaling})
it follows that $\rho \propto a^{-\lambda^2}$
so in the limit $\lambda \rightarrow \sqrt{6}$
we recover the scaling of an exactly flat potential.
When $\lambda>\sqrt{6}$ the pre-factor 
cannot be written in this way. There is 
no single attractor solution, and the ratio $\zeta$ rather
than being fixed approaches zero asymptotically.

In this analysis we have assumed a simple roll down a 
potential. Another possibility is that a field oscillates 
about a minimum. It is easy to see from (\ref{eq: scaling})
the well-known result that an oscillation in a quadratic
potential gives an energy scaling like matter $\propto 1/a^3$
since $\zeta$ can be replaced by its average 
$\langle\zeta\rangle={1}/{2}$ 
over a time-scale of the expansion time.
The analysis for the $\phi^n$ potential is 
given in \cite{Turner}. The result is 
that an oscillating mode scales as $a^{-{6n}/{(n+2)}}$ 
and, correspondingly, $\langle\zeta\rangle={2}/{(n+2)}$,  
so that the kinetic energy becomes more
dominant as $n$ increases. 

We also require that, if such a mode is excited, the energy 
remains in it, {\it i.e.\/} that it does not leak out by 
decay of the coherent mode into particle excitations of
itself or other fields to which it is coupled. In
the present context we want to maximize the effect
of the mode and therefore need the energy to stay in the
kinetic mode from before the electroweak scale
until shortly before nucleosynthesis. Potentials which
support coherent modes which are so weakly coupled
to other fields (or self-coupled in the case of massless
fields) that they do not decay before nucleosynthesis
are in fact commonplace in particle physics - they are
the source of problems like the Polonyi problem. In
particular, exponential potentials which arise in
theories involving compactifications are typically
extremely weakly coupled to other sectors. 
Accordingly, these sectors are termed ``hidden,'' 
as they interact with the ``visible'' matter 
only (or predominantly) gravitationally. 
We will thus
assume that perturbative decay is negligible. On the
other hand non-perturbative decay mechanisms like 
parametric resonance which have been much discussed
recently in the context of the problem of reheating
after inflation 
\cite{TraschenBrandenberger, GreeneKofmanLindeStarobinskii,
KofmanLindeStarobinskii, 
Boyanovskyetal, KhlebnikovTkachev, ProkopecRoos},
must be considered in the case of oscillating modes 
of non-linear potentials. 
This is a possibility we will consider explicitly below, and
the requirement that such a mode survive until nucleosynthesis
will place constraints on the potentials we consider.

The {\it second\/} question above concerned how such a mode would come
to dominate over radiation. In the analysis just given of a field
rolling down a potential we set $\rho_r=0$, and the solutions are
therefore valid only if $\rho_{\phi}\gg\rho_r$ (where $\rho_{\phi}$ is
the total energy in the scalar field).  What happens if
$\rho_r\gg\rho_{\phi}$? Can we end up rolling into the kinetic energy
dominated mode with $\rho_{\phi}\gg\rho_r$ if we are in an exponential
with such a mode?  In a radiation dominated universe $a\propto
t^{1/2}$ ($H={1}/{2t}$) so that the damping is stronger than in
the kinetic energy dominated universe where $a \propto t^{1/3}$
($H=1/3t$).  If the scalar field is in a kinetic energy
dominated mode, we find, assuming radiation domination, that
\begin{equation}
\dot\phi(t)= \dot\phi_o (\frac{a_o}{a})^3=\dot\phi_o 
\left(\frac{t_o}{t}\right)^{\frac{3}{2}}\,,
\qquad \phi(t)= \phi_o +2\dot\phi_o t_o\left( 1- (\frac 
{t_o}{t})^{\frac{1}{2}}\right)\,.
\label{eq: phiradn}
\end{equation}
The result is that the exponential potential energy will always 
``catch up'' 
with the kinetic energy and the field will be driven into a mode 
which scales
much slower than radiation, until $\rho_{\phi} \sim \rho_r$. 
These
quite different behaviours in the two limits $\rho_\phi\gg
\rho_r$ 
($\rho_\phi \propto 1/a^6$) and $\rho_\phi\ll\rho_r$ ($\rho 
\sim const$)
which tend to drive the system from one regime to the other 
suggests that
there may be an attractor solution with $\rho_\phi\sim 1/a^4$. 
That such a solution exists and is an attractor has been noted 
in \cite{Wetterich} and \cite{RatraPeebles}. 
It can in
fact be generalized to the case that the non-scalar component 
scales as  $\rho_m\propto 1/a^m$  
({\it e.g.\/} non-relativistic matter with $m=3$), 
and all components
scale as it does with the ratios of their contributions given by
\begin{equation}
 \zeta=1-\frac{m}{6}\,, \qquad
\frac{\rho_\phi}{\rho_\phi + \rho_m}= \frac{m}{\lambda^2}\,.
\label{attractor}
\end{equation}
In this case one need not assume that 
the pre-factor in the exponential
can be written in the special form required for
the solution (\ref{eq: jhsoln})
and $\lambda$ can take on any value 
$\lambda>\sqrt{m}$,
which is just the requirement that the attractor mode 
in (\ref{eq: jhsoln}) with $\rho_m=0$ scale faster than 
$1/a^m$. 
 
The existence of this attractor means that if we start in a 
radiation dominated universe (or, more generally, in a universe
dominated by energy scaling  as $1/a^m$) 
we will always end up in this solution (\ref{attractor}) rather
than in the kinetic energy dominated mode of the exponential. 
In order to realize kination in this potential we must therefore 
satisfy the condition at the end of inflation, or some time 
after it, 
that the kinetic energy dominate over the radiation. The 
dynamics of 
the simple exponential alone will not produce kination if we have
a radiation dominated universe after inflation. We will examine
two possibilities: (i) A non-standard theory of re-heating in 
which the
radiation in the Universe is that created by the expansion
of the background during inflation, and radiation is 
naturally subdominant after inflation, and (ii) standard reheating
with a slightly different exponential potential which can first
cool the radiation with a short inflationary phase and then roll
into a kinetic mode.

We will concentrate on this first kind of model, because in it 
there must by construction be a phase of kination, and in our 
view it offers a very attractive (and unjustifiably neglected) 
alternative to the standard re-heating scenario. 
What sort of model would lead to this phase ending as late 
as nucleosynthesis will be the question 
which interests us in the specific context of electroweak 
cosmology.

In the oscillatory potential things are slightly different.
The scaling was predicated on the assumption that the
field oscillated on a time-scale short compared to the expansion
time, but not on any assumption about the time dependence of the 
expansion rate ({\it i.e.\/}
about which component dominates the energy
density). Thus if the Universe is radiation dominated
when we enter the oscillatory mode of a potential with $n>4$, 
it will always be radiation dominated since the energy in the 
scalar field red-shifts away faster. To realize kination in this
potential we therefore require the radiation to be sub-dominant
when the oscillatory phase begins. Just as for the exponential
potential we will discuss in this case how this condition can 
be realized in two ways: 
(i) in the same alternative 
standard theory of reheating after a period of inflation
driven by the power-law potential itself, and (ii) with 
ordinary reheating by another field followed by a brief subsequent 
period of inflation as the field with the power-law potential
rolls before  it begins oscillating. Again most of our attention
will focus on the first case, in which a single field is both
inflaton and `kinaton'. 

\subsection{Reheating by Kination}

Reheating after inflation is required in order to match the 
`cold empty'  Universe left behind by inflation onto the 
radiation dominated one 
which must in the standard cosmology be established by 
nucleosynthesis at the latest, (and usually, it is assumed, at some
temperature high enough to support some theory of baryogenesis).
In the standard theory  this is 
achieved by the decay of the
inflaton into particles in an oscillatory phase after inflation, 
the 
zero entropy coherent state producing the enormous entropy of the
radiation dominated universe. That there exists a simple 
alternative to this scenario has been pointed out by Spokoiny in 
\cite{spokoiny}. The Universe is {\it not\/} in fact in an
exactly cold zero entropy state 
after inflation -- besides the energy in the inflaton, there is 
some energy in the particles created by the accelerated 
expansion. The process
which gives rise to the perturbations from homogeneity required
for structure formation on large scales creates an energy 
density which is peaked at the scale $H$, where $H$ is the 
expansion rate during inflation, with energy density 
$\delta\rho^{H}=\epsilon_\rho H^4$, where the 
superscript {\it H\/} denotes that this energy
density is dominated by the scale $k\sim H$,
and  $\epsilon_\rho\simeq (\pi^2 g_*^{eff}/30)/(2 \pi)^4$
\cite{brandenberger},  
where $g_*^{\rm eff} \sim 10^2$ 
is the effective number of light  
(mass $m < H$) degrees of freedom\footnote{This estimate
assumes the same contribution from all particles
as from the scalar particles analysed in
\cite{brandenberger}.}. 
In a typical inflationary model with energy density 
$\rho_{\rm i\,,\; end}$ 
at the end of inflation
\begin{equation}
\frac{\delta\rho^H_{\rm i\,,\; end}}
{\rho_{\rm i\,,\; end}} 
\simeq \frac{\epsilon_\rho}{3}
\frac{H_{\rm i\,,\; end}^2}{M_{\rm P}^2} 
\simeq \frac{\epsilon_\rho}{9}
\frac{\rho_{\rm i\,,\; end}}{M_{\rm P}^4}\,,
\end{equation}
which is very  small since the energy scale associated with 
inflation
is typically required to be well below $M_{\rm P}$,
{\it e.g.\/} for
chaotic inflationary model in a potential
$\lambda \phi^4$, ${\delta\rho^H_{\rm i\,,\; end}}
/{\rho_{\rm i\,,\; end}}\sim \epsilon_\rho\lambda$, 
while the requirement
that one gets density perturbations
of the correct magnitude
(on COBE scales  ${\delta \rho}/{\rho}\sim 5\times 10^{-5}$) 
gives $\lambda \sim  10^{-13}$.
In the context of ordinary re-heating this small fraction is 
irrelevant as it is swamped by the radiation created by inflaton decay.
The possibility envisaged in \cite{spokoiny} is one which is easy to
see given the observations of the preceeding section on kination:
If, instead of decaying, the inflaton rolls into a potential in 
which its energy density scales as $1/a^s$ with $s>4$, the 
energy density
$\delta\rho^H$ will come to dominate at some time 
$t_{\rm k\,,\; end}$ after inflation 
when the scale factor has evolved to $a_{\rm k\,,\; end}$
from $a_{\rm i\,,\; end}$ 
at the end of inflation with 
\begin{equation}
\frac{a_{\rm k\,,\; end}}{a_{\rm i\,,\; end}} \approx 
\left(
\frac{9 }{\epsilon_\rho}
\frac{M_P^4}{\rho_{\rm i\,,\; end}}
\right)^{\frac{1}{s-4}}=
\left(
\frac{3}{\epsilon_\rho}\right)^{\frac{1}{s-4}}\,
\left(\frac{M_P}{H_{\rm i\,,\; end}} \right)^{\frac{2}{s-4}}\,,
\label{endkination}
\end{equation}
where $a_{\rm k\,,\; end}$ is the scale factor at the end of 
kination (during which $\rho \propto 1/a^s$), the phase 
which interpolates between inflation and radiation domination. 
The energy in the inflaton simply red-shifts away instead 
of decaying. 
As discussed in \cite{spokoiny}, in order to accommodate 
nucleosynthesis there are two requirements which must be 
fulfilled: 
(i) the radiation must thermalize at a temperature above
$1$MeV and, (ii) the transition to radiation dominance must 
occur sufficiently long before nucleosynthesis to satisfy 
the appropriate
constraints at that time on the expansion rate. 
Taking $k_{\rm eff}(a)$ to be the typical energy of the 
created radiation as a function of scale factor, we have 
$k_{\rm eff}= H_{\rm i\,,\; end} {a_{\rm i\,,\; end}}/{a}$.
Assuming that the dominant form of this radiation 
is in standard model degrees of freedom, 
the interaction rate for processes coupling them is
$\sim \alpha^2 k_{\rm eff}$ (for $k_{\rm eff}\gg M_W$ and 
$\alpha \sim {1}/{30} - {1}/{50}$).
Comparing this to the expansion rate \footnote{A far from equilibrium system 
may in fact need many rescatterings ({\it i.e.\/} $N_{\rm scatt}\gg 1$) to
fully thermalize. Modifying the estimate in (\ref{thermtemp}) to incorporate
this gives  $T_{\rm reheat}$ smaller by a factor 
$N_{\rm scatt}^{-2/(s-2)}$.}
$H \approx {2}/{st} \approx H_{\rm i\,,\; end} 
({a_{\rm i\,,\; end}}/{a})^{{s}/{2}}$
(in kination), we get an estimate for the thermalization
temperature $T_{\rm reheat}$
\begin{equation}
T_{\rm reheat} \sim H_{\rm i\,,\; end} 
\left(\frac{30 \epsilon_\rho}
{\pi^2 g_{* \,,\; \rm i \,,\; \rm end}}\right)^{1/4} 
\alpha^{\frac{4}{s-2}}
\label{thermtemp}
\end{equation}
where $g_{* \,,\; \rm i \,,\; \rm end}$ 
is the number degrees of freedom which are relativistic at 
$k \sim H_{i \,,\; \rm end}$, and we have defined 
$\delta\rho^H_{\rm i\,,\; end}= g_{* \,,\; \rm i \,,\; \rm end} \pi^2
T_{\rm i \,,\; \rm end}^4/30$, and taken $T \propto 1/a$
 \footnote{Here and below we neglect the effect of possible 
decouplings between $T_{\rm reheat}$
and ${T_{\rm k\,,\; end}}$, {\it i.e.\/}
we assume the number of relativistic
degrees of freedom to be fixed.}. 
Assuming this temperature $T_{\rm reheat}$ to be attained before the transition
to radiation dominance, it follows from (\ref{endkination}) 
that $T_{\rm k\,,\; end}$, the temperature at the beginning of 
radiation domination, is given approximately by 
\begin{equation}
\frac{T_{\rm k\,,\; end}}{M_P} 
\simeq 
\left(\frac{\epsilon_\rho}{3}\right)^{\frac{1}{s-4}}
\left(
\frac{30\epsilon_\rho}{\pi^2  g_{* \,,\; \rm i \,,\; \rm end}}
\right)^{\frac{1}{4}}
\left(\frac{H_{\rm i\,,\; end}}{M_P} 
\right)^{\frac{s-2}{s-4}}
\label{kination-endtemp}
\end{equation}
Requiring this to be above the nucleosynthesis temperature 
$1$MeV places a lower bound on $H_{\rm i\,,\; end}$. For $s=6$
we find that $H_{\rm i\,,\; end} > 10^{7}$ GeV, which 
corresponds to $T_{\rm reheat} > 10^{6}$GeV, consistent with 
the assumption that $T_{\rm reheat} > T_{\rm k\,,\; end}$. 
For $s=5$ both $H_{\rm i\,,\; end}$ and $T_{\rm reheat}$ 
are greater by a factor of $\sim 10^4$. 
In both cases a late transition to radiation dominance
implies that the energy scale at the end of inflation and
thermalization scale are well below the GUT scale.

In standard inflationary models the usual constraint on 
$H_{\rm i\,,\; end}$ or
the energy density at the end of inflation comes from the 
requirement that the amplitude of perturbations be that required
for structure formation. In the models which we discuss below we
will consider how this non-trivial constraint is satisfied in
this model of reheating (a question not considered in 
\cite{spokoiny}), 
and in particular how it fits with the particular type of 
realization 
of this model we are interested in, where 
the transition to radiation domination does actually occur 
close to nucleosynthesis with the potentially important 
consequences
for electroweak baryogenesis discussed in the first part of this
paper.

\subsection{Inflation-Kination in an Exponential Potential}

As discussed in section 5.1, a simple exponential 
which gives rise to the kinetic energy 
dominated mode required for kination does not itself accommodate 
an inflationary solution. We need to have a potential which is 
flatter in some region (for inflation) and sufficiently steep
(for kination) in the part of the potential the field rolls into
after inflation. 
An example is an exponential $\sim e^{-\lambda\phi}$ where 
$\lambda$ varies  as a function of $\phi$. As a simple case of this,  
which we can treat analytically, we consider\footnote{We could of course
consider any potential which accommodates inflation in some region 
and is asymptotically a sufficiently steep exponential. Motivation 
for an exponential with varying $\lambda$ is given in \cite{Wetterich}.}
\begin{eqnarray}
V(\phi)=V_o {\rm e}^{-\lambda\phi}\,,\quad {\rm where} 
\quad & \lambda
< \sqrt{2} \quad \rm{for} 
\quad  \phi < \phi_{\rm i\,,\; end} \nonumber \\
& \lambda\equiv\lambda^\prime
> 2 \quad \hbox{\rm for} \quad \phi > 
\phi_{\rm i\,,\; end}
\label{twoexponentials}
\end{eqnarray}
where we set $M_{\rm P}=1$.
As discussed above, one solution to the 
equations of motion for this potential is 
a power-law inflationary attractor 
(\ref{eq: jhsoln}) with  
$\phi=\frac{1}{\lambda}
\ln\left(\frac{V_o \lambda^4 t^2}{2(6-\lambda^2)}\right)$,
and $a \propto t^{2/\lambda^2}$. 
We assume the field to evolve in this attractor 
in inflation from $\phi\ll\phi_{\rm i\,,\; end}$.
When the field reaches $\phi_{\rm i\,,\; end}$ inflation
ends and after a transient period it
will roll, for $\lambda ' < \sqrt{6}$,  into the
new attractor in the steeper potential.
 If $\lambda^\prime\ge \sqrt{6}$, there is 
no single attractor, but the field will run
after a few expansion times into a solution
in which the kinetic energy is very dominant.
In either case we will neglect the details of the 
few expansion times in which this transition occurs.

We calculate first the cosmological perturbations generated
in the usual way by the amplification of quantum fluctuations
during inflation. The amplitude of the perturbation in a mode
with comoving momentum $k$ when it re-enters the horizon after
inflation at time $2X$, is given by the usual formula
\begin{equation}
\frac{\delta\rho}{\rho}(k)\approx 
\epsilon_\delta\frac{H_{1X}^2}{\dot\phi_{1X}}\,,
\label{eq:cos-per I}
\end{equation}
where $1X$ denotes the time when the perturbation $k$ exits 
the horizon in inflation, and 
$\epsilon_\delta=3/5\pi\simeq 0.2$ \cite{inflation} 
is a constant. The formula is valid provided the slow-roll
approximation holds at this time, which in the case of the 
exponential potential corresponds to $\lambda \ll \sqrt{2}$.
\begin{figure}[htb]
\epsfxsize=5.5in
\centerline{\epsfbox{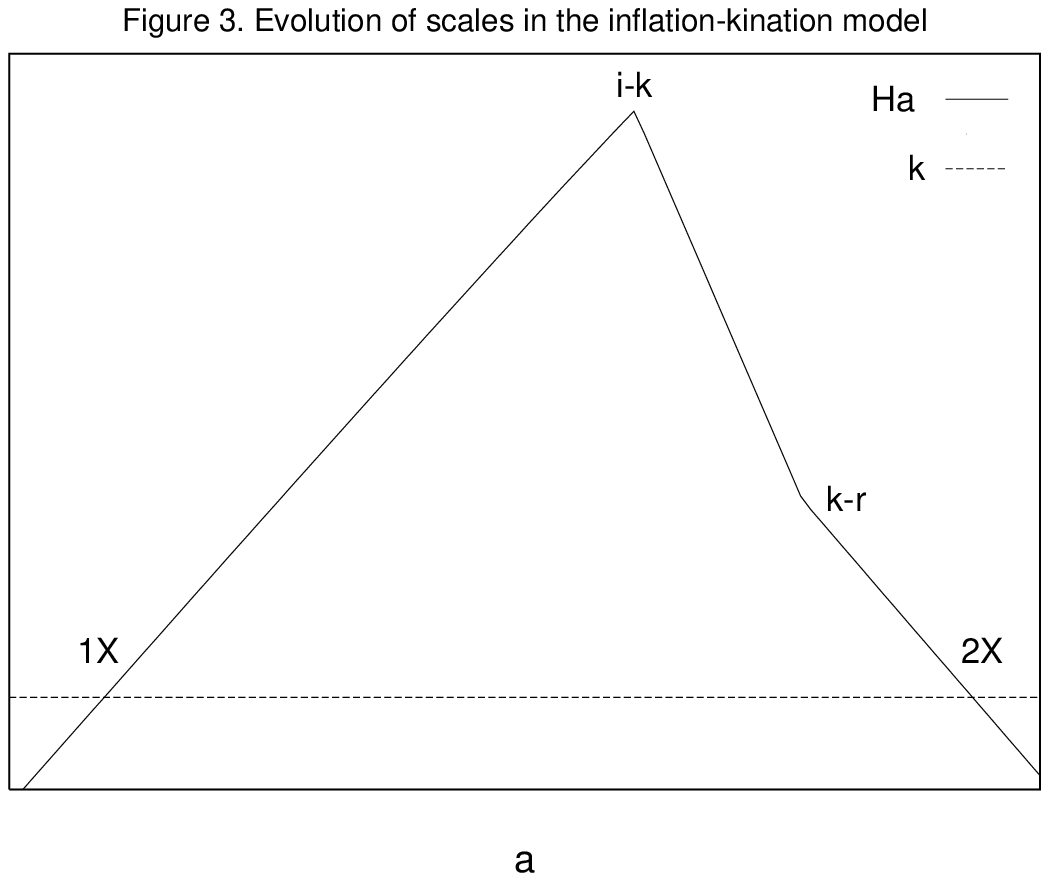}}
\end{figure}
The evolution of scales is illustrated on figure 3, where
$k=k_{\rm physical}a$ is plotted versus $Ha$
(both on the logarithmic scale). 
Since the comoving scale is fixed, it follows that 
\begin{equation}
k\equiv k_{1X}=k_{2X}\quad \Leftrightarrow 
\quad(Ha)_{1X}=(Ha)_{2X}\,,
\label{eq:cos-per II}
\end{equation}
and therefore
\begin{equation}
\frac{(Ha)_{1X}}{(Ha)_{\rm i\,,\; end}}\,
\frac{(Ha)_{\rm k\,,\; beg}}{(Ha)_{\rm k\,,\; end}}\,
\frac{(Ha)_{\rm r\,,\; beg}}{(Ha)_{\rm r\,,\; end}}\,
\frac{(Ha)_{\rm m\,,\; beg}}{(Ha)_{2X}}=1\,,
\label{eq:cos-per III}
\end{equation}
where the indices {\it i, k, r, \rm and \it m\/} denote 
{\it inflation, kination, radiation \rm and \it matter\/}, 
respectively,  and we have assumed that 
$(Ha)_{\rm i\,,\; end}=(Ha)_{\rm k\,,\; beg}$,
$(Ha)_{\rm k\,,\; end}=(Ha)_{\rm r\,,\; beg}$,
$(Ha)_{\rm r\,,\; end}=(Ha)_{\rm m\,,\; beg}$, and that 
the relevant perturbation enters the horizon in the 
matter era. 
In writing (\ref{eq:cos-per III}) we assumed sharp 
transitions $i\rightarrow k\rightarrow r\rightarrow m$. 
Within our approximation we keep $a$ continuous, but 
its derivative exhibits a jump
($H=2/\lambda^2t$ in inflation matches onto 
$H=2/\lambda^{\prime\; 2}t$ in kination). 
With the attractor solutions for 
(\ref{twoexponentials}) in (\ref{eq:cos-per I}) we obtain  
\begin{equation}
\frac{\delta \rho}{\rho}(k)= \epsilon_\delta
\frac{2}{\lambda^3}
\frac{1}{t_{1X}
}\,.
\label{deltarho}
\end{equation}
Using (\ref{eq:cos-per III})
and calculating $Ha$ in each of
the eras, we find that  
\begin{equation}
\frac{t_{1X}}{t_{\rm i\,,\; end}} =\left[\left(
\frac{a_{\rm i\,,\; end}}
{a_{\rm k\,,\; end}}\right)
^{\frac{\Lambda^{\prime\; 2}-2}{2}}
\left(\frac{a_{\rm k\,,\; end}}{a_{2X}}
\right)\right]^{\frac{\lambda^2}{2-\lambda^2}}\,,
\qquad \Lambda^\prime={\rm min}\left[
\sqrt{6},\lambda^{\prime}\right]\,,
\label{time-ratio}
\end{equation}
where, for simplicity, we take $2X$ to be in the radiation
era. The behavior of (\ref{deltarho}) is the usual one, with
an overall
amplitude set by the expansion rate at the end of inflation and,
for sufficiently small $\lambda$ (required for consistency 
of our slow-roll approximation), a fairly flat spectrum of 
perturbations over the scales relevant to structure formation.
Indeed, using the standard expression {\tt n\/}  $\approx 1+ 
\left[-3(V^\prime/V)^2+2V^{\prime\prime}/V\right]M_{\rm P}^2$
\cite{Lyth}, we obtain {\tt n\/}  $\approx 1-\lambda^2$.
Assuming the reheating scenario
of \cite{spokoiny} discussed in the previous section, we can 
express (\ref{deltarho}) in terms of the radiation 
temperature at the end of kination, $T_{\rm k\,,\; end}$.
Taking the radiation energy density
at the inflation-kination transition to be 
$\rho_{\rm i\,,\; end}=\epsilon_\rho H_{\rm i\,,\; end}^4$,
and using (\ref{endkination}) and (\ref{kination-endtemp})
allows us to express the temperature and time at the end
of inflation as follows 
\begin{equation}
T_{\rm i\,,\; end}^{\Lambda^{\prime\, 2}-2}=
\left(\frac{270}{\pi^2 g_{* \,,\; \rm i \,,\; \rm end} \epsilon_\rho}
\right)^{\frac{1}{2}}
T_{\rm k\,,\; end}^{\Lambda^{\prime\; 2}-4}\,,\qquad
H_{\rm i\,,\; end}=
\left(\frac{\pi^2 g_{* \,,\; \rm i \,,\; \rm end}}{30\epsilon_\rho}
\right)^{\frac{1}{2}} T_{\rm i\,,\; end}
=\frac{2}{\lambda^2t_{\rm i\,,\; end}}\,,
\label{eq: T and t at inflation end}
\end{equation}
so that 
\begin{equation}
\frac{\delta \rho}{\rho}(k_{2X}) \approx 
\frac{\epsilon_\delta}{\lambda}
\left(
\frac{270}{\pi^2  g_{* \,,\; \rm i \,,\; \rm end}\epsilon _\rho}
\right)^{\frac{1}{4}\left[
\frac{2}{\Lambda^{\prime\; 2}-2}+
\frac{\lambda^2}{2-\lambda^2}
 \right]}
T_{\rm k\,,\; end}^
{\frac{\Lambda^{\prime\; 2}-4}{\Lambda^{\prime\; 2}-2}}
\; T_{2X}^{-\frac{\lambda^2}{2-\lambda^2}}\,.
\label{expfluctuations}
\end{equation}
This is the desired expression for 
the amplitude of fluctuations at the comoving scale
$k_{2X}$ which re-enters the horizon when the temperature
is $T_{2X}$.

This result depends on three unknown parameters - the temperature
at the end of kination $T_{\rm k\,,\; end}$, and the parameters
$\lambda$ and $\lambda'$ in the potential. The COBE experiment
provides us with a constraint on the amplitude 
($\delta\rho/\rho(k) \approx 5\times 10^{-5}$) and the spectral
index of density perturbations ($0.7\le$ {\tt n\/} $\le 1.3$). 
The extra constraints we impose are those required by our consideration
of electroweak baryogenesis: The phase of kination must persist well
past the electroweak phase to have an important effect on the expansion
at that scale.
For example, for $T_{\rm k\,,\; end}=T_{\rm ns}$, and 
$T_{2X}=1$eV, we find $\lambda=\sqrt{0.33}=0.57$ for 
$\lambda^\prime\ge \sqrt{6}$, 
and $\lambda=\sqrt{0.11}=0.33$ 
for $\lambda^\prime= \sqrt{5}$. 
These lead to the tilt in the power spectrum
{\tt n\/} $\approx 0.7$ for $\lambda^\prime\ge \sqrt{6}$, 
and {\tt n\/} $\approx 0.9$ for 
$\lambda^\prime= \sqrt{5}$, which are consistent with
the constraint from COBE. Following the discussion in
section 4 we know that in order to create the observed BAU
at a second order or cross-over electroweak phase transition 
we need to have very close to $1/a^6$ scaling in kination,
{\it i.e.\/} $\lambda^\prime\ge \sqrt{6}$. This requirement
therefore leads in this model to a prediction of the
spectral index {\tt n\/} $\approx 0.7$. 
Using (\ref{eq: T and t at inflation end}) we can also 
compute $T_{\rm i\,,\; end}\sim H_{\rm i\,,\; end}$, 
$t_{\rm i\,,\; end}$, $\phi_{\rm i\,,\; end}$, {\it etc\/}.
In particular, for $\lambda^\prime\ge \sqrt{6}$ we have
$T_{\rm i\,,\; end}=6\times 10^7$GeV, 
and for $\lambda^\prime= \sqrt{5}$ we have
$T_{\rm i\,,\; end}=2\times 10^{11}$GeV 
(independent of $\lambda$ in inflation).

What we have illustrated with this analysis is the observational
adequacy (and even potential predictivity) of a model of this
type. The `prediction' we derived here is of course particular to
a model we have invoked in its specific form in an {\it ad hoc\/}
way. It would be of interest to study models which are derived in 
detail from a well motivated particle physics model. We will limit 
ourselves here to one qualitative comment on the sort of model which
motivated our choice (see \cite{Wetterich}) in which the parameter
$\lambda$ varies slowly (logarithmically) as a function of $\phi$.
It is not difficult to see that the constraint on the spectral index
may be much weaker when $\lambda$ interpolates between our 
limiting values: We were constrained to increase $\lambda$
as $\Lambda'$ increased (to give $1/a^6$ scaling) to make
the amplitude of perturbations sufficiently large. The effect
of having an interpolating scaling between that in inflation and
$1/a^6$ scaling will be to increase $H_{\rm i\,,\; end}$ 
(at fixed $T_{\rm i\,,\; end}$), which also increases
the density perturbations, allowing a spectral index
closer to one.
  
\subsection{Inflation-Kination in a Power-Law Potential}

In this section we consider another one field model in which
an inflation-kination-radiation domination cosmology can be 
realized. 
Again we assume the mechanism of reheating through 
particle creation in inflation 
discussed in section 5.2.
The potential we study is 
simply the non-renormalizable power-law potential 
\begin{equation}
V(\phi)=\frac{\lambda_n}{n}M_P^4\left (\frac{\phi}{M_P}\right )^n
\label{eq: potential for phi n}
\end{equation}
where $n>4$ is an integer, which is taken to be
even for stability reasons. 
As discussed in section 5.1 this potential has an
oscillatory solution in which the energy density in
the field scales as $a^{-{6n}/({n+2})}$, {\it i.e.\/}
faster than radiation for $n>4$. It also has, as we will
discuss below, ``slow-roll'' inflationary solutions
which will precede such an oscillatory phase for appropriate 
initial conditions, just as in ``chaotic'' inflation
in a $\phi^4$ potential. As mentioned in the introduction
potentials of this type have been studied in the context 
of inflation motivated by $F$ and $D$ flat directions of 
supergravity theories (see, for example, \cite{LazaridesShafi,Lyth}). 
Lower order perturbative terms are forbidden by 
a discrete symmetry imposed on the superpotential. 

There are several important differences between the 
exponential we have considered in the previous section
and this potential. There the potential was made up of two 
pieces, one with an inflationary attractor the other with
a `kinationary' attractor. Here we also have solutions of the 
two types in different regions of the potential, but the
cross-over from one to the other is dynamically determined 
rather than an independent input ({\it i.e.\/} specified
by $\phi_{\rm i\,,\; end}$). Therefore once 
$\lambda_n$ and $n$ are specified,
the potential and $T_{\rm k\,,\; end}$ and 
$\delta\rho/\rho(k)$ are completely determined.
 
The second difference between the two potentials is that
in the power-law potential kination is associated with 
an oscillatory mode, which can decay non-perturbatively 
{\it via\/} parametric resonance \cite{TraschenBrandenberger}.
Only if such decay occurs after the transition to radiation
domination, is the scenario we have envisaged possible. If
it occurs a little earlier, but still sufficiently close to 
nucleosynthesis that the reheat temperature resultant from 
the decay of the field is below the electroweak temperature,
there will be some minor effect on the predictions of
electroweak
cosmology. We will not consider this marginal case and simply 
require the stability of the oscillatory mode until after
the transition to radiation domination, which we require
below the electroweak scale in order to have an effect on
electroweak cosmology. 
Later in this section we will investigate in more 
detail the consequences of the resonant inflaton decay.

In analogy to our treatment of the exponential potential
in section 5.3, we now determine how the potential 
(\ref{eq: potential for phi n}) is constrained by 
the requirement that $\phi$ generates cosmological 
perturbations of the required magnitude for
structure formation, and that kination -- driven by the
oscillatory mode -- ends in radiation domination 
(by the mechanism of \cite{spokoiny}) 
before the temperature $T_{\rm ns}$ at which nucleosynthesis 
occurs.  We will see that these two requirements cannot be
simultaneously satisfied by a suitable choice of the two variables 
in the potential $\lambda_n$ and $n$.  

To determine the amplitude of the cosmological perturbations
we follow exactly the analysis of the previous section, 
taking the perturbations to be given by
(\ref{eq:cos-per I}). Furthermore, we will make
use of (\ref{eq:cos-per II})
and (\ref{eq:cos-per III}) to 
determine $Ha$ in inflation and 
kination.

The equations of
motion for the homogeneous mode can be written as 
\begin{eqnarray}
\ddot \phi &+&3H\dot \phi+\lambda_n\phi^{n-1} = 0
\label{eq:cos-per eoma}\\
H^2 &\equiv&\left(\frac{\dot a}{a}\right)^2 = \frac{\rho_\phi}{3}
\,,\qquad 
\rho_\phi=\frac{1}{2}\dot\phi^2+\frac{\lambda_n}{n}\phi^n
\label{eq:cos-per eomb}
\end{eqnarray}
setting $M_{P}=1$, {\it i.e.\/} 
with the rescaling
\begin{equation}
\phi\rightarrow\frac{\phi}{M_{\rm P}}
\,,\quad t\rightarrow tM_P\,,\quad
H\rightarrow  \frac{H}{M_{\rm P}}\,. 
\label{eq:cos-per M rescaling}
\end{equation}
In the standard ``slow roll'' approximation we take
\begin{equation}
\ddot \phi\ll 3H\dot \phi,\lambda_n\phi^{n-1}
\qquad \frac{1}{2}\dot\phi^2 \ll \frac{\lambda_n}{n}\phi^n \,,
\label{eq:cos-per slow roll}
\end{equation}
and find from (\ref{eq:cos-per eoma}) that
\begin{equation}
\dot \phi=-\left[\frac{n\lambda_n}{3}\right]^{\frac{1}{2}}
\phi^{\frac{n}{2}-1}\,.
\label{eq:cos-per slow roll eoma}
\end{equation}
Putting this 
expression back in (\ref{eq:cos-per slow roll}) 
it is easy to show that the ``slow roll'' 
condition is 
\begin{equation}
\phi^2\gg \frac{n^2}{6}\,.
\label{eq:cos-per slow roll II}
\end{equation}
Furthermore during the slow roll 
(\ref{eq:cos-per eomb}) gives
\begin{equation}
\frac{H}{\dot\phi}\equiv\frac{d\ln a}{d\phi}=-\frac{\phi}{n}
\label{eq:cos-per slow roll eomb}
\end{equation}
and hence 
\begin{equation}
a=\exp-\frac{\phi^2}{2n}\,,
\label{eq:cos-per a}
\end{equation}
where we chose for the integration constant
$a_0=\exp [-\phi_0^2/2n]$. 
For completeness, we also write the solution to 
(\ref{eq:cos-per slow roll eoma})
\begin{equation}
\phi=\left[\frac{12}{(n-4)^2n\lambda_n}\right]^{\frac{1}{n-4}}
t^{-\frac{2}{n-4}}\,,
\label{eq:cos-per phi}
\end{equation}
where we chose $t=0$ such that $\phi\rightarrow \infty$ as  
$t\rightarrow 0$.

We can now write the desired expression in inflation:
\begin{equation}
Ha=\left(\frac{\lambda_n}{3n}\right)^{\frac{1}{2}}
\phi^{\frac{n}{2}}\exp-\frac{\phi^2}{2n}\, \qquad
\hbox{\rm (inflation)}\,,
\label{eq:cos-per Ha}
\end{equation}
which allows us to compute the first term 
in (\ref{eq:cos-per III}).

\bigskip

Next we compute $Ha$ during  kination. 
It is convenient for this analysis to change
the field and time variables to rescaled 
variables $\tau$ and $\varphi$ given by 
\begin{equation}
dt=\left(\frac{a}{a_0}\right )^{\frac{3(n-2)}{n+2}} 
d\tau\,,\qquad
\phi=\varphi\left (\frac{a_0}{a}\right )^{\frac{6}{n+2}}
\label{eq: redefinitions of t and a}
\end{equation}
in terms of which (\ref{eq:cos-per eoma})
and (\ref{eq:cos-per eomb}) become
\begin{eqnarray}
\frac{d\varphi}{d\tau^2}
&+&\frac{6}{n+2}\left[
-\frac{1}{a}\frac{d^2 a}{d\tau^2}+\frac{n-4}{n+2}
\left(
\frac{1}{a}\frac{da}{d\tau}
\right)^2
\right]
+\lambda_n\varphi^{n-1}=0
\label{eq:cos-per rescaled eoma}
\\
{\cal H}^2 &\equiv & \left(
\frac{1}{a}\frac{da}{d\tau}\right)^2=
\frac{\rho_0(\tau)}{3}
\left(\frac{a_0}{a}\right)^{\frac{12}{n+2}}\,,
\label{eq:cos-per rescaled eomb}\\
\rho_0(\tau)&=&\frac{1}{2}\left(
\frac{d\varphi}{d\tau}-\frac{6}{n+2}{\cal H}\varphi
\right)^2\!+\frac{\lambda_n}{n}\varphi^n\,.
\nonumber
\end{eqnarray}
The approximation of a sharp transition from inflation to
kination consists in ignoring the 
explicit time dependence of $\rho_0(\tau)$, which is equivalent 
to 
\begin{equation}
{\cal H}\ll \frac{n+2}{6} 
\frac{1}{\varphi}
\frac{d\varphi}{d\tau}
\simeq \frac{n+2}{6}\omega_n\,, 
\label{eq:cos-per sharp transition approx}
\end{equation}
where $\omega_n^2\sim \lambda_n\varphi_0^{n-2}$ is the average
frequency squared of $\varphi$
(see (\ref{eq: period}) below), and $\varphi_0=\phi_0$
is the inflaton amplitude at the beginning of kination. 
Within this approximation, $d^2 a/ad\tau^2=(n-4){\cal 
H}^2/(n+2)$, 
and hence the term in the square brackets of 
(\ref{eq:cos-per rescaled eoma})  vanishes.
It is this feature of the damping term in these variables which
made their choice appropriate.
Thus all of the time dependence in 
(\ref{eq:cos-per rescaled eoma})
and~(\ref{eq:cos-per rescaled eomb}) drops out and  
the equations can be easily integrated. 
The first integral of 
(\ref{eq:cos-per rescaled eoma})
leads to
\begin{equation}
\rho_0\equiv V(\varphi_0)=\frac{1}{2} 
\left (\frac{d\varphi}{d\tau}\right )^2
+\frac{\lambda_n}{n}\varphi^n\,,
\end{equation}
which is just the energy conservation law for $\varphi$.

The oscillatory solution for $\varphi$  can be then
expressed in terms of an elliptic integral 
with the frequency
\begin{eqnarray}
\omega_n & \equiv & 
\frac{2\pi}{\tau_n}=\frac{\pi}{\sqrt{2n}c_n}
\lambda_{\rm eff}^{\frac{1}{2}}\; \phi_0\,, 
\label{eq: period}
\\
\lambda_{\rm eff} & =  & \lambda_n
\phi_0^{n-4}\,,
\qquad c_n=\int_0^1\frac{dx}{\sqrt{1-x^n}}
\nonumber
\end{eqnarray}
where $\tau_n$ is the oscillation period. 
Note that  $\omega_n$ decreases exponentially with $n$, 
when the initial amplitude $\phi_0< 1$. 
In the limit of a large $n$, $c_n$ approaches unity. 
Finally, the solution to (\ref{eq:cos-per rescaled eomb})
is 
\begin{equation}
\frac{a}{a_0}=\left[
\frac{6}{n+2}H_0\tau
\right]^{\frac{n+2}{6}}\,,
\label{eq:cos-per solution to eomb}
\end{equation}
(where we used ${\cal H}_0=H_0$)
and 
\begin{equation}
\frac{Ha}{H_0a_0}=
\left(\frac{a_0}{a}\right)^{\frac{2(n-1)}{n+2}}
\qquad\hbox{\rm (kination)}\,.
\label{eq:cos-per Ha kination}
\end{equation}
The results for radiation and matter era 
are given by setting $n=4$ and $n=2$ in
(\ref{eq:cos-per Ha kination}) respectively,
so that (\ref{eq:cos-per III}) can be recast as  
\begin{equation}
\phi_{1X}^2-\phi_{\rm i\,,\; end}^2=2n\left\{
\ln\left[\left(\frac{a_{\rm k\,,\; end}}{a_{\rm k\,,\; beg}}
\right)^{\frac{2(n-1)}{n+2}}
\frac{a_{\rm r\,,\; end}}{a_{\rm r\,,\; beg}}
\left(\frac{a_{\rm 2X}}{a_{\rm m\,,\; beg}}
\right)^{\frac{1}{2}}\right]
+\frac{n}{2}\ln\frac{\phi_{1X}}{\phi_{\rm i\,,\; end}}
\right\}\,.
\label{eq:cos-per estimate of phi_1X}
\end{equation}

The ratio of the scale factors at the beginning and the end
of kination is given by (\ref{endkination}) in section 5.2.
Using $s=6n/(n+2)$ we have (with $M_P=1$)
\begin{equation}
\left(\frac{a_{\rm k\,,\; end}}{a_{\rm k\,,\; beg}}
\right)^{2\frac{n-4}{n+2}}
=\frac{\rho_{\phi\;\rm i\; end}}{\rho_{\rm i\; end}}
= \left(
\frac{270}{\epsilon_\rho  g_{* \,,\; \rm i \,,\; \rm end} \pi^2}
\right)^{\frac{1}{2}}\frac{1}{T_{\rm i\; end}^2}\,.
\label{eq: ratio of a's}
\end{equation}
and recalling that $Ta=const$, we have
\begin{equation}
\left(\frac{a_{\rm k\,,\; end}}{a_{\rm k\,,\; beg}}
\right)^{2\frac{n-1}{n+2}}
= \left(
\frac{270}{\epsilon_\rho  g_{* \,,\; \rm i \,,\; \rm end}\pi^2}
\right)^{\frac{1}{4}}\frac{1}{T_{\rm k\; end}}
\label{eq: ratio of a's II}
\end{equation}
so that (\ref{eq:cos-per estimate of phi_1X}) becomes 
\begin{equation}
\phi_{1X}^2=\phi_{\rm i\,,\; end}^2+2n\left\{
\ln\left[\left(
\frac{270}{\epsilon_\rho  g_{* \,,\; \rm i \,,\; \rm end} \pi^2}
\right)^{\frac{1}{4}}
\frac{1}{T_{\rm r\,,\; end}}
\left(\frac{T_{\rm m\,,\; beg}}{T_{\rm 2X}}
\right)^{\frac{1}{2}}\right]
+\frac{n}{2}\ln\frac{\phi_{1X}}{\phi_{\rm i\,,\; end}}
\right\}\,.
\label{eq:cos-per estimate of phi_1X II}
\end{equation}
Taking this expression with 
\begin{equation}
\frac{\delta\rho}{\rho}(k_{1X}) =
\epsilon_\delta\,\frac{\lambda_n^{1/2}}{3^{1/2}n^{3/2}}
\;\phi_{1X}^{(n+2)/{2}}\,,
\quad \epsilon_\delta=\frac{3}{5\pi}
\label{eq: COBE normalization}
\end{equation}
specifies the  amplitude of density perturbations in the model
implicitly in terms of the parameters in the potential
$\lambda_n$ and $n$. Comparing this to the requirement of
COBE provides the first constraint on the model.
The second constraint is the requirement that 
kination ends before nucleosynthesis, {\it i.e.\/} at a temperature 
$T_{\rm k\,,\; end} > T_{\rm ns}$.
After some algebra we obtain the simple relation  
\begin{equation}
T_{\rm k\,,\; end} = \left(
\frac{270}{\epsilon_\rho  g_{* \,,\; \rm i \,,\; \rm end} \pi^2}
\right)^{\frac{1}{4}}
\left(
\frac{\epsilon_\rho}{9}
\frac{\lambda_n}{n}\phi_{\rm i\,,\; end}^n
\right)^{\frac{n-1}{n-4}}\,.
\label{eq: second constraint}
\end{equation}
\begin{figure}[htb]
\epsfxsize=5.5in
\centerline{\epsfbox{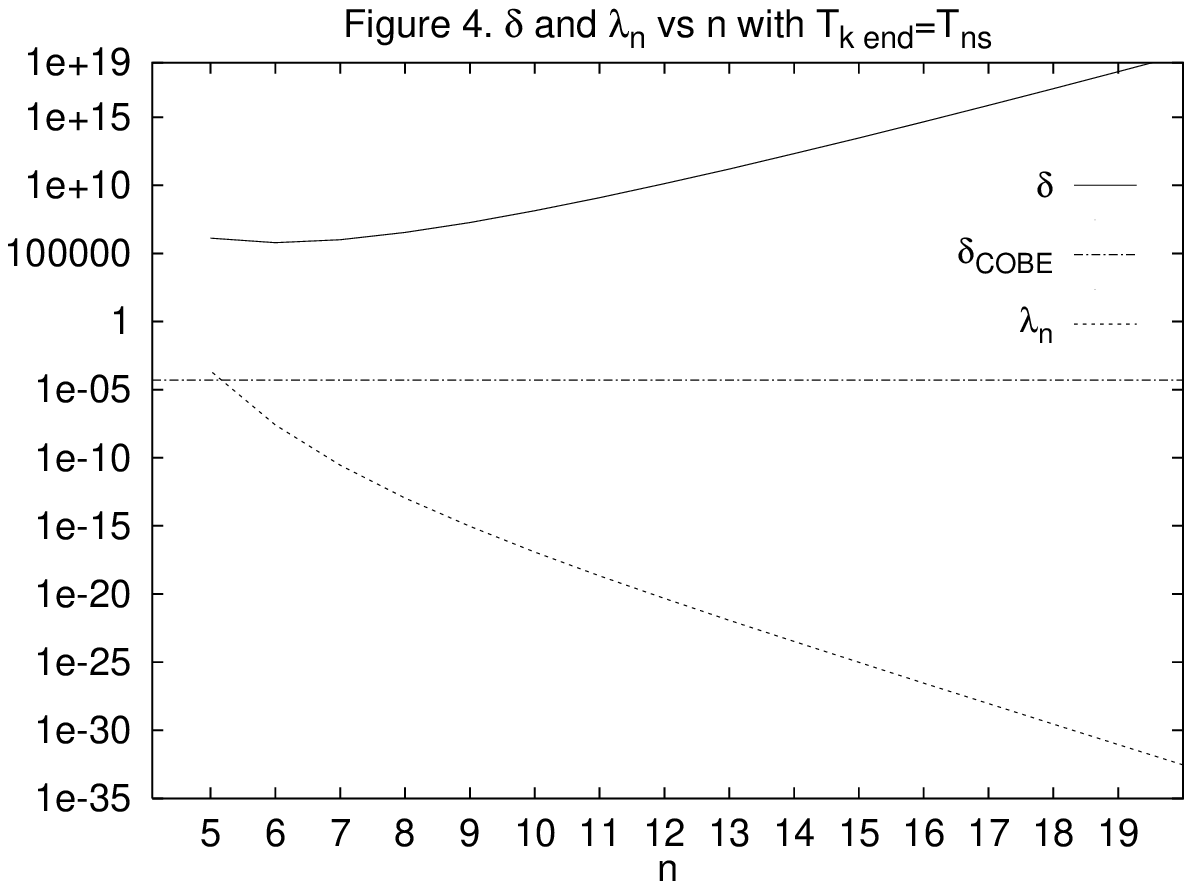}}
\end{figure}
On figure 4 we show a plot of 
$\delta\equiv \delta\rho/\rho(k_{1X})$ and of 
$\lambda_n$ as a function of $n$ with 
$T_{\rm k\,,\; end} = T_{\rm ns}$
and $T_{2X}=1$eV. It is 
clear that everywhere the amplitude is too large (by several orders of
magnitude) to satisfy the constraint from COBE. Further, 
to increase $T_{\rm k\,,\; end}$ at fixed $n$ we require
a larger $\lambda_n$, which results in a larger $\delta$.
That kination end by nucleosynthesis thus forces the energy density
in radiation at the end of inflation to be sufficiently large,
which forces $\lambda_n$ to be so large that the density
perturbations produced are too large, for any $n$. A single field
model of this type is therefore ruled out.

This analysis neglects the possible decay of the 
oscillating mode by parametric resonance into either its own 
fluctuations, or other fields that it couples to.
If such decay occurs when the energy density of
the oscillating mode has red-shifted to be sub-dominant 
relative to the radiation, then such decay is irrelevant
and the model is simply ruled out by nucleosynthesis
constraints (and the requirement that density perturbations
not be too large). If, on the other hand, it decays when
the energy density of the oscillating mode is still dominant 
over the radiation, the model may be viable if the 
decay products can thermalize with ordinary matter.
In this case, however, there will always be 
large production of entropy (with corresponding dilution of the
baryon to entropy ratio) and hence
the model is not of much interest
in the context of the question of how an increased
expansion rate at the electroweak could lead to production 
of the BAU at that scale in scenarios when it is usually
assumed to be impossible.

It is nevertheless of interest to study the non-perturbative decay
of this mode (i) to see whether such models are really ruled out
by the observations above, and (ii) because this decay channel
is relevant to a scenario where another field plays the role
of inflaton, which we will discuss briefly below. 

We wish to compare the resonant decay time $\tau_{\rm decay}$ of the
inflaton-kinaton to the time at the end of kination 
$\tau_{\rm i\,,\; end}$. As we show in the appendix, 
the field can decay either into its own fluctuations,
or to other fields that it couples to. In both cases
the decay  time $\tau_{\rm decay}$ can be approximated by
\begin{equation}
\tau_{\rm decay}\sim\frac{1}{2\omega_n\mu}
\ln\frac{n_{\rm scatt}}{n_0}\,,
\label{eq: decay time}
\end{equation}
where $n_0\simeq 1/2$ is the initial occupation 
number, and $n_{\rm scatt}$ is the `late time'
occupation number. The resonance is usually terminated
by the back reaction effects from the created particles 
\cite{ProkopecRoos}, and $n_{\rm scatt}$ can be 
estimated to be $n_{\rm scatt}\sim 
1/\lambda_{\rm eff}=
1/(\lambda_n\phi_{\rm i\,,\; end}^{n-4})$ for the 
inflaton-kinaton decay into its own fluctuations, and 
$n_{\rm scatt}\sim 1/g$ when $\phi$ couples to  
a scalar field $\zeta$, where $g$ is the coupling 
constant of the interaction term $g\zeta^2\phi^2/2$.
In the appendix we also show that $\mu=\mu_n\simeq 0.16/n$ for 
the decay into its own fluctuations, and 
for the decay into another field $\mu\simeq 0.1-0..2$ 
when $q\ge 1$, and $\mu\ll 1$ when $q\ll 1$. 

In order to get an expression for $\tau_{\rm k\,,\; end}$,
we make use of (\ref{eq:cos-per sharp transition approx}),
(\ref{eq:cos-per solution to eomb}), and 
(\ref{eq: ratio of a's}) to get 
\begin{equation}
\tau_{\rm k\,,\; end}\simeq \frac{1}{\omega_n}\left(
\frac{9}{\epsilon_\rho}
\frac{n}{\lambda_n}
\frac{1}{\phi_{\rm i\,,\; end}^n}
\right)^{\frac{3}{n-4}}\,.
\label{eq: tau kination end}
\end{equation}
Setting $\tau_{\rm k\,,\; end}=\tau_{\rm decay}$
results in the following constraint on the 
instability coefficient
\begin{equation}
\mu^{\rm constr}\simeq \frac{1}{2}
\left(
\frac{\epsilon_\rho}{9}
\frac{\lambda_n}{n}
\phi_{\rm i\,,\; end}^n
\right)^{\frac{3}{n-4}}
\ln\frac{n_{\rm scatt}}{n_0}\,,
\label{eq: mu constraint}
\end{equation}
such that when the inflaton decays only into its own
fluctuations, $n_{\rm scatt}/n_0 \sim
1/(\lambda_n\phi_{\rm i\,,\; end}^{n-4})$,
and $\mu_n^{\rm constr}\simeq 0.049$, 
significantly larger than
any $\mu_n$ for $n>4$. This implies
$\tau_{\rm decay}\simeq 
(\mu_n^{\rm constr}/\mu_n)\tau_{\rm k\,,\; end}$,
typically greater than $\tau_{\rm ns}$, 
where we took $T_{\rm k\,,\; end}\simeq 2 T_{\rm ns}$. 
When $\phi$ couples to another 
scalar field {\it via\/} a quartic term of form
$g\phi^2\zeta^2/2$, then
$n_{\rm scatt}/n_0\sim 1/g\simeq 1/(4q\lambda_{\rm eff})$,
where $q=g\phi^2/4\omega_n^2$ ({\it cf.\/} 
(\ref{eq: Hills equation II}) in the appendix), and we see 
that for $q\ge 1$, $1/g<1/\lambda_{\rm eff}$, and hence 
$\mu^{\rm constr}<0.049$. Recall that in this case $\mu$ 
is typically of order $0.1$. This means that when $q>1$,
the inflaton decays somewhat before
$\tau_{\rm k\,,\; end}$ {\it via\/} parametric 
resonance. If, on the other hand, 
$q\ll 1$, then $\mu\ll 1$, implying a late inflaton
decay, $\tau_{\rm decay}>\tau_{\rm ns}$.  

In summary, the oscillating mode in the power-law potential
decay {\it via\/} parametric resonance before nucleosynthesis if
it couples to another scalar field with $q\ge 0.1$, or equivalently 
$g\ge 10^{-20}$. When $g\ge 10^{-20}$ it is not immediately obvious 
whether the resonance shift slows down the decay or not, and,
although the discussion in the appendix suggests that it
does not, further analysis is required to establish this definitively.
In any case we can conclude that this single power-law potential
driving inflation with reheating of the type we have discussed
(as in \cite{spokoiny}), is therefore only ruled out as a viable
cosmological model for $g\le 10^{-20}$.

\subsection{Two Field Models}

Finally we consider briefly models with ordinary re-heating
(through the decay of the inflaton). In this case the
field which supports the kinetic energy dominated mode cannot
also be the inflaton, but is a second field which comes to be
important after inflation and ordinary re-heating.
Again we consider the two cases of an exponential with its
rolling mode and the power-law potential with its oscillating
mode. 

{\bf Case 1: Inflaton + Exponential Potential } 

As discussed in section 5.1 a simple exponential 
$V(\phi)=V_o e^{-\lambda \phi/M_P}$ with $\lambda$ such that
it supports, when dominant over radiation, a mode scaling 
faster than radiation,  will not come to dominate over radiation
irrespective of the initial conditions on the field. (If the
initial condition gives a scaling slower than radiation,
it will bring the system to the attractor (\ref{attractor}) 
in which the scalar field contributes at most an amount 
comparable to the radiation.) Therefore, just as in the one-field case, a
potential is required which is only asymptotically this simple
exponential. If the field lies initially in a part of the 
potential
which is flatter - flat enough to support an inflationary type 
solution - a period of inflation will occur once the radiation
cools so that its energy density is comparable to that in the
scalar field. The initial  conditions and details of the
potential will determine what the final ratio of the energy in 
the scalar field
and radiation energy is when the scalar field enters its
asymptotic kinetic mode. If this second period of inflation 
occurs
at intermediate energy scales (after `full' inflation at the GUT
scale, say) and is of a small number of e-foldings, the ratio 
will be such that the kinetic energy domination may end before
nucleosynthesis.

Such a short period of inflation at an intermediate scale occurs
in so-called `thermal' inflation \cite{Yamamoto}. A scalar field 
is trapped in a false
minimum by its coupling to the plasma and comes to dominate for a
short period until the inflation it drives cools the plasma and
allows it to roll away. In the present context all that is 
required is that the field, rather than rolling into an oscillating mode 
and decaying, roll into a potential which is asymptotically exponential.

Another way in which such a transient period of inflation which 
cools the radiation and leaves a kinetic mode dominant could 
occur is by special initial conditions in certain potentials,
{\it e.g.\/} if the field
$\phi$ with potential $V(\phi)=V_o e^{-\lambda \phi^2/M_P^2}$
sits initially close to $\phi=0$, a period of inflation will
occur when $\rho_{\rm rad}$ becomes comparable to $V_o$, the 
duration of which will depend on how close to $\phi=0$ the 
field is initially.
Without significant fine-tuning there will be a few e-foldings of
inflation followed by a period of kination.  

And lastly, we mention a variant of hybrid inflation. 
Recall that in hybrid inflation, one field ($\phi$) is 
held at the false vacuum minimum by a large 
expectation value of a second field ($\psi$),
and hence it drives inflation.
When $\psi$ becomes sufficiently small, $\phi$ rolls down to 
its true minimum. The roll in a steep potential,
{\it e.g.\/} $g\phi^2\psi^2/2+V_oe^{ -\lambda\phi}$, leads 
to kination. Since the shape of the $\psi$ potential 
determines the amplitude of density perturbations, we have 
more freedom to tune parameters of the model than in the one 
field case. In particular there is no need for variation 
of $\lambda$.

{\bf Case 2: Inflaton + Power-Law Potential}

The various examples just given can be carried over in
an obvious way to the case of a power-law potential.
The difference is parallel to that in the one-field case:
If the field is initially sub-dominant relative to the
radiation, the oscillatory mode about the minimum could
end up being dominant depending on the initial conditions.
If the field lies initially at $\phi>nM_{\rm P}$ there
will be a period of inflation which brings the field
to dominance over the radiation. For a small number of e-foldings
the radiation produced by ordinary re-heating (by decay of
the inflaton) may be dominant over any radiation produced
by particle production as in the mechanism we discussed 
in the one-field models. The constraints which we derived in
the one field model in section 5.4, and which we found could
not be satisfied, are circumvented simply because the initial
radiation density is not specified by the potential,
and the relation (\ref{eq: second constraint}) no longer
holds. For the model to work we also require that the field
decay {\it via\/} parametric resonance occur after the mode
has become sub-dominant relative to the radiation, {\it i.e.\/} 
after nucleosynthesis, which will translate into some
upper bound on the couplings of the field. The precise bound
would have to be derived in analogy to the treatment given above
for any particular model (which will specify
$\tau_{\rm k\,,\; end}$).

\section{Testability: Connections to Other Observables}

We have shown in the preceeding section that there are
simple post-inflationary cosmologies quite different from the
standard scenario in which radiation domination begins
straight after inflation. Instead a phase interpolates
between the two  in which a kinetic energy dominated mode, 
red-shifting more rapidly than radiation, drives the expansion 
of the Universe. As we discussed in the earlier sections of the
paper, if such a phase continues until after the electroweak
scale, the implications for electroweak cosmology would
potentially be extremely important. To conclude we wish
to discuss some of the broader implications of the observations
we have made. In particular, we began this paper with the
usual motivation for the consideration of electroweak baryogenesis:
It promises to follow nucleosynthesis in making firm and
observable predictions about the cosmological remnants from
an epoch at which temperatures are such that we can have
experimental knowledge of the relevant physics. It promises
to be a testable theory. What is left of this testability 
now that we have effectively turned one crucial parameter, which
is usually assumed to be known, into an unknown?

In contrast to nucleosynthesis there is in this case only 
one ``observable'' - the baryon to entropy ratio -  produced
by a calculation.  Our analysis shows that, at least
in certain particle physics theories, it will be possible
to ``fit'' the observed asymmetry by an appropriate
expansion rate. 
Does making $H$ a variable make the theory intrinsically
untestable? The answer is 
negative for two reasons.
Firstly,  it is an extremely non-trivial requirement that 
one can produce the observed BAU in any given electroweak
model, even if the expansion rate is a variable. In 
a first order phase transition, for example,  the requirement 
of various parameters - most importantly on CP violating parameters -
are typically extremely strong, independently of the expansion
rate (without the sphaleron bound). As we have seen in this paper, 
it is conceivable that it could turn out that the scalar sector
indicates an analytic cross-over 
or a weakly first order transition and CP violation sufficiently large
that the BAU could be produced if the expansion rate is 
greater by about five orders of magnitude than usually assumed.
Would we then take it this to tell us that this is the case or
that we are unlikely to be able to draw a definite conclusion
as to whether the BAU was created at this scale? 
This brings us to the second answer to the question:
The theory is truly testable only if we can find other 
observables which depend on pre-nucleosynthesis cosmology.
If we do indeed find that the BAU can be generated with a different
cosmology, that would provide a major incentive to pursue this
possibility.

One possibility is exactly the relic densities of
dark matter particles discussed in \cite{Barrow} and
\cite{KamionkowskiTurner}. The discovery of a 
candidate dark-matter particle would allow one
to determine the expansion rate at its time of decoupling
from the requirement that it be the cold dark matter in
the Universe. For example, from figure 2 in 
\cite{KamionkowskiTurner} we see that the relic density of
a Majorana neutrino changes by several orders of magnitude 
as the expansion rate at its decoupling does. If this
indicated an expansion rate different from the standard
value and consistent with that required at the electroweak
scale for generation of the observed BAU, one would 
have compelling evidence that cosmology is indeed
different. Another possible way of probing pre-nucleosynthesis
cosmology is with magnetic fields, which in certain models are 
produced at or before the electroweak scale, or at the QCD 
phase transition. This seems a more remote possibility for
a firm constraint in that the connection to observed fields
is itself very indirect. However, it is one worth bearing
in mind. For example, in the mechanism discussed in \cite{mjms}
in which fields are generated by an instability related to 
the abelian anomaly, the expansion rate enters in determining
when perturbative processes come into equilibrium. This depends
on the expansion rate, and for a significantly different expansion
rate the results would be different.  

Further there is also the possibility of probing cosmology
at the electroweak scale indirectly by its connections to
other epochs. A good example of this is in fact the scalar
field cosmology we have discussed, in particular the exponential
potential. In this case the same coherent mode which dominated
in kination can in fact play an important role again at later
times. We noted the existence of an attractor solution with
energy densities given as in (\ref{attractor}) for 
the exponential in the presence of a component of matter
or radiation. How soon this will be established after
the end of kination depends essentially how much the
ratio of kinetic to potential energy at the end of kination
differs from its value in the attractor (\ref{attractor}),
and this will vary depending on the model. In 
\cite{pgfmj} the case is treated 
in which this transient period between the two attractors 
is assumed to end well before the beginning of matter domination,
and details of the observable consequences on structure formation
in a flat CDM dominated universe are studied; in \cite{vl}
the case of entry into the attractor well into the matter era 
at a red-shift $z \sim 70$ is treated. With the forthcoming
satellite experiments which will measure the properties of the
microwave sky, such models will become testable in detail.

\section{Appendix: Resonant inflaton decay in the  
$\phi^n$ potential}

In this appendix we study the decay of the
inflaton  via parametric resonance. 
First we address the decay 
into its own fluctuations, and then we discuss how it 
decays into other fields. 
We start with writing
the evolution equation for small fluctuations around the 
inflaton zero momentum mode: 
$\phi\rightarrow \phi+\delta\phi$ in 
(\ref{eq:cos-per eoma}). After
a Fourier transform and  setting 
$M=M_{\rm P}=1$ as in (\ref{eq:cos-per M rescaling}), 
we get the following linearized mode equation
\begin{equation}
\frac{d^2\delta\phi_{\vec k}}{dt^2}+3H\frac{d\delta\phi_{\vec k}}{dt}+
\left [\left (\frac{a_0}{a}\right )^2\vec k^2+\lambda_n (n-1) 
\phi^{n-2} \right ] \delta \phi_{\vec k}=0\,.
\label{eq: mode equation expanding}
\end{equation}
Rescaling to the new variables as in 
(\ref{eq: redefinitions of t and a}) and assuming pure
kination, {\it i.e.\/}
that the field amplitude is small 
(\ref{eq:cos-per sharp transition approx}), we obtain
\begin{equation}
\frac{d^2\delta\varphi_{\vec k}}{d\tau^2}+
\left [\left (\frac{a}{a_0}\right )^{\frac{4(n-4)}{n+2}}\vec k^2
+\lambda_n (n-1) \varphi^{n-2}
 \right ] \delta \varphi_{\vec k}=0\,.
\label{eq: mode equation expanding II}
\end{equation}
Notice the scale dependence next to $\vec k^2$  which means that,  
even though the zero-mode equations are  time independent, 
the mode equations are {\it not.}
Assuming adiabatic variation of $a/a_0$,
Eq.~(\ref{eq: mode equation expanding II}) becomes 
the famous Hill equation
\begin{eqnarray}
\frac{d^2\delta\varphi_{\vec k}}{d\tau^{\prime\;2}}
+\left [A+2qf(\tau^\prime)\right ] 
\delta \varphi_{\vec k} &=& 0\,,
\nonumber\\
 A =\left (\frac{a}{a_0}\right 
)^{\frac{4(n-4)}{n+2}}
\frac{k^2}{\omega_n^2}+2q\,,\quad
q & = & 
\frac{(n-1)\lambda_n\varphi_0^{n-2}}{4\omega_n^2}\,,
\label{eq: Hills equation}
\end{eqnarray}
where $\tau^\prime=\omega_n\tau$
and $f =2(\varphi(\tau^\prime)/\varphi_0)^{n-2}-1$ is defined so
that ${\rm max}|f|=1$, $\langle f \rangle =0$, 
$f(\tau^\prime+\pi)=f(\tau^\prime)$. 
The general solution of Hill's equation
is of the form $e^{\pm\mu\tau^\prime}P(\tau^\prime)$, 
where $P(\tau^\prime+\pi)=P(\tau^\prime)$,
and it is often given as the stability chart. 
The unstable regions 
are specified by the curves of constant $\mu$ in the $\{q,A\}$
plane, and the stable regions are bounded by $\mu=0$. 
The instability chart is important since the field
decays exponentially into the modes with $\mu>0$, preferably so
to the ones with large $\mu$.
The special case of the Hill equation when $n=4$ -
the Lam\'e equation - is extensively studied in the literature
on inflaton decay
\cite{Boyanovskyetal,KofmanLindeStarobinskii},
\cite{GreeneKofmanLindeStarobinskii}. 
The instability chart exhibits unstable regions which branch
off from $A= n^2$ at $q=0$. For $q\ll A$ one is in
the narrow resonance regime, since the bands are narrow and 
$\mu\ll 1$.
The chart is symmetric under $q\rightarrow -q$. On the other 
hand, for $1<2q\le A$ the resonance bands
become broad and $\mu$ ``large.'' Typically, when 
$2q\simeq A$, $\mu$ peaks at $\sim (2\pi)^{-1}$.
In this case the field decays very fast, characteristically in 
a few dozens oscillations. 

Notice that in general for a given $n$,
$q_n=(n-1)\varphi_0^{n-2}/(4\omega_n^2)$ is specified.
Consequently, to get a rough estimate of
the decay time, it suffices to plot the one dimensional slice
$q=q_n$ of the chart. As the field decays,
$q$ stays constant, unless the backreaction of the created 
particles is large enough to change $\omega_n$. 
Numerical simulations \cite{KhlebnikovTkachev,ProkopecRoos} 
show that for the $\lambda_4\varphi^4$ potential the 
backreaction from created particles grows to about 
$\delta m^2\equiv 3\lambda\langle\delta\varphi^2\rangle
\sim \lambda_4\varphi_0^2/4$, 
changing the effective frequency 
$\omega_n^2\rightarrow\omega_n^2+\delta m^2$, and 
consequently reducing $q$ to about half of its original value
and $A\rightarrow A+\delta m^2/(\omega_n^2+\delta m^2)$.
The growth of $\delta m^2$ is terminated by narrowing the
resonance as a consequence of the 
backreaction on $A$ and $q$, and intensifying 
scatterings of the resonant particles off the zero mode, as 
a consequence of increasing resonant amplitudes. 
By then a significant portion of the field has decayed. 
We will assume that a similar scenario holds for a generic 
$\phi^n$ case.
This is plausible since, as we will see below, the instability 
charts are quite similar. 

\begin{figure}[htb]
\epsfxsize=5.5in
\centerline{\epsfbox{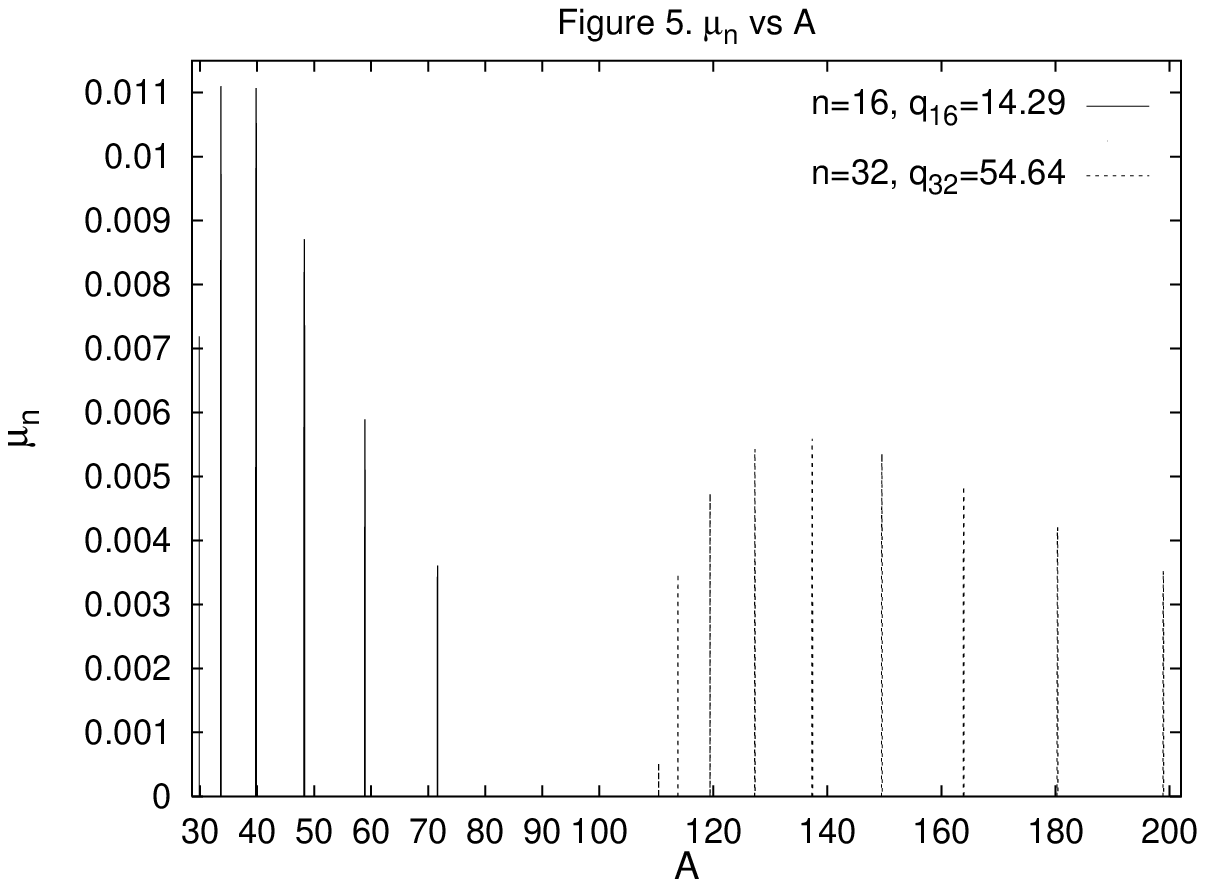}}
\end{figure}
We have evaluated numerically \cite{ThomasRoos} 
the instability charts for some of the models. The results for 
$n=4,8,16,32$ are plotted in figures 5. 
The corresponding initial values for $q_n$ 
are $1.045, 3.84, 14.29, 54.64$.  
Note that in all cases to a very good approximation the 
first instability band terminates at $A=2q$, so that 
the field decays into the second instability band. The values 
for $\mu$ are $\mu_4=0.0425$ 
\footnote{Note that this value is a bit higher 
from $\mu_4=0.0359$, the value quoted in  
\cite{GreeneKofmanLindeStarobinskii}.},
$\mu_8=0.023$, $\mu_{16}=0.011$, 
$\mu_{32}=0.0056$,
so that $\mu_n\simeq 4\mu_4/n\simeq 0.16/n$. 
For the following analysis the details of the chart
are not that important. It is sufficient to keep in mind
that for $n$ larger, $\mu_n$ decreases. 

As discussed in section 5, an inflaton that decays into 
its own fluctuations and does not couple to other fields
leads to disastrous consequences for nucleosynthesis. 
Indeed, since the inflaton decay products scale as 
radiation and 
eventually dominate the energy density of the Universe, 
and also decouple from the rest of matter, 
they behave effectively as many additional massless 
degrees of freedom, leading to a very different
expansion rate than predicted by nucleosynthesis. 
If, on the other hand, the inflaton predominately decays 
into another scalar field, which consequently thermalizes,
producing thus standard radiation and matter particles, 
nucleosynthesis may be unaffected by late inflaton-kinaton 
decay, as long as the decay occurs comfortably before
nucleosynthesis. 

Before we start discussing the inflaton decay time,
we outline the physics of the inflaton 
decay into another scalar field $\zeta$. 
We assume a standard quartic coupling $g$ 
to $\zeta$ (that itself couples to 
Standard Model particles) of the form $g\zeta^2\phi^2/2$
such that the linearized mode equations of motion are
\begin{equation}
\frac{d^2\zeta_{\vec k}}{dt^2}+3H\dot{\zeta}_k+
\left [\left (\frac{a_0}{a}\right )^2\vec k^2+
g\phi^{2} \right ]\zeta_{\vec k}=0\,.
\label{eq: fluctuation for quartic g}
\end{equation}
With the rescaling 
(\ref{eq: redefinitions of t and a})\footnote{
In this case conformal rescaling
might seem more appropriate since it would get rid of all
dependence on $a$. Nevertheless, we stick to the rescaling
in (\ref{eq: redefinitions of t and a}), in order to be able 
to make direct comparison of decay times.} and 
$\zeta=\tilde\zeta(a_0/a)^{6/(n+2)}$, this becomes
\begin{equation}
\frac{d^2\tilde\zeta_{\vec k}}{d\tau^2}+
\left [\left(
\frac{a}{a_0}\right )^{\frac{4(n-4)}{n+2}}\vec k^2
+\left(\frac{a}{a_0}\right )^{\frac{6(n-4)}{n+2}}g\varphi^2
 \right]\tilde\zeta_{\vec k}=0\,,
\label{eq: fluctuation for quartic g II}
\end{equation}
and can be recast as
\begin{eqnarray}
\frac{d^2\tilde\zeta_{\vec k}}{d\tau^{\prime\;2}}
+\left [A_\zeta+
2q_\zeta f_\zeta\right ] 
\tilde\zeta_{\vec k} &=& 0
\nonumber\\
 A_\zeta=\left(\frac{a}{a_0}\right 
)^{\frac{4(n-4)}{n+2}}
\frac{k^2}{\omega_n^2}+2q_\zeta\,,\quad
q_\zeta & = & 
\left(\frac{a}{a_0}\right )^{\frac{6(n-4)}{n+2}}
\frac{g\varphi_0^2}{4\omega_n^2}
\label{eq: Hills equation II}
\end{eqnarray}
where $\tau^\prime=\omega_n\tau$
and $f_\zeta =2(\varphi(\tau^\prime)/\varphi_0)^{2}-1$ is
defined so that 
${\rm max}|f_\zeta|=1$, $\langle f_\zeta \rangle =0$, 
$f_\zeta(\tau^\prime+\pi)=f_\zeta(\tau^\prime)$. 
As above in (\ref{eq: Hills equation}), 
in adiabatic limit, this reduces to Hill's equation.
There are however two differences: first, $q_\zeta$ can
assume a wide range of values depending on $g$, 
and, second, 
$q_\zeta$ is a (growing) function of $a$.
The corresponding Mathieu equation 
for $n=2$ is studied in great detail in the literature, and 
shows that for $q_\zeta>1$ the field decays with an average
value $\mu\sim 0.1$ \cite{KofmanLindeStarobinskyIII}. 
For the conformal case with $n=4$
a similar value for $\mu$ is obtained. 
Here we will assume that, for any $n>4$, $\mu\sim 0.1$ as well.

Now we present an estimate of the decay time
$\tau_{\rm decay}$. 
For a moment we assume that the resonance shift
does not drastically affect particle production. 
Later on we comment on the plausibility of this
assumption. The typical initial mode amplitudes are 
such that the corresponding initial 
``occupation numbers'' 
$n_k\propto \omega_k\varphi_{\vec k}\varphi_{-\vec k}$
(where $\omega_k$ is the energy of the mode $k$)
are of order $n_k^{\rm initial}\simeq n_0\simeq 1/2$. 
Since the resonant mode amplitudes grow 
as $\delta\varphi_k\propto \exp \mu_k\omega_n \tau$, one 
can estimate the field decay time as follows. 
The field decays when the energy in fluctuations become 
comparable to the energy in the zero momentum mode, 
{\it i.e.\/} when the occupation numbers 
$n_k\simeq n_0\exp 2\mu_k\omega_n \tau$ become of order
$n_k\sim 1/\lambda_{\rm eff}$
($\lambda_{\rm eff}=\lambda_n \phi_0^{n-4}$), 
implying that the decay time can be approximated by
\begin{equation}
\tau_{\rm decay}\sim \frac{1}{2\omega_n\mu_k}
\ln\frac{n_{\rm scatt}}{n_0}\,.
\label{eq:decay time}
\end{equation}
This same equation applies for the field $\phi$ 
decaying into other scalar fields. The only 
difference is that the maximum 
occupation number is in this case 
$n_{\rm scatt}\sim 1/g$. 
As a caveat to (\ref{eq:decay time}),
the authors of \cite{ProkopecRoos} showed that 
one should expect longer decay times
if the self-coupling of the second field ($\zeta$) 
is large, {\it i.e.\/} $\lambda_\zeta\gg g$. 
In this paper we do not 
dwell on these complications, and assume the couplings
such that the simple estimate (\ref{eq:decay time}) 
is valid.   

Finally we comment on how the time dependence  
of $A$ in (\ref{eq: Hills equation}) and 
(\ref{eq: Hills equation II})
can affect the decay time (\ref{eq:decay time}).
We first discuss the decay into a second scalar
field. The (comoving) resonant momentum is specified 
by $\delta A\sim \sqrt{q}$, which in 
(\ref{eq: Hills equation II}) gives 
\begin{equation}
k_{\rm res}^2\simeq \frac{\sqrt{g}\phi_0\omega_n}{2}
\left(
\frac{a_0}{a}
\right)^{\frac{n-4}{n+2}}\,.
\label{eq: resonant momentum shift}
\end{equation}
This agrees with the well known result
that for $n=4$ the resonance is static.
On the other hand, for $n=2$ (Mathieu case) the resonant
momenta grow rather fast as the Universe expands and, for
$q$ large, adiabatic approximation breaks down, leading to 
``stochastic resonance'' \cite{KofmanLindeStarobinskyIII}.
However, it turns out that the instability exponent is 
rather robust and maintains the value $\mu\sim 0.1$. 
What happens when $n>4$? In this case the resonant 
momentum decreases with time and again for $q>1$
we expect breakdown of adiabatic approximation. 
Just like in the $n=2$ case we expect $\mu$ to be 
robust and be of order $\mu\sim 0.1$. This should not in any 
case be considered as proof, but conjecture.

In the case when the field decays into its own fluctuations, 
$\delta A\sim \sqrt{q}$ gives ({\it cf.\/}
 (\ref{eq: Hills equation})) 
\begin{equation}
k_{\rm res}^2\simeq 
\frac{1}{2}\sqrt{(n-1)\lambda_n}\;\phi_0^{\frac{n-2}{2}}\omega_n
\left(
\frac{a_0}{a}
\right)^{\frac{4(n-4)}{n+2}}
\label{eq: resonant momentum shift II}
\end{equation}
which again leads to a shift in the resonant momentum.
Unfortunately, the conformal case ($n=4$), in which 
the resonance is static,
is the only case studied in the literature, so we cannot 
make any analogy as we did in the former case. 
Since in this case 
the resonance is rather narrow, the resonant momentum 
redshift may significantly slow down the decay. 
An implication would be that the effective $\mu$ decreases,
leading to somewhat less stringent bounds on 
$\lambda_n$ and $n$ than indicated in 
(\ref{eq: mu constraint}).

\section{Acknowledgements}

We thank C. Korthals-Altes and M. Shaposhnikov for useful
conversations. TP acknowledges funding from the U.S. NSF, and
hospitality of Columbia University and Cornell's LNS where part of
this work was done. MJ is supported by an Irish Government (Department
of Education) post-doctoral fellowship.  We are grateful to Thomas
Roos who provided code for determination of the instability
bands.

\end{document}